\documentclass[aps, prd, twocolumn, superscriptaddress, nofootinbib]{revtex4}
\usepackage{graphicx}
\usepackage{dcolumn}
\usepackage{amssymb,amsmath,amsthm,mathrsfs}
\usepackage{epstopdf}
\usepackage{lipsum}
\usepackage{hyperref}

\begin{document}

\newcommand{\Eq}[1]{Eq. \ref{eqn:#1}}
\newcommand{\Fig}[1]{Fig. \ref{fig:#1}}
\newcommand{\Sec}[1]{Sec. \ref{sec:#1}}

\newcommand{\PHI}{\phi}
\newcommand{\vect}[1]{\mathbf{#1}}
\newcommand{\Del}{\nabla}
\newcommand{\unit}[1]{\mathrm{#1}}
\newcommand{\x}{\vect{x}}
\newcommand{\ScS}{\scriptstyle}
\newcommand{\ScScS}{\scriptscriptstyle}
\newcommand{\xplus}[1]{\vect{x}\!\ScScS{+}\!\ScS\vect{#1}}
\newcommand{\xminus}[1]{\vect{x}\!\ScScS{-}\!\ScS\vect{#1}}
\newcommand{\diff}{\mathrm{d}}

\newcommand{\be}{\begin{equation}}
\newcommand{\ee}{\end{equation}}
\newcommand{\bea}{\begin{eqnarray}}
\newcommand{\eea}{\end{eqnarray}}
\newcommand{\vu}{{\mathbf u}}
\newcommand{\ve}{{\mathbf e}}
\newcommand{\vk}{{\mathbf k}}
\newcommand{\vx}{{\mathbf x}}
\newcommand{\vy}{{\mathbf y}}
\newcommand{\bx}{{\bf x}}
\newcommand{\bk}{{\bf k}}
\newcommand{\br}{{\bf r}}

\newcommand{\uden}{\underset{\widetilde{}}}
\newcommand{\den}{\overset{\widetilde{}}}
\newcommand{\denep}{\underset{\widetilde{}}{\epsilon}}

\newcommand{\nn}{\nonumber \\}
\newcommand{\dd}{\diff}
\newcommand{\fr}{\frac}
\newcommand{\del}{\partial}
\newcommand{\eps}{\epsilon}
\newcommand\CS{\mathcal{C}}

\def\be{\begin{equation}}
\def\ee{\end{equation}}
\def\ben{\begin{equation*}}
\def\een{\end{equation*}}
\def\bea{\begin{eqnarray}}
\def\eea{\end{eqnarray}}
\def\bal{\begin{align}}
\def\eal{\end{align}}

\def\TT{{\rm TT}}
\def\GW{{_{\rm GW}}}

\title{The Decay of the Standard Model Higgs after Inflation}

\newcommand{\addressIFT}{Instituto de F\'isica Te\'orica UAM-CSIC, Universidad Auton\'oma de Madrid, Cantoblanco, 28049 Madrid, Spain}
\newcommand{\addressCERN}{CERN, Theory Division, 1211 Geneva, Switzerland.}
\newcommand{\addressGeneva}{D\'epartement de Physique Th\'eorique and Center for Astroparticle Physics, Universit\'e de Gen\`eve, 24 quai Ernest Ansermet, CH--1211 Gen\`eve 4, Switzerland.}

\author{Daniel G. Figueroa}
\affiliation{\addressCERN}
\affiliation{\addressGeneva}

\author{Juan Garc\'ia-Bellido}
\affiliation{\addressIFT}

\author{Francisco Torrent\'i\,}
\affiliation{\addressCERN}
\affiliation{\addressIFT}

\date{\today}

\begin{abstract}
We study the nonperturbative dynamics of the standard model (SM) after inflation, in the regime where the SM is decoupled from (or weakly coupled to) the inflationary sector. We use classical lattice simulations in an expanding box in (3+1) dimensions, modeling the SM gauge interactions with both global and Abelian-Higgs analogue scenarios. We consider different postinflationary expansion rates. During inflation, the Higgs forms a condensate, which starts oscillating soon after inflation ends. Via nonperturbative effects, the oscillations lead to a fast decay of the Higgs into the SM species, transferring most of the energy into $Z$ and $W^{\pm}$ bosons. All species are initially excited far away from equilibrium, but their interactions lead them into a stationary stage, with exact equipartition among the different energy components. From there on the system eventually reaches equilibrium. We have characterized in detail, in the different expansion histories considered, the evolution of the Higgs and of its dominant decay products, until equipartition is established. We provide a useful mapping between simulations with different parameters, from which we derive a master formula for the Higgs decay time as a function of the coupling constants, Higgs initial amplitude and postinflationary expansion rate. 
\end{abstract}

\keywords{cosmology, nonperturbative effects, Standard Model Higgs, early Universe}

\maketitle

\section{Introduction}
\label{sec:I}

Inflation, an early period of accelerated expansion, is the leading framework to explain the initial conditions of the Universe. The concrete particle physics realization of inflation has eluded any clear identification so far, so the inflationary dynamics is often described in terms of a scalar field, the inflaton, with a vacuum-like energy density. Furthermore, the confirmed discovery of the standard model (SM) Higgs in the {\it Large Hadron Collider} (LHC)~\cite{ATLAS2012,CMS2012} has initiated the quest for understanding the cosmological implications of the Higgs, and in particular, its possible role during and after inflation. Intriguingly, the SM Higgs could play the role of the inflaton, if a nonminimal coupling to gravity is introduced, appropriately fixed to fit the observed amplitude of the cosmological perturbations~\cite{Bezrukov:2007ep}. This model, known as Higgs-inflation, constitutes undoubtedly one of the most attractive and economical scenarios for realizing inflation, though it is not free of criticism~\cite{Barbon:2009ya,Burgess:2010zq}; see, however, Ref.~\cite{Bezrukov:2010jz}.

In this paper, we will rather explore a different route for the possible role of the Higgs during and after inflation. 
We will merely assume that inflation was driven by a very slowly evolving energy density, without specifying the nature of the field responsible for it. Inflation can then be seen effectively as a quasi-de Sitter background with a slowly changing Hubble rate. 
We will assume that the SM Higgs is not coupled directly to the inflationary sector~\cite{Espinosa:2007qp,DeSimone:2012qr,Enqvist:2013kaa,Enqvist:2014tta}. Under these circumstances, the Higgs behaves simply as a spectator field living in a (quasi-)de Sitter background, with the effective potential of the Higgs ultimately dictating its behavior. Let us note that even if there is  no direct coupling, it is likely that effective operators will connect the Higgs with the inflaton, via some possible mediator field(s). Moreover, the need to reheat the Universe after inflation requires somehow the presence of such coupling, though there is no particular constraint on this. As we will see, the Higgs decays very fast after inflation into all SM species, so one can safely assume that the effect of an inflaton-Higgs coupling is negligible, unless that coupling is significantly large. Therefore, even if such coupling is present, we will consider it weak enough so that any Higgs-inflaton interaction does not affect the dynamics of the latter during or after inflation. 

\begin{figure}[t]
    \begin{center}
        \includegraphics[width=8cm]{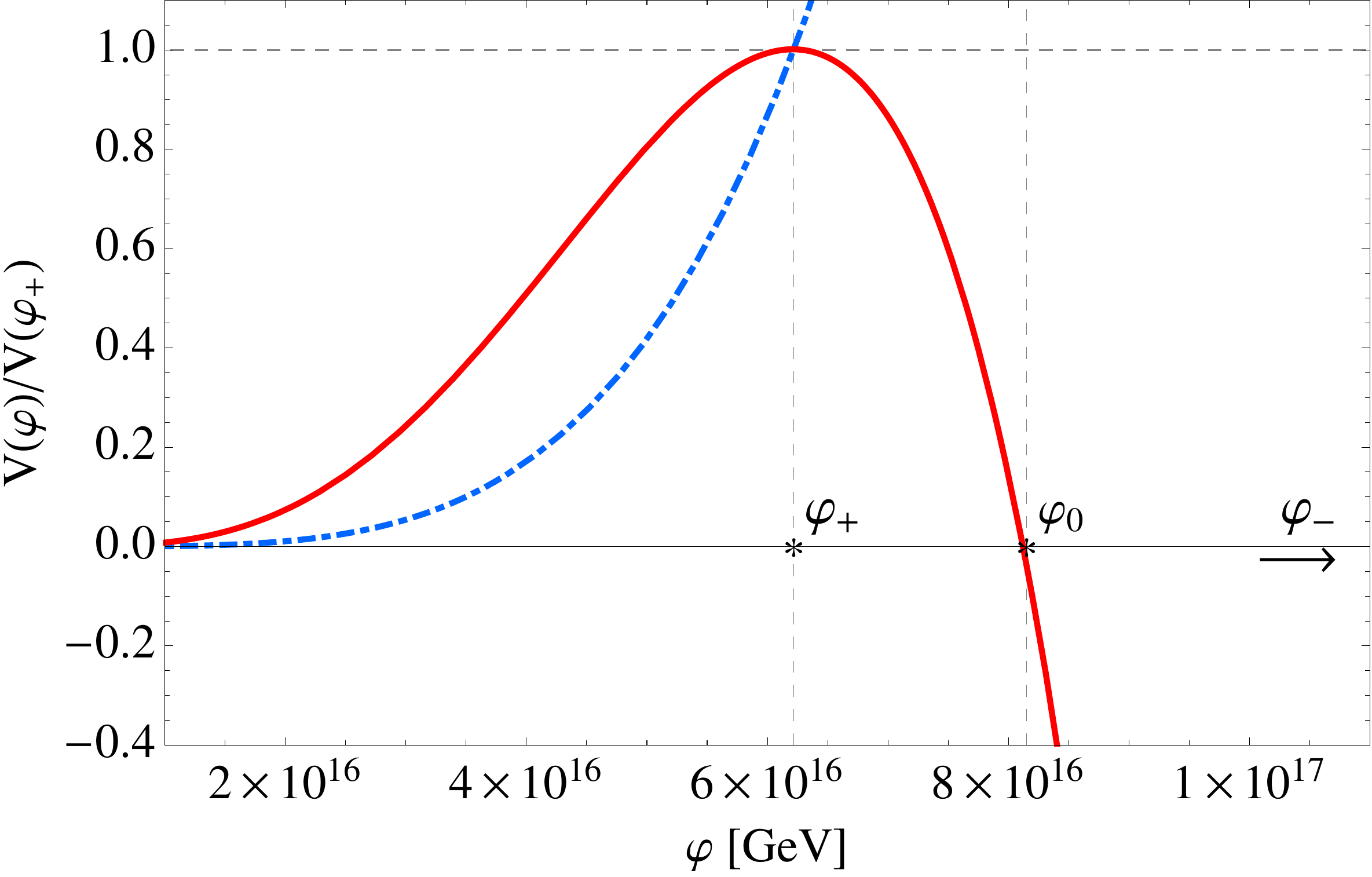}
    \end{center}
    \caption{Improved renormalized Higgs potential at next-to-next-to-leading order (red continuous line) computed for $\alpha_s = 0.1184$, $m_H = 125.5$ GeV, and $m_t = 171.2$ GeV ($\lesssim 2\sigma$ below CMS central value). Also shown, for comparison, the function ${1\over 4}\lambda_+\varphi^4$ (blue dashed line), where $\lambda_+ \equiv \lambda(\mu_+) \simeq 3\times 10^{-5}$.
    }\label{fig:Figure0}
\end{figure}

The improved renormalized Higgs potential has been computed at next-to-next-to-leading order~\cite{Degrassi:2012ry,Bezrukov:2012sa}. It is characterized by the running of the Higgs self-coupling $\lambda(\mu)$, which decreases with energy ${d\lambda/d\mu} < 0$, and becomes negative above a certain critical scale $\mu_0$, $\lambda(\mu \geq \mu_0) \leq 0$. Equivalently, the effective potential develops a barrier at large field amplitudes, reaching a maximum height at some scale $\mu_{+} < \mu_0$, so that at higher energies $\mu > \mu_{+}$ the effective potential goes down, crosses zero at $\mu = \mu_0$ and becomes rapidly negative, possibly reaching a (negative) minimum at some scale $\mu_{-}  \gg \mu_0$. This can be seen in Fig.~\ref{fig:Figure0}. These scales depend sensitively on the Higgs mass $m_H$, the strong coupling constant $\alpha_s$, and especially on the top Yukawa coupling $y_t$. For the SM central values, $\alpha_s = 0.1184$, $m_H = 125.5$ GeV, and the most recent measurement of the top quark mass by CMS, $m_t = 172.38$ GeV~\cite{CMS-PAS-TOP-14-015}, one finds $\mu_+  \simeq 2\times 10^{11}$ GeV and $\mu_0 \simeq 3 \times 10^{11}$ GeV. If one takes the world average top quark mass $m_t = 173.36$ GeV~\cite{ATLAS:2014wva}, then $\mu_+, \mu_0$ are reduced by a factor $\sim 1/30$. However, by considering a value of $y_t$ merely 2-3 sigma smaller than its central one, we can push the critical scales to $\mu_+, \mu_0 \gtrsim 5\times 10^{16}$ GeV. Besides, minimal additions to the SM such as a scalar singlet coupled to the Higgs~\cite{EliasMiro:2012ay,Ballesteros:2015iua}, or even a small nonminimal coupling of the Higgs to gravity~\cite{Herranen:2014cua}, can also modify the running of $\lambda(\mu)$ and stabilize the effective potential. In such a case, the Higgs self-coupling may remain always positive $\lambda(\mu) > 0$.\vspace*{0.25cm} 

We will consider that the Higgs amplitude during inflation remains always in the 'safe' side of the effective potential, where $\lambda(\mu)$ is positive. This can be guaranteed if $\mu_+$ is sufficiently large (compared to the inflationary scale), or alternatively, if beyond-the-SM physics stabilizes the potential at high energies. With these considerations, the Higgs fluctuates during inflation, like any light degree of freedom. The fluctuations then pile up at super-Hubble scales, creating a condensate~\cite{Starobinsky1986,Linde:2005ht}. The amplitude of the Higgs condensate, however, will not grow unbounded with the numbers of e-folds, as it happens in the case of a massless free field. On the contrary, the Higgs self-interactions provide an effective (sub-Hubble) mass to the fluctuations, which eventually saturates the growth of the condensate amplitude~\cite{StarobinskyYokoyama1994}. The distribution of the Higgs amplitude at super-Hubble scales enters very fast, within a few e-folds, into a self-similar regime, which continues until the end of inflation. The Higgs condensate acquires this way a fixed physical correlation length (exponentially larger than the Hubble radius) and a fairly large amplitude. This will set up the initial condition for the behavior of the Higgs after inflation. \vspace*{0.25cm}

Notice, however, that our analysis will not really depend on the condition $\lambda(\mu) > 0$ during inflation. The possible implication of the Higgs self-coupling becoming negative, $\lambda(\mu) < 0$, during inflation has indeed been analyzed in detail in Refs.~\cite{Espinosa:2007qp, Kobakhidze:2013tn, Enqvist:2014bua, Hook:2014uia, Kobakhidze:2014xda, Spencer-Smith:2014woa, Kamada:2014ufa, Shkerin:2015exa, Espinosa:2015qea}. In this case, if the scale of inflation is sufficiently high, the second minimum can be reached
and anti-de Sitter bubbles are formed during inflation. The consequence of this does not need to be catastrophic, but rather indicative that either the condition that our Universe is in the electroweak (EW) vacuum is something very special (very improbable), or either some new physics beyond the SM is necessary to stabilize the Higgs potential. The crucial ingredient for our analysis is, therefore, not that the Higgs self-coupling $\lambda(\mu)$ remains positive during inflation, but the fact that the Higgs develops a vacuum expectation value (VEV) during inflation much larger than the electroweak (EW) scale $\sim \mathcal{O}(10^2)$ GeV. The way such condensate is attained is mostly irrelevant. The case $\lambda(\mu) > 0$ all through inflation provides simply a reference case, where the formation of a large VEV during inflation is unavoidable, and its typical amplitude can be easily calculated. 

In this paper we investigate in detail the Higgs's decay into its most energetically dominant decay products, the SM electroweak gauge bosons, during the immediate stages following the end of inflation. Our work represents a complementary analysis to that of Enqvist et al.~\cite{Enqvist:2013kaa,Enqvist:2014tta}, where analytical techniques were employed to study the same problem. We use instead lattice simulations in an expanding box in (3+1) dimensions, modeling the SM interactions with global and Abelian-Higgs setups, which go beyond the assumptions behind any analytical calculation. Besides this, we also consider different Higgs initial amplitudes and postinflationary expansion rates. 

The paper is organized in such a way that we  increase progressively the complexity of the different approaches used to describe the dynamics of the system, approximating the structure of the SM interactions better and better at each new step. In Section~\ref{sec:II} we first present a brief analysis of the behavior of the Higgs after inflation, ignoring its coupling to the rest of the SM species. In Section~\ref{sec:III} we switch on the coupling to the SM fields, but ignore the gauge nature of the interactions. We obtain analytical estimates for a later comparison with numerical simulations. In Section~\ref{sec:iv} we present the first set of lattice simulations, where we follow the Higgs and its decay products, yet under the assumption that the gauge nature of the SM interactions can be neglected. In Section~\ref{sec:v}, we finally incorporate gauge interactions into the simulations, by modeling the SM with an Abelian-Higgs setup. Although this is just an approximation to the gauge structure of the SM, the non-Abelian nonlinearities can arguably be neglected. Therefore, the outcome of these simulations represents the most precise calculation of the dynamics of the SM after inflation, fully incorporating the nonlinear and nonperturbative effects of the SM, while considering the gauge nature of its interactions. In Section~\ref{sec:vi} we present a useful mapping between simulations with different parameters, from which we obtain a characterization of the Higgs decay width as a function of the coupling constants, initial Higgs amplitude, and postinflationary expansion rate. In Section~\ref{sec:vii} we conclude and discuss some of the possible cosmological implications of our results.

All through the text $\hbar = c = 1$, and $m_p \simeq 2.44\times10^{18}$ GeV is the reduced Planck mass. We take the flat Friedmann-Robertson-Walker (FRW) line element $ds^2 = a^2(t) (-dt^2 + dx^{i} dx^{i})$ for the background metric, with $a(t)$ the scale factor and $t$ the conformal time.

\section{Higgs Oscillations after Inflation}
\label{sec:II}

Let us characterize inflation as a de Sitter period with Hubble rate $H_{*} \gg M_{\rm EW}$, where $M_{\rm EW} \sim \mathcal{O}(10^2)$ GeV is the EW scale. In reality, we know that inflation cannot be exactly a de Sitter background, since inflation must end after a finite number of e-folds. The curvature perturbation spectral index $n_s = 0.968 \pm 0.006$~\cite{Planck2015}, constrained by Planck to be smaller than unity at more than 7 sigma, is actually interpreted as an indication of the quasi-de Sitter nature of Inflation. For our purposes, however, the distinction between de Sitter and quasi-de Sitter is irrelevant. 

With a gauge transformation, the SM Higgs doublet can be parametrized in the unitary gauge by a single scalar real degree of freedom, $\Phi = \varphi/\sqrt{2}$. The renormalized improved potential for large-field amplitudes $|\varphi| \gg M_{\rm EW}$ is just given by the quartic part
\begin{eqnarray}
    V(\varphi) = {\lambda(\mu)\over4}\varphi^4,
\end{eqnarray}
with $\lambda(\mu)$ the Higgs self-coupling at the renormalization scale $\mu = \varphi$. Radiative corrections to the potential are encoded in the running of $\lambda(\mu)$, which to date has been computed to three loops when the Higgs is minimally coupled to gravity~\cite{Degrassi:2012ry,Bezrukov:2012sa}. 

We ignore the nature of the sector responsible for inflation, so a priori there is no need for the Higgs to be coupled directly\footnote{Here we refer to a particle physics coupling, not the gravitational coupling.} to the inflationary sector. We will just consider that the Higgs field simply 'lives' on the $de~Sitter$ background, playing no dynamical role during inflation, and behaving simply as a spectator field~\cite{Espinosa:2007qp,DeSimone:2012qr,Enqvist:2013kaa}. As mentioned in Section~\ref{sec:I}, the need to reheat the Universe after inflation requires somehow a coupling between the SM and the inflationary sector, though there is no particular constraint on this. Therefore, effective operators are expected to connect the Higgs with the inflaton when integrating out some possible mediator field(s). However, as we will show in the following sections, the Higgs decays very fast after inflation into all SM species. Hence, even if there is an inflaton-Higgs effective coupling, we will assume in practice that its effect is negligible, with the possible Higgs-inflaton interactions not affecting the Higgs dynamics during or after inflation. 

Under these circumstances, the Higgs amplitude during inflation 'performs' a random walk at superhorizon scales, reaching very quickly, within few e-folds, the equilibrium distribution~\cite{StarobinskyYokoyama1994}
\begin{eqnarray}\label{eq:ProbEQ}
    P_{\rm eq}(\varphi) = \mathcal{N}\exp\left\lbrace{-{2\pi^2 \over 3}{\lambda\varphi^4\over H_{*}^4}}\right\rbrace,~~~
    \mathcal{N} \equiv  {2^{1\over 4}\lambda^{1\over 4}\sqrt{4\pi}\over 3^{1\over 4}\Gamma({1\over 4})H_*} \ .
\end{eqnarray}
The correlation length, i.e.~the physical scale above which the Higgs amplitude $\varphi$ fluctuates according to Eq.~(\ref{eq:ProbEQ}), is given by $l_* \approx \exp\lbrace3.8/\sqrt{\lambda}\rbrace\,H_*^{-1}$~\cite{StarobinskyYokoyama1994}, so it is exponentially larger than the inflationary Hubble radius $H_*^{-1}$. After the equilibrium distribution is reached at some point during inflation, the correlation length remains invariant until the end of the exponential expansion. Hence, immediately after inflation, the Higgs amplitude $\varphi$ can be safely considered homogeneous within any volume of size $l \ll l_*$. The Higgs amplitude varies randomly according to Eq.~(\ref{eq:ProbEQ}), but only when we compare it at scales $l \gg l_*$, much larger than the correlation length. 
For convenience, we define 
\begin{eqnarray}
    \alpha \equiv \lambda^{1\over 4}{\varphi / H_*}\,,
\end{eqnarray}
so that the distribution probability expressed over this dimensionless variable reads $P_{\rm eq} \propto \exp[ -2\pi^2\alpha^4/3]$.  
The roots of the moments of $P_{\rm eq}$ are then given by
\begin{eqnarray}\label{eq:cumulants}
    c_n \equiv  \left\langle \alpha^n \right\rangle^{1/n} = \lambda^{1/4} {\left\langle (\varphi/H_*)^n \right\rangle}^{1/n}\,,
\end{eqnarray}
where $\left\langle ... \right\rangle$ denotes statistical average over the equilibrium distribution in Eq.~(\ref{eq:ProbEQ}). One finds 
$$c_2 \simeq 0.363,~~~ c_4 \simeq 0.442,~~~c_6 \simeq 0.497, ~~~....$$
whereas $c_1 = c_3 = c_5 = ... = 0$. We find $\alpha \in$ 
$[0.001,1]$ with $99.8\%$ probability, whereas $\alpha < 0.001$ holds only with a $0.17\%$ probability, and $\alpha > 1$ is yet further suppressed with a $0.03\%$ probability. 

A typical amplitude of the Higgs at the end of inflation is given by its {\it root mean square} (rms) value 
\begin{eqnarray}
    \varphi_{\rm rms} = c_2\,H_{*}/\lambda^{1/4}\simeq 1.15\,H_*/\lambda_{001}^{1/4},
\end{eqnarray}
where we have defined the self-coupling normalized to a canonical value $\lambda_c \equiv 0.01$,
\begin{eqnarray}
    \lambda_{001} \equiv {\lambda/\lambda_c} \equiv 100\lambda \ .
\end{eqnarray}
As we explain later, reasonable values of $\lambda$ are taken within the interval $[10^{-2},10^{-5}]$. Hence, for $\lambda = 10^{-2}, 10^{-3}, 10^{-4}, 10^{-5}$ ($\lambda_{001}^{1/4} = 1, 0.562, 0.326, 0.178$), we conclude that the typical Higgs amplitudes are of the order $\varphi_{\rm rms} \sim H_*$, independently of the value of $\lambda$. 
We do not know the actual value of $\varphi$ within the 'progenitor' patch from which our visible Universe grew up. Actually, we do not know the value of the Higgs condensate within any patch, we just know that typically $\varphi/H_* \sim \mathcal{O}(0.01)-\mathcal{O}(1)$ for reasonable values of $\lambda$. That means that just after inflation, within any patch of size $l \lesssim l_*$, the Higgs has a nonzero amplitude that could be really large, almost as big as $H_*$  depending on its realization. The most updated upper bound for the inflationary Hubble rate is~\cite{Planck2015} 
$$H_* \leq H_*^{\rm (max)} \simeq 8.4\times10^{13} {\rm GeV}\,,$$
so the Higgs amplitude at the end of inflation could be ranging around $|\varphi| \lesssim (10^{12}-10^{14})$ GeV $\times (H_*/H_*^{\rm (max)})$. 

In order to analyze the dynamics of the Higgs after inflation, it is necessary first to fix the postinflationary expansion rate. Since we do not specify the nature of the inflationary sector here, we can parametrize the scale factor after inflation like
\begin{eqnarray}\label{eq:ExpRate}
    a(t) = a_*\left(1+{1\over p}a_*H_*(t-t_*)\right)^{p}\,,~~~ p \equiv {2\over(1+3w)}
\end{eqnarray}
with $a_*$ being the scale factor at the initial time $t = t_*$ (i.e. at the end of inflation), and $w$ being the equation of state of the Universe characterizing the expansion rate of the period following inflation. For instance, if the inflationary sector is described by an inflaton with a quadratic potential, the Universe expands as in a matter-domination (MD) regime during the inflaton oscillations following the end of inflation, so $w = 0$ and $p = 2$. If it is an inflaton with a quartic potential, the Universe expands as in a radiation-domination (RD) regime, with $w = 1/3$ and $p = 1$. Since we do not really specify the inflaton sector, we are also free to consider other possibilities, including more `exotic' scenarios where the background energy density decays faster than relativistic degrees of freedom, i.e.~with $w > 1/3$ and $p < 1$. The paradigmatic example of this is a kination-domination (KD) regime, with $w = 1$ and $p = 1/2$, obtained when an abrupt drop of the inflaton potential takes place at the end of inflation, transferring all the energy into kinetic degrees of freedom~\cite{Spokoiny:1993kt,Joyce:1996cp}.

\subsection{Higgs oscillations}

The amplitude of the Higgs after the end of inflation is nonzero, and given that the Higgs potential is symmetric, the Higgs condensate is 'forced' to oscillate around its minimum at $\varphi = 0$. The larger the Higgs amplitude, the sooner the oscillations will start after the end of inflation. The EOM (equation of motion) of the Higgs just after inflation is
\begin{eqnarray}\label{eq:HiggsEOM}
    \ddot\varphi + 2\mathcal{H}\dot\varphi + a^2\lambda \varphi^3 = \nabla^2\varphi\,,
\end{eqnarray}
where $^\cdot \equiv d/dt$, and $\mathcal{H}$ is the comoving Hubble rate, given by
\begin{eqnarray}
    \mathcal{H}(t) \equiv {\dot a\over a} = {a_*H_*\over [1+p^{-1}a_*H_*(t-t_*)]} \equiv {a_*H_*\over\sqrt[p]{a(t)/a_*}} \ .
\end{eqnarray}
We will consider the evolution of the Higgs in an arbitrary patch, inside which its amplitude [randomly drawn from Eq.~(\ref{eq:ProbEQ})] can be regarded as homogeneous. The correlation length is exponentially bigger compared to the Hubble radius, $l_* \simeq e^{38.2/\lambda_{001}^{1/2}}H_*^{-1} \gg H_*^{-1}$, so if we just follow the Higgs within a causal domain of initial size $l \sim 1/H_* \ll l_*$, then we can drop the Laplacian term on the $rhs$ of Eq.~(\ref{eq:HiggsEOM}). The only scale in the problem is therefore $a_*H_*$, so it is convenient to define a dimensionless conformal time $z \equiv a_*H_*(t-t_*)$. 
We can then write the scale factor as $a(z) = a_*(1+p^{-1}z)^{p}$. Introducing the variable
\begin{eqnarray}
    h(z) \equiv {a\over a_*}{\varphi\over\varphi_*}\,,
\end{eqnarray}
with $\varphi_*$ being the initial amplitude of the Higgs, we can rewrite the Higgs EOM in a more convenient form as
\begin{eqnarray}\label{eq:HiggsEOMtransformed}
    h'' + \beta^2 h^3 = {a''\over a}h\,,~~~~~~\beta^2 \equiv {\lambda\varphi_*^2\over H_*^2} = \sqrt{\lambda}\alpha^2\,,
\end{eqnarray}
where $' \equiv d/dz$, and $\beta$ characterizes the frequency of oscillations. The term on the rhs scales as $a''/a \sim (a_*/a)^{2/p}$, and hence it becomes irrelevant very soon, since it decays as $a''/a \sim z^{-2/p} \ll 1$.

The initial condition for the Higgs amplitude in the new variables is, by construction, 
\be h_* \equiv 1 \ .\ee
The initial condition for the derivative $h_*' \equiv dh_*/dz = 1+{\dot\varphi(t_*)/(a_*H_*\varphi_*)}$, taking into account that the Higgs was in slow roll during inflation [i.e.~$\dot\varphi(t_*) = -{\lambda a_*^2\varphi_*^3/ 2H_*}$], reads out
\begin{eqnarray}
    h'_* \equiv 1-{\beta^2\over2}\,.
\end{eqnarray}

\begin{figure}[t]
    \begin{center}
        \includegraphics[width=8cm]{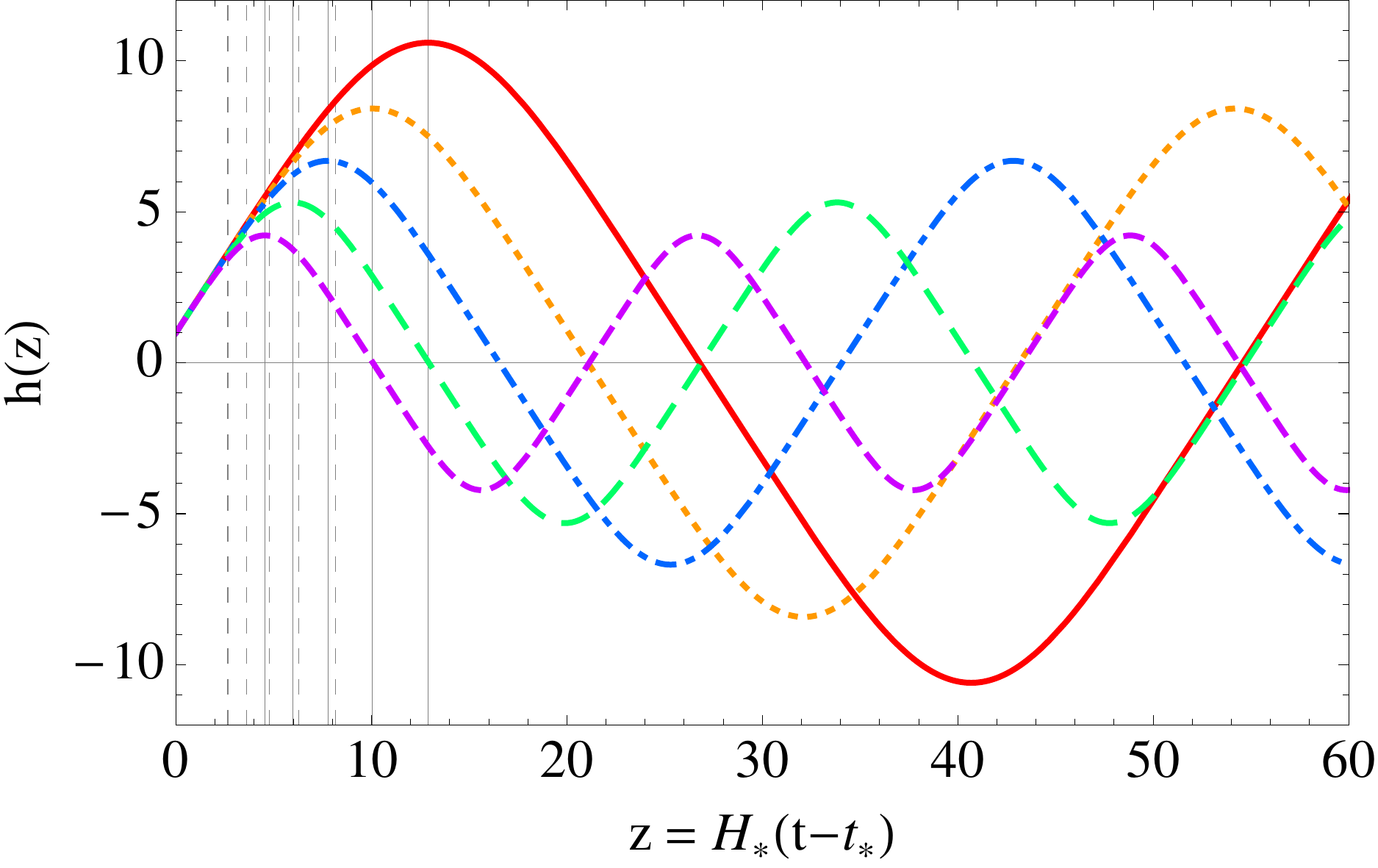}\vspace*{5mm}
        \includegraphics[width=8cm]{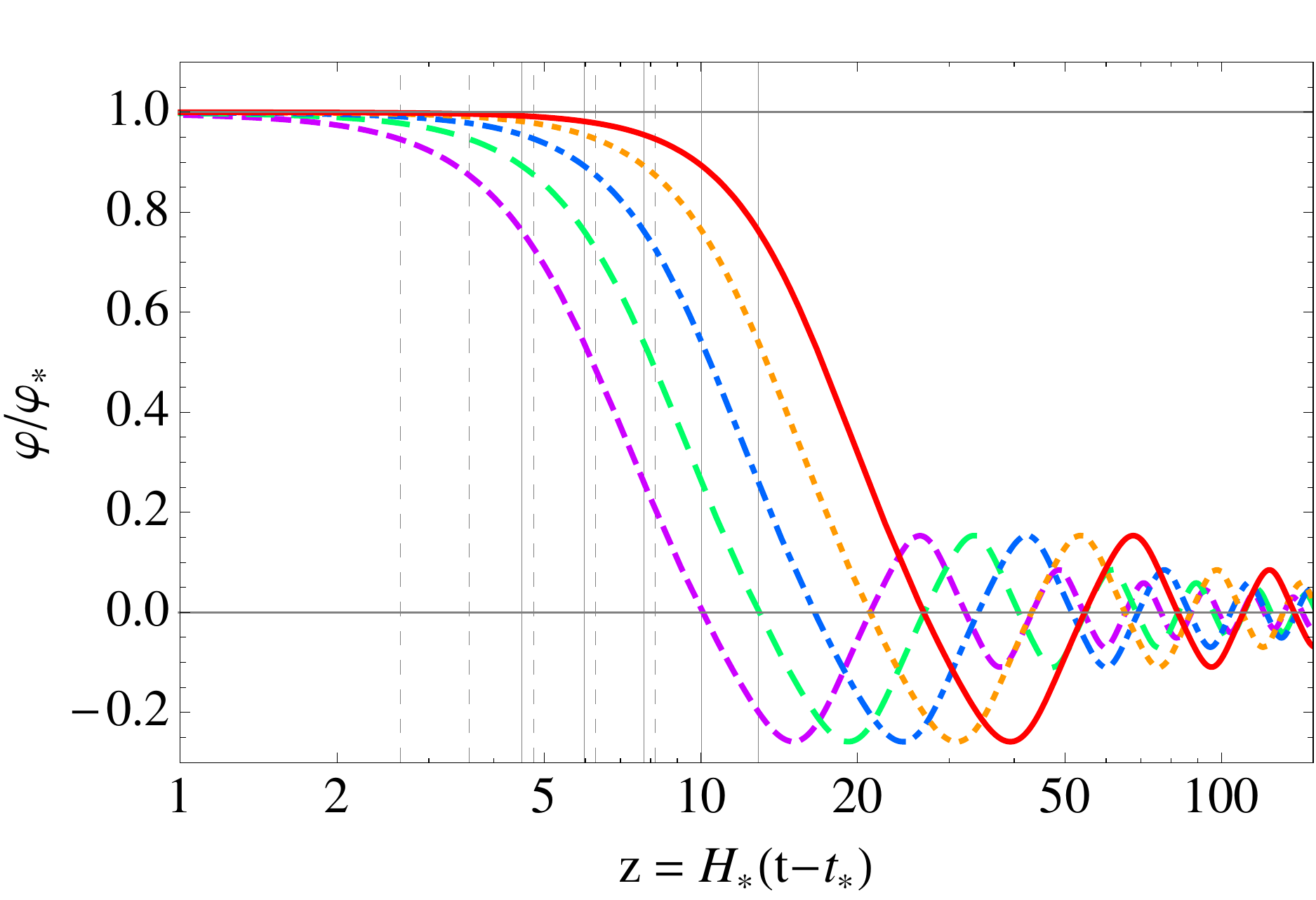}
    \end{center}
    \caption{Evolution of the Higgs field for $\beta = 10^{-2}, 2.5\times 10^{-2}, 5.0\times 10^{-2}, 7.5\times 10^{-2}$ and $10^{-1}$ (corresponding to the red solid, orange dotted, blue dotted-dashed, green long-dashed and purple short-dashed lines, respectively). The background is RD, so $w = 1/3$. Dashed vertical lines mark the time $z_{\rm osc}(\beta)$ when the oscillation condition is attained, $a\sqrt{\lambda}\varphi \equiv \mathcal{H}$, whereas continuous vertical lines mark the time $z_{M}(\beta)$ when the first maximum in the oscillations is reached, characterized by the condition $h'(z_{M}) \equiv 0$. Top: Evolution of $h(z)$. Lower: Evolution of the physical Higgs $\varphi/\varphi_*$, which initially is frozen until the oscillations start, and then decreases as $\propto 1/a$ afterwards, as it oscillates. Similar plots are obtained for MD and KD backgrounds, whereas for other values of $\beta$ the scale in the horizontal axis changes quite significantly.}\label{fig:Figure1}
\end{figure}

The initial velocity of the Higgs and the frequency of its oscillations (in the dimensionless variables) both depend, through $\beta$, on the initial amplitude of the Higgs $\varphi_*$, and the actual value of $\lambda$. Therefore, at different patches of the Universe (separated at distances larger than the correlation length $l \gg l_*$), the Higgs will start oscillating with different amplitudes, and the oscillation frequency will also be different, see Fig.~\ref{fig:Figure1}.

Depending on the amplitude of $\beta$, the Higgs will start oscillating around the minimum of its potential sooner or later. This can be clearly seen in Eq.~(\ref{eq:HiggsEOMtransformed}), where the effective squared frequency of the oscillations of $h(z)$ scales as $\propto \beta^2$. For the canonical value of $\lambda = \lambda_c = 0.01$ ($\lambda_{001} = 1$), the probability for the Higgs to start oscillating immediately at the end of inflation (i.e. that $ \beta \geq 1$) is extremely suppressed as $10^{-287} \%$, being even smaller for $\lambda < \lambda_c $ ($\lambda_{001} < 1$). 

Therefore, at the end of inflation, the Higgs has, within any arbitrary patch of size smaller than $l_*$, an initial velocity in slow roll and a nonzero amplitude as large as $\varphi/H_* \sim \mathcal{O}(0.01)-\mathcal{O}(1)$. This amplitude remains 'frozen' for a finite time until the start of the oscillations. Looking at Eq.~(\ref{eq:HiggsEOM}), and denoting as $z_{\rm osc}(\beta)$ the time at which oscillations start at each patch, we see that the condition for the onset of oscillations is $a(z_{\rm osc})\sqrt{\lambda}\varphi(z_{\rm osc}) = \mathcal{H}(z_{\rm osc})$. For simplicity, we will set the initial value of the scale factor to unity $a_* \equiv a(t_*) = 1$, so that $\mathcal{H}_* \equiv H_*$, $z \equiv H_*(t-t_*)$, and $a(z) = (1+z/p)^p$. We will also denote any quantity evaluated at $z_{\rm osc}$ with the suffix $_{\rm osc}$, so for example $a_{\rm osc} \equiv a(z_{\rm osc})$. It follows that $a_{\rm osc}\sqrt{\lambda}\varphi_{\rm osc} = a_{\rm osc}H_{\rm osc} = H_*/a_{\rm osc}^{1/p}$, from which we find
\begin{eqnarray}\label{eq:OscCondition}
    \varphi_{\rm osc} \equiv {H_*\over \sqrt{\lambda}}{1\over (a_{\rm osc})^{1+{1\over p}}}~~\Rightarrow~~\sqrt[p]{a_{\rm osc}}\,\beta\,h_{\rm osc} = 1 \ .
\end{eqnarray}

For a given expansion rate (characterized by the postinflationary equation of state $w$), the period of oscillations depends sensitively on $\beta$, since the period is fixed when the oscillation condition $a\sqrt{\lambda}\varphi = \mathcal{H}$ is attained at the time $z_{\rm osc}$, which is itself a function of $\beta$ and $w$. The time scale $z_M$ at which $h(z)$ reaches its first maximum, characterized by $h'(z_{M}) = 0$, also depends consequently on $\beta$ and $w$. The period of oscillation can be easily obtained from the case of a field with quartic potential, initial amplitude $\varphi_*$, zero initial velocity $\dot\varphi_* = 0$, and RD background. In conformal time, when the scale factor at the onset oscillations is set to unity, it is given by~$T = 7.416/(\sqrt{\lambda}\varphi_*)$~\cite{Greene:1997fu}. In our case, we just need to count the oscillations from the first maximum at $z = z_M$, taking into account that in our convention, $a(z_M) \neq 1$. The period, in units of $z$, is then found to be
\begin{eqnarray}\label{eq:Period}
    Z_T \equiv  {7.416 \,H_*\over \sqrt{\lambda}\varphi(z_M)a(z_M)}
    = {7.416\over \beta\,h(z_{\rm M})}\,.
\end{eqnarray}
Let us note that the factor $7.416$ is only exact for RD. For MD or KD, one expects a similar though somewhat different number, simply due to the term $a''/a$ in Eq.~(\ref{eq:HiggsEOMtransformed}), which affects the very early stages of the Higgs dynamics (even if it decays very fast after the onset of oscillations). 

We have obtained fits for $z_{\rm osc}$, $h_{\rm osc}$, $h(z_M)$ and $Z_T$ as a function of $\beta$ and for each postinflationary expansion rate, characterized by the equation of state $\omega$. These fits will turn out to be useful later on. We find at the onset of oscillations
\begin{eqnarray}\label{eq:VariablesAtOsc}
    h_{\rm osc} &=& {0.98\,\beta^{-{2\over 3(1+w)}}}\\
    z_{\rm osc} &=& {2\over(1+3w)}\left({1.02\,\beta^{-{(1+3w) \over 3(1+w)}}}-1\right)\,.
\end{eqnarray}
On the other hand, we find the field amplitude at $z=z_M$, and the oscillation period (measured from $z=z_M$ onwards), as
\begin{eqnarray}\label{eq:VariablesAtMax}
    h(z_{\rm M}) = {A h_{\rm osc}}\,,~~~~ Z_{\rm T} = B\beta^{-{(1+3w) \over 3(1+w)}}\,,
\end{eqnarray}
where $(A,B) \simeq (1.28,6.30)$, $(1.22, 6.25)$, $(1.17, 6.25)$ for $w = 0, 1/3$, and $1$, respectively. 

\begin{figure}[t]
    \begin{center}
        \includegraphics[width=8cm]{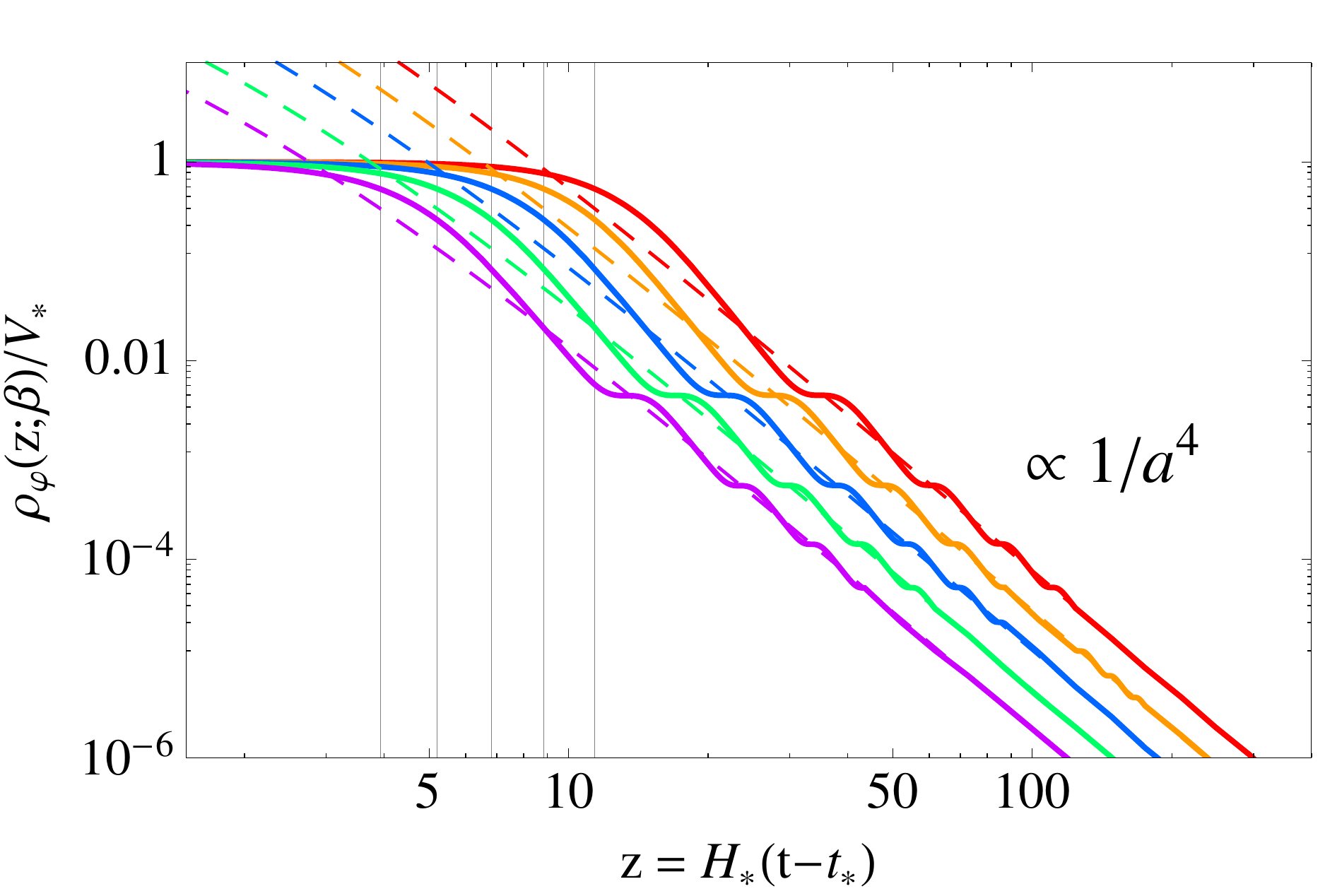}\vspace*{5mm}
        \includegraphics[width=8cm]{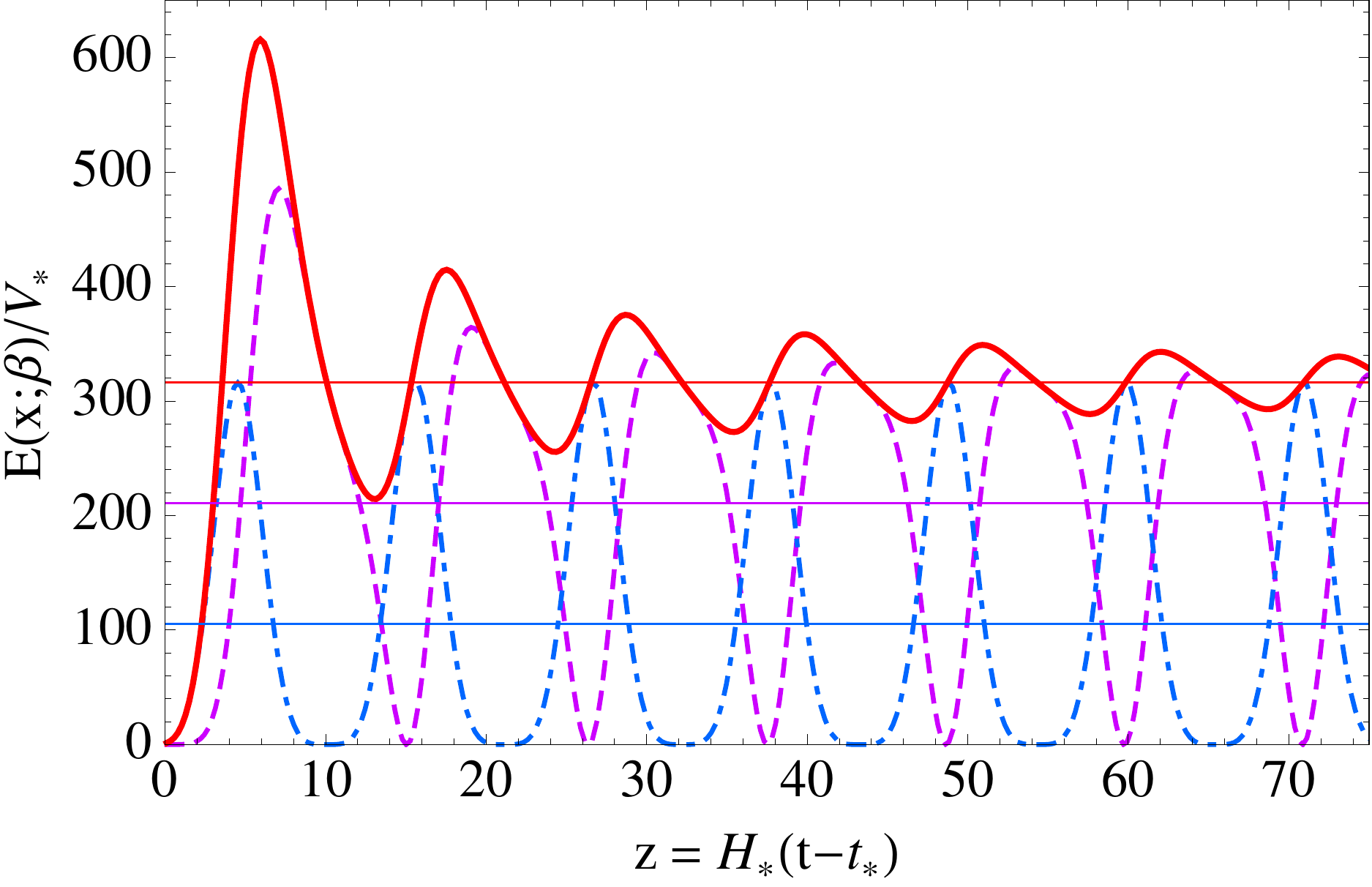}
    \end{center}
    \caption{Top: Total energy $\rho_\varphi/V_*$ (continuous lines) and its oscillation-averaged value $\bar{\rho}_\varphi/V_*$ (dashed lines), for $\beta = 10^{-2}, 2.5\times 10^{-2}, 5.0\times 10^{-2}, 7.5\times 10^{-2}, 10^{-1}$ (from right to left, red, orange, blue, green, purple). Vertical grey lines mark $z_{M}(\beta)$, signaling from what point the averaged curves should be considered valid. Bottom: The functions $E_K(z,\beta)$ (dashed, purple), $E_V(z,\beta)$ (dotted-dashed, blue) and $E(z,\beta) \equiv E_K(z,\beta) + E_V(z,\beta)$ (solid, red), and their averages $\bar{E}_K(\beta)$ (purple), $\bar{E}_V(\beta)$ (blue) and $\bar{E}(\beta) \equiv \bar{E}_K(\beta) + \bar{E}_V(\beta)$ (red), for $\beta = 10^{-1}$. The figure shows that $\bar{E}_V(\beta) = {1\over 2}\bar{E}_K(\beta) =  {1\over 3}\bar{E}(\beta)$. All plots obtained for a RD background.
    }\label{fig:3}
\end{figure}

At the end of inflation, the Higgs energy density at a given patch is mostly dominated by its potential energy,
\begin{eqnarray}
    V_* \equiv {\lambda \varphi_*^4\over4}
\end{eqnarray}
which represents a very small contribution of the total energy budget at that moment, $\rho_* = 3m_p^{2}H_*^{2}$. Averaging over realizations, we find
\be {\langle V_* \rangle \over 3m_p^2H_*^2} = c_4^4 \left( \frac{H_*}{m_p} \right)^2 \simeq 4\times 10^{-12}\left(H_*\over H_*^{\rm (max)}\right)^2 \ . \label{eq:higgs-energy-infl} \ee
At the onset of oscillations, part of the potential energy will become kinetic, with the two contributions -- kinetic and potential -- becoming of the same magnitude. In order to see this, let us first write the total energy density of the Higgs as
\begin{eqnarray}\label{Eq:HiggsEnergyDensity}
    \rho_\varphi &=& {1\over 2 a^2}{\dot\varphi}^2 + {\lambda\over4}{\varphi}^4 = {V_*\over a^4(z)}E(z,\beta)\,,
\end{eqnarray}
with the kinetic and potential contributions given by
\begin{eqnarray}
    E(z,\beta) &=&
    {1\over \beta^2}\left(h'-{a'\over a}h\right)^2 ~  + ~~ h^4
    \,, \nonumber\\
    &\equiv& ~~~~ E_K(z,\beta) ~~~ + ~~~ E_V(z,\beta)\,.
\end{eqnarray}
We can then take the average over the Higgs oscillations as\footnote{Note that we are not including in the average the prefactor ${1/a^4(t)}$ factorized out in Eq.~(\ref{Eq:HiggsEnergyDensity}), since the scale factor changes only marginally during each oscillation. Therefore, we are only averaging the contribution due to the Higgs oscillatory behavior.}
\begin{eqnarray}
    \overline{\rho}_{\varphi}(z,\beta) &=&  {V_*\over a^4(z)}\overline{E}(\beta)\\
    \overline{E}(\beta)  &=& {1\over Z_T(\beta)}\int_z^{z+Z_T(\beta)}\hspace*{-7mm}dz' E(z',\beta)\,, 
\end{eqnarray}
and again split the result into potential and kinetic contributions, $\overline{E}(\beta) = \overline{E}_K(\beta) + \overline{E}_V(\beta)$, where
\begin{eqnarray}
    \overline{E}_V(\beta) &\equiv& {1\over Z_T(\beta)}\int_z^{z+Z_T(\beta)} h^4(z')dz' = {1\over 3}\overline{E}(\beta) \ , \\
    \overline{E}_K(\beta) &\equiv&   {1\over Z_T(\beta)}\int_z^{z+Z_T(\beta)} \hspace*{-6mm}dz'{1\over \beta^2}\left(h'-{a'\over a}h\right)^2 = {2\over 3}\overline{E}(\beta) \ . \nonumber\\
\end{eqnarray}
In Fig.~\ref{fig:3} we can see the total energy density of the Higgs for different values of $\beta$, with and without averaging. Of course, the oscillation-averaged expressions are only valid when the Higgs has started oscillating at $z \gtrsim z_M$, as clearly appreciated in the plot. The figure also shows very nicely the fact that the averaged components verify $\overline{E}_V(\beta) = {1\over 3}\overline{E}(\beta)$ and $\overline{E}_K(\beta) = {2\over 3}\overline{E}(\beta)$. Possibly, the most relevant aspect to be remarked is the well-known fact that the Higgs energy density scales as $a^{-4}$ with the expansion of the Universe~\cite{Turner:1983he}, behaving as if it were a fluid of relativistic species.

\section{Higgs decay: Analytical estimates}
\label{sec:III}

As just explained, the Higgs oscillates everywhere in the Universe, although the time to start the oscillations depends sensitively on the initial condensate amplitude, which varies from patch to patch according to $P_{\rm eq}(\varphi)$. Once the oscillations have begun within a given patch, all fields coupled directly to the Higgs are excited every time the Higgs goes through the minimum of its potential. In the case of bosonic species, this is known as parametric resonance, since a cumulative effect takes place, producing a resonant growth of the number density of species~\cite{Traschen:1990sw,Kofman:1994rk,Kofman:1997yn,Greene:1997fu,VaccaHiggs,Enqvist:2013kaa,Enqvist:2014tta}. Although there is no parametric resonance in the case of fermionic species, yet an interesting effect occurs, since modes with successively higher momenta are excited as the oscillations carry on~\cite{Greene:1998nh,Giudice:1999fb,GarciaBellido:2000dc,Greene:2000ew,Peloso:2000hy,Figueroa:2014aya}. For a review of parametric excitation of fields in the similar context of preheating, see~\cite{Allahverdi:2010xz,Amin:2014eta}.

All charged leptons of the SM are directly coupled to the Higgs via a Yukawa interaction, so all fermions of the SM will be excited during the oscillations of the Higgs~\cite{Figueroa:2014aya}, with the possible exception of neutrinos. Among the SM fermions, the top quark has the largest coupling to the Higgs,
so most of the energy transferred into fermions goes into top quarks. More importantly, the $SU(2)_L$ gauge bosons are also coupled directly to the Higgs, and indeed the strength of their coupling is very similar to that of the Yukawa top quark. When two species, one fermionic and another bosonic, are coupled with the same strength to an oscillatory homogeneous field, the first burst of particle production is actually spin independent, and hence an equal number of bosons and fermions are created~\cite{GarciaBellido:2001cb}. However, the successive particle creation bursts at each Higgs zero crossing take place on top of an already existing number density of previously created species. The spin statistics becomes then crucial, differentiating bosons from fermions in a noticeable way: bosonic occupation numbers start growing exponentially as the oscillations accumulate, whereas the fermion occupation numbers are always Pauli-blocked, forcing the transfer of energy into modes with higher and higher momenta. Both bosonic and fermionic excitations represent a sizable transfer of energy from the Higgs condensate. However, for equal coupling strength [as it is the case between top quarks and $SU(2)_L$ gauge bosons], the transfer of energy is much more efficient into the bosonic species~\cite{GarciaBellido:2000dc}. Besides, in the context under study here -- the decay of the Higgs after inflation --, the subdominant production of the SM charged leptons has been already addressed in~\cite{Figueroa:2014aya}. Therefore, in this paper we will only focus on the production of the most energetically dominant species among the Higgs decay products, the $W^{\pm}$ and $Z$ gauge bosons. 

In order to study the dynamics after inflation of the Higgs and its most energetic decay products, one should in principle consider the full $SU(2) \times U(1)$ gauge structure of the SM electroweak sector. However, one can make reasonable approximations for both analytical and computational purposes. In this work we have considered the following approximate schemes, mimicking the structure of the SM interactions:

\begin{list}{}{}
    \item $i)$ {\it Abelian model}. This consists in modeling the interactions between the electroweak gauge bosons and the Higgs with an Abelian-Higgs analogue. Since gauge fields are initially excited by the Higgs from the vacuum, it is clear that nonlinearities due to the truly non-Abelian nature of $SU(2)$ are expected to be negligible during the initial growth of the gauge field occupation numbers~\cite{GarciaBellido:2003wd}. The authors of Ref.~\cite{Enqvist:2014tta} have shown that using the Hartree approximation, the effective contribution induced by the created gauge bosons onto themselves (due to the non-Abelian nonlinearities) can be neglected as long as the backreaction from the gauge fields onto the Higgs does not become significant. In principle, this fact fully justifies ignoring the non-Abelian structure of the SM interactions, while maintaining only the Abelian dominant part.
    
    \item $ii)$ {\it Global model}. A more crude approximation can yet be done, by ignoring the gauge structure of the interactions. This does not mean that we ignore the interactions themselves, but rather that we consider them as if they were dictated by a global symmetry, instead of a gauge one. In this scenario, one simply solves the mode equations of various scalar fields coupled to the Higgs with a quadratic interaction. Each of these scalar fields mimics a component of the gauge fields, with the quadratic interactions reproducing the coupling of the gauge bosons and the Higgs obtained from the SM gauge covariant derivative terms. This way, one can presumably capture the initial stages of the parametric resonance of $W^{\pm}$ and $Z$ bosons. 
\end{list}

The approach $i$ is our most precise modeling of the SM interactions, but also the most involved one. We thus postpone its implementation for later on in Section~\ref{sec:v}. The approach $ii$, though less accurate, has a clear advantage versus the gauge case: it allows not only for a lattice implementation (which we introduce in Section~\ref{sec:iv}), but also for an analytical treatment (which we present in the remaining of this section). The analytical estimates represent only an approximation to the system described by the scenario $ii$, but yet provide a valuable insight into the understanding of the dynamics. The order of presentation in the paper of our different approaches is thus based on increasing progressively the degree of proximity to the real system. First, in the remainder of this section, we start with the analytical treatment of the global modeling, ignoring all nonlinearities of the system. In Section~\ref{sec:iv} we implement the global scenario $ii)$ on the lattice. This way, we fully capture all nonlinearities within this modeling, even if we yet neglect the gauge nature of the interactions.
Finally, in Section~\ref{sec:v}, we present a lattice implementation of an Abelian modeling of the system. This fully captures the nonlinearities within such modeling, while preserving at the same time the gauge-invariant nature of the interactions. 

\subsection{Analytical approach to the Higgs decay}
\label{subsec:III.a}

In principle, we can follow the initial stages of the parametric resonance of the $W^{\pm}, Z$ bosons by simply solving the mode equation for a scalar field $\chi$, coupled to the Higgs with an interaction term of the form ${e^2\over2}\chi^2\varphi^2$. Analytical results following this approach have indeed been presented in~\cite{Enqvist:2013kaa}, so our work in this section should be understood only as complementary to such reference. We develop nevertheless some new formulas which will be useful later on, in order to assess the reliability of this analytical approximation when compared to the fully nonlinear numerical lattice simulations.

\begin{figure}[t]
    \begin{center}
        \includegraphics[width=8.5cm]{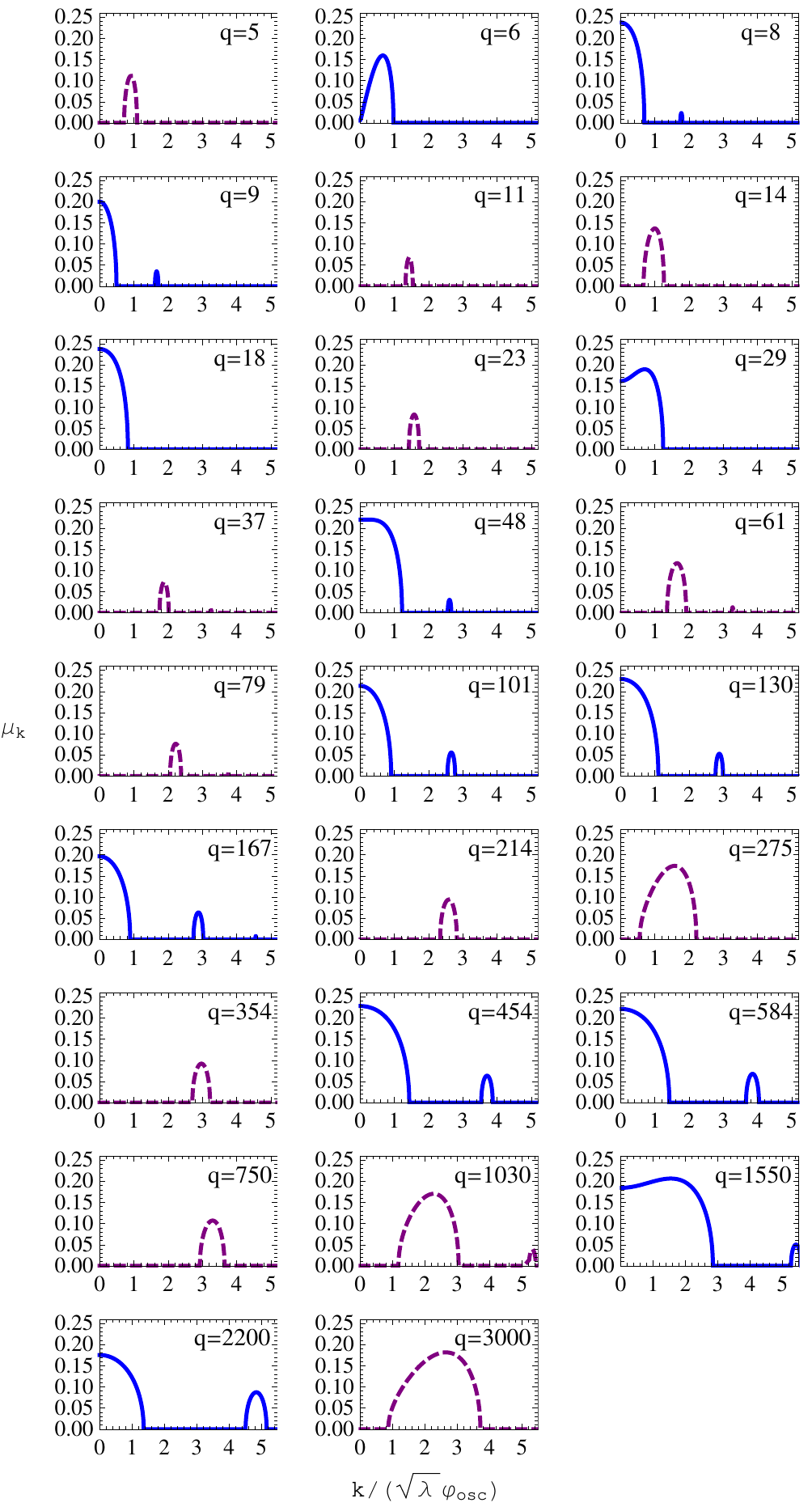}\vspace*{5mm}
        \caption{This shows the band structures of the Lam\'e equation, Eq.~(\ref{eq:modeEQ}),  for several resonance parameters ranging between $q=5$ and $q=3000$. In each panel, we plot the corresponding Floquet index $\mu_{\kappa}$ (where $\chi_{\kappa} \propto e^{\mu_\kappa z}$) as a function of the momentum $\kappa$. We can divide the different $q$ into two groups: those which contain a resonance at $\kappa=0$ (blue lines) and those which not (purple dashed lines).}\label{fig:FloquetVarious}
    \end{center}
\end{figure}

The equation for the Fourier modes of the field $\chi$, after an appropriate conformal redefinition $\chi_k \equiv X_k /a$, and assuming RD, can be mapped into~\cite{Greene:1997fu}
\begin{eqnarray}\label{eq:modeEQ}
    X_k''+ \left(\kappa^2+q(h/h_{\rm osc})^2\right)X_k = 0~,~~~~~q \equiv {e^2\over \lambda}\,,
\end{eqnarray}
with $q$ being the resonance parameter, $\kappa \equiv k/(\sqrt{\lambda}\varphi_{\rm osc})$, $' \equiv d/dz$, and $z \equiv \mathcal{H}_{\rm osc}t$. Given the behavior of $h(z)$, dictated by the Higgs quartic potential, this equation corresponds indeed to the Lam\'e equation~\cite{Greene:1997fu}, which has a well-understood structure of resonances. Whenever $q \in {1\over2}[n(n+1),(n+1)(n+2)]$, with $n = 1, 3, 5, ...$ (i.e. $q \in$ [1, 3], [6, 10], ...), there is an infrared band of resonance $0 \leq k \lesssim k_* \equiv {1\over \sqrt{2}\pi}q^{1/4}\mathcal{H}_{\rm osc}$, for which $X_k \propto e^{\mu_kz}$, with $\mu_k$ bigger the smaller the $k$ (therefore maximum at $k = 0$). 
If the resonance parameter $q > 1$ is not within one of the resonant bands, but lies in between two adjacent bands, then there is still a resonance of the type $X_k \propto e^{\mu_kz}$, but within a shorter range of momenta $0 < k_{\rm min} \leq k \lesssim k_*$, and hence with a smaller Floquet index $\mu_k$. There is a theoretical maximum value for the Floquet index given by $\mu_k^{\rm (max)} \equiv 0.2377...$~\cite{Greene:1997fu}, so that any $\mu_k$ is always constrained as $\mu_k \leq\mu_k^{\rm (max)}$ for $q > 1$. For resonant parameters $q \gg 1$, $\mu_k$ is typically of order $\sim \mathcal{O}(0.1)$; see Fig.~\ref{fig:FloquetVarious}.

For simplicity, we will consider until the end of this section that the resonance parameter $q = e^2/\lambda$ always falls within one of the resonant bands, $ q \in [1, 3]$, $[6, 10]$, $[15, 21], ...$. As a matter of fact, in order to identify $e^2$ with the gauge coupling $g^2$ between the Higgs and a gauge field, we need to make the identification $e^2 \rightarrow g^2/4$, with $g^2$ the gauge coupling $g^2_Z$ or $g_W^2$ of either the $Z$ or the $W^{\pm}$ gauge bosons. This matches correctly the interaction derived from the covariant gauge derivative of the electroweak sector of the SM. The gauge couplings of the $Z$ and $W^{\pm}$ gauge bosons verify $g_Z^2 \approx 2g_W^2 \approx 0.6$ at very high energies. Due to this relation, it is likely that either $q_W \equiv g_W^2/4\lambda$ or $q_Z \equiv g_Z^2/4\lambda \approx 2q_W$, will fall within one of the instability bands. Let us note, however, that we cannot predict this, since the value of the Higgs self-coupling $\lambda$ at high energies is quite sensitive to the uncertainties in the Higgs mass $m_H$, the top quark mass $m_t$, and the strong coupling constant $\alpha_s$. Consequently, we cannot really know the exact value of these resonance parameters. However, in order to guarantee that during inflation the Higgs fluctuations remain below the critical scale $\mu_+$ (above which the self-coupling starts decreasing, $d\lambda/d\mu \leq 0$), and taking into account that the inflationary Hubble rate is constrained from above as $H_* \lesssim 10^{14}$ GeV~\cite{Planck2015}, the Higgs self-coupling value at high energies can then only be within the range $10^{-2} \lesssim \lambda \lesssim 10^{-5}$. Pushing $\lambda$ to smaller values is in principle possible, but it represents a fine-tuning and requires some of the parameters $m_t, m_H, \alpha_s$ to be more than 3 sigma away from their central values. We will consider therefore the range $10^{-2} \lesssim \lambda \lesssim 10^{-5}$ as the only acceptable one (with $\lambda \sim 10^{-5}$ only marginally valid). If beyond the SM physics affects the running, say stabilizing the Higgs potential at high energies, then $\lambda$ remains positive and typically of the order $\lambda \sim 10^{-2}-10^{-3}$. Considering the range $10^{-2} \lesssim \lambda \lesssim 10^{-5}$, and taking into account the strength of the $W^\pm, Z$ gauge couplings at high energies, we obtain that the resonant parameters can only possibly be within the range $\mathcal{O}(10) \lesssim q \lesssim \mathcal{O}(10^3)$. In particular, since at high energies $g^2 = g^2_W \simeq 0.3$ for $W$ gauge bosons, we obtain $q = 7.5$ for $\lambda = 10^{-2}$, and $q = 3000$ for $\lambda = 2.5\times10^{-5}$. For $Z$ bosons we obtain similar resonance parameters, but twice as big. For completeness, we have sampled resonance parameters within the interval $q \in [5,3000]$, which corresponds to a range $\lambda = 1.5\times 10^{-2}-2.5\times 10^{-5}$ for $W$ bosons and $\lambda = 3.0\times 10^{-2}-5.0\times10^{-5}$ for $Z$ bosons.

Let us then consider just one particle species $A_{\mu}$, representing either the $Z$ or one of the $W$ gauge bosons, that will be parametrically excited during the Higgs oscillations. Let $g^2$ be the coupling strength to the Higgs and let us represent the gauge field as if it were simply a collection of three scalar fields (one for each spatial component), all coupled with the same strength to the Higgs. The growth of the fluctuations in the initial stages of resonance is described by the linearized Eq.~(\ref{eq:modeEQ}). As long as the linear regime holds, even if the amplitude of the fluctuations grows exponentially, the use of three scalars should represent a good mapping of the real problem of gauge field excitation. Of course, one is ignoring this way the backreaction of the created bosons into the Higgs, as well as certain contributions in the gauge fields' EOM, which should be present if the gauge symmetry was restored. 

The energy density of the created particles due to the resonance is then given by
\be
    \rho_{A} = {3\over 2\pi^2a^3}\int dk k^2 n_k \omega_k~,~~~\omega_k^2 \equiv {k^2\over a^2} + {g^2\over 4}\overline{\varphi^2}\,,
\ee
with the factor $3$ accounting for the three spatial components of a gauge field, and where we have introduced an oscillation-averaged effective mass for the gauge boson,
\begin{eqnarray}
    m_A^2 &=& {g^2\over 4} \overline{\varphi^2} = {g^2\over 4}{\varphi_*^2\over a^2} \overline{h^2} \nonumber\\
    &\equiv & {g^2\over 4} {\varphi_*^2\over a^2} {1\over Z_T(\beta)}\int_{z}^{z+Z_T(\beta)} dz' h^2(z') \ .
\end{eqnarray}
For $q \gg 1$, the maximum (comoving) momentum possibly excited in broad resonance is given by 
\begin{eqnarray}
    k_*^2 &\equiv& {q^{1/2}\over 2\pi^2}a_{\rm osc}^2\lambda\varphi_{\rm osc}^2 = {q^{1/2}\over 2\pi^2}h_{\rm osc}^2\lambda\varphi_{*}^2 \ ,
\end{eqnarray}
from which, given that $\overline{h^2} \sim h_{\rm osc}^2$, we see that
\begin{eqnarray}
    {m_A^2\over (k_*/a)^2} \sim \mathcal{O}(10)q^{1/2} \gg 1\,.
\end{eqnarray}
In other words, in broad resonance $q \gg 1$, the decay products are always nonrelativistic, and correspondingly we can approximate the effective mode frequency as $\omega_k \simeq m_A \sim {g\over2}{\varphi_*\over a}h_{\rm rms}$, where $h_{\rm rms} \equiv \sqrt{\overline{h^2}}$. It turns out that $h_{\rm rms} \simeq h_{\rm osc}$ independently\footnote{For $\beta \gtrsim 0.3$ there is some dependence, but still $\sqrt{\overline{h^2}}/h_{\rm osc} \sim \mathcal{O}(1)$.} of $\beta$.
If $q$ is within a resonant band, then all modes with momenta $0 \leq k \lesssim k_*$ are excited with some Floquet index varying within $[0,\mu_k^{\rm (max)}(q)]$. This corresponds to the cases with blue solid lines in Fig.~\ref{fig:FloquetVarious}. We can therefore model the occupation number of the excited modes simply as a step function $n_k = e^{2\overline{\mu}_ky}\Theta(1-k/k_*)$, where $\overline{\mu}_k \sim \mathcal{O}(0.1)$ and $y \equiv \mathcal{H}_{\rm osc}(t-t_{\rm osc}) = a_{\rm osc}\sqrt{\lambda}(\varphi_{\rm osc}/H_*)(z-z_{\rm osc}) = (a_{\rm osc})^{-{1\over p}}(z-z_{\rm osc})$, with $z = H_*t$. It follows that
\begin{eqnarray}\label{eq:GaugeEnergyAtEff}
    \rho_{A}(z) &\simeq& {(h_{\rm rms}/h_{\rm osc})^2\over 4\pi^2}{1\over a^4} e^{{2\overline{\mu}_k\over\sqrt[p]{a_{\rm osc}}}(z-z_{\rm osc})}\,g\varphi_* h_{\rm osc} k_*^3 \nonumber\\
    &\simeq & q^{5/4}{(h_{\rm rms}/h_{\rm osc})^2\over 2^{5/2}\pi^5} e^{{2\overline{\mu}_k\over\sqrt[p]{a_{\rm osc}}}(z-z_{\rm osc})}{H_*^4\over (a\sqrt[p]{a_{\rm osc}})^4},\nonumber\\
\end{eqnarray}
where we have used the fact that $\beta h_{\rm osc} = 1/\sqrt[p]{a_{\rm osc}}$.

This is how the energy density of the gauge bosons (those fully within a resonant band) will grow, at least as long as their backreaction into themselves and/or into the Higgs remains negligible. Using this linear approximation we can estimate the moment $z_{\rm eff}$ at which an efficient transfer of energy has taken place from the Higgs into the gauge bosons, characterized by $\rho_A(z_{\rm eff}) = \rho_\varphi(z_{\rm eff})$. This will be just a crude estimate of the time scale of the Higgs decay, since by then backreaction and rescattering effects have become important, invalidating the linear approach. However, the nonlinear effects due to backreaction/rescattering of the decay products, simply tend to shut off the resonance. Hence, using the linear regime for inferring the Higgs time scale should provide, at least, a reasonable estimate of the order of magnitude. More importantly, it provides the parametric dependences of both the time when the resonance is switched off, and the moment when the energy has been efficiently transferred into the gauge bosons. 

The energy of the Higgs, since the onset of the oscillations, decays as
\begin{eqnarray}\label{eq:HiggsEnergyAtEff}
    \rho_{\varphi}(z) &=& V_*{1\over a^4} 3\overline{E}_V(\beta) = {3\over4}{\lambda\varphi_*^4\over a^4}h_{\rm osc}^4\overline{(h/h_{\rm osc})^4} \nonumber\\
    &=& {3\over 4\lambda}\overline{(h/h_{\rm osc})^4}{H_*^4\over (a\sqrt[p]{a_{\rm osc}})^4}\,,
\end{eqnarray}
where $\overline{(h/h_{\rm osc})^4} \sim \mathcal{O}(1)$. We can now find $z_{\rm eff}$ by simply equating Eqs.~(\ref{eq:GaugeEnergyAtEff}) and (\ref{eq:HiggsEnergyAtEff}),
\begin{eqnarray}
    {q^{1/4}\over 2^{5/2}\pi^5}\sqrt{\overline{(h/h_{\rm osc})^2}}e^{{2\overline{\mu}_k\over\sqrt[p]{a_{\rm osc}}}(z-z_{\rm osc})} = {3\over g^2}\overline{(h/h_{\rm osc})^4}\,,
\end{eqnarray}
so that
\begin{eqnarray}\label{eq:EffEnergyTransferTimeScale} 
    z_{\rm eff} = z_{\rm osc} + {\sqrt[p]{a_{\rm osc}}\over 2\overline{\mu}_k}\left[\log\left({\overline{(h/h_{\rm osc})^4}\over(h_{\rm rms}/h_{\rm osc})}\right) ~~~~~~~\right. \\
    \left.  ~~~~~~~~~~~~~~+ \log\left({3\cdot 2^{5/2}\pi^5\over g^2}\right) - {1\over4}\log q\right] \ . \nonumber
\end{eqnarray}
Let us recall that $g^2 \simeq 0.3,0.6$ at large energies, and $q \equiv g^2/(4\lambda) \sim \mathcal{O}(10)-\mathcal{O}(10^3)$, depending on the value of $\lambda$. Taking this into account, we find that the first term in the brackets of the rhs is always irrelevant, the second term is constant and of the order $\simeq 9$, and the last term is of order $\sim -1$. Therefore, we can approximate the above expression, using $\sqrt[p]{a_{\rm osc}} = (1+{1\over p}z_{\rm osc})$, as
\begin{eqnarray}
    z_{\rm eff} &\simeq& z_{\rm osc} + {8\over2\overline{\mu}_k}\sqrt[p]{a_{\rm osc}} = {4\over \overline{\mu}_k} + \left(1 + {4\over  p\overline{\mu}_k}\right)z_{\rm osc} \nonumber\\
    &\simeq & {4\over p\overline{\mu}_k}z_{\rm osc} \ .
\end{eqnarray}
Looking at Fig.~\ref{fig:FloquetVarious}, we see that the Floquet index of the modes $0 \leq k \lesssim k_*$ for which $q$ is within a resonant band (blue solid lines of the figure), can be well approximated by a simple step function $\mu_k \approx \overline{\mu}_k\Theta(1-k/k_*)$, with a mean Floquet index $\overline{\mu}_k \simeq 0.2$. Taking this into account and using the fit of Eq.~(\ref{eq:VariablesAtOsc}) for the time scale at the onset of oscillations $z_{\rm osc}(\beta)$, we find 
\begin{eqnarray}\label{eq:EffEnergyTransferTimeScaleApprox}
    z_{\rm eff} \sim 20\times\left({0.2\over\overline{\mu}_k}\right)\beta^{-{(1+3w)\over 3(1+w)}} \ .
\end{eqnarray}
The scale factor at $z=z_{\rm eff}$ is then given by
\begin{eqnarray}\label{eq:EffEnergyTransferScaleFactorApprox}
    a_{\rm eff} \equiv a(z_{\rm eff}) \sim (20(1+3w))^{2\over(1+3w)}\cdot\beta^{-{2\over 3(1+w)}} \ .
\end{eqnarray}

It is clear that depending on how small the initial value of $\beta$ is within a given path of the Universe, the longer it takes for the Higgs to transfer energy efficiently into the gauge bosons, simply because the longer it takes (since the end of inflation) to start oscillating. Since $\beta_{\rm rms} \sim \mathcal{O}(0.1)$, we see that typically the Higgs decays at a time $z_{\rm eff}(\beta_{\rm rms}) \sim \mathcal{O}(10^2)$.
Although the time varies from patch to patch depending on the values of $\beta$, it is clear that the Higgs tends to decay really fast after inflation, within a few dozens of oscillations. In the following sections we will check the validity of this estimate by comparing it with the outcome obtained directly from lattice simulations.

\section{Lattice simulations, Part 1: Global modeling}
\label{sec:iv}

In this section, we continue modeling the SM interactions with a set of scalar fields. More specifically, we consider the Lagrangian 
\be \label{eq:action-scalar}
-\mathcal{L} = {1\over2}\partial_\mu\varphi\partial^\mu\varphi + {1\over2}\partial_\mu\chi_i\partial^\mu\chi_i + \frac{\lambda}{4}{\varphi}^4 + \frac{e^2}{2} {\varphi}^2 \sum_i {\chi}_i^2\,,
\ee
with $i = 1,2,3$. Varying the action $S = \int d^4x\,\mathcal{L}$ leads to the classical EOM
\begin{eqnarray}\label{eq:ScalarEOM}
\ddot \varphi + 2\mathcal{H}\dot \varphi - \nabla^2\varphi  + a^2(\lambda\varphi^2 + e^2\sum_i {\chi}_i^2)\varphi = 0\,,\\
\ddot \chi_i + 2\mathcal{H}\dot \chi_i - \nabla^2\chi_i + a^2e^2{\varphi}^2\chi_i = 0 \ . \label{eq:ScalarEOMbis}
\end{eqnarray}
The term $e^2{\varphi}^2\chi_i$, under the identification $e^2 = g^2 /4$, mimics precisely the interaction term from the covariant derivative of the EW gauge bosons, ${g^2\over2}\Phi^\dag\Phi A_\mu$, where $A_\mu$ stands for either $Z_\mu$ or $W_\mu^{\pm}$, and $\Phi$ is the Higgs doublet. More concretely, choosing the unitary gauge for the Higgs $\Phi = (0,\phi/\sqrt{2})$, and fixing $A_0 = 0$, we can identify each $\chi_i$ with each spatial component of the gauge boson $A_i$, and $\varphi$ with the unitary representation of the Higgs. This way, by solving the system of scalar field equations (\ref{eq:ScalarEOM}) and (\ref{eq:ScalarEOMbis}), we can study the properties of the Higgs interactions with gauge bosons in an approximative way. 

In Section~\ref{subsec:III.a} we studied this scenario, following the fluctuations of the fields $\chi_i$ with the help of the analytical solutions of the Lam\'e equation, Eq.~(\ref{eq:modeEQ}). We exploited the band structure of this equation and used some approximations in order to arrive at our analytical results, summarized in Eqs.~(\ref{eq:EffEnergyTransferTimeScale})-(\ref{eq:EffEnergyTransferScaleFactorApprox}). In reality, the scalar fields $\chi_i$ follow the Lam\'e equation only initially, in the regime when the nonlinearities (due to their small backreaction onto the Higgs) can be neglected. The fluctuations of the $\chi_i$ fields grow exponentially during the linear regime, and as we will show, it does not take long until they start to impact onto the Higgs dynamics. At that moment, the system becomes nonlinear, and only by following in parallel the coupled EOM of the Higgs and $\chi_i$ fields, can we really understand the field dynamics within this modeling. The aim of this section is, therefore, to solve numerically in a three-dimensional lattice the system of equations (\ref{eq:ScalarEOM}) and (\ref{eq:ScalarEOMbis}). Only in that way can we fully capture the nonlinear behavior of this system beyond the simpler linear regime of the Lam\'e equation. 

We now present the main results of the lattice simulations carried out for this scenario. We start with the following change of field variables:
\begin{eqnarray}
h \equiv {a\over a_*}{\varphi\over\varphi_*}\,,~~~ X_i \equiv {\chi_i\over H_*}{a\over a_*}\ .
\end{eqnarray}
It is also convenient to redefine new spacetime coordinates $z^{\mu} = (z^0, z^i)$ with respect to the conformal ones $x^{\mu} = (x^0,x^i) \equiv (t,x^i)$, as 
\be z \equiv z^0 = H_* t \,,~~~ z^i = H_* x^i \label{def-st} \ . \ee
With these field and coordinate redefinitions, we eliminate the friction terms in Eqs.~(\ref{eq:ScalarEOM}) and (\ref{eq:ScalarEOMbis}), and produce an equivalent set of dynamical equations, written in terms of the new dimensionless variables:
\bea 
h'' - \nabla^2 h + \beta^2 h^3 + e^2 h \sum_{j} X_j^2 = \frac{a''}{a} h\,, \label{eq:eoms1a} \\ 
X_i'' - \nabla^2 X_i + q \beta^2 h^2 X_i = \frac{a''}{a} X_i \ , \label{eq:eoms2} 
\eea
with $' \equiv d/dz$, and the spatial derivatives taken with respect to the $z^i$ variables. A lattice version of these equations is presented in Appendix \ref{App-Lattice}. As already mentioned, we will identify $e^2 \rightarrow g^2/4$, with $g^2$ being either $g_W^2$ or $g_Z^2$. The resonance parameter that appears naturally in Eq.~(\ref{eq:eoms2}), $q \equiv {e^2\over \lambda}$, should therefore be interpreted as $q \equiv {g^2\over 4\lambda}$. 

We have solved Eqs.~(\ref{eq:eoms1a}) and (\ref{eq:eoms2}) in three-dimensional lattices with periodic boundary conditions. We consider initial conditions given by a homogeneous Higgs mode (as described in Section~\ref{sec:II}),
\begin{eqnarray}
h(0) \equiv 1\,,~~~h(0)' \equiv 1 - {\beta^2\over 2}\,,
\end{eqnarray}
and a null zero mode for the scalar fields coupled to the Higgs,
\begin{eqnarray}
X_i(0) = 0\,,~~~ X_i'(0) = 0\,.
\end{eqnarray}
We add, on top of the homogeneous contributions, a set of Fourier modes with spectrum $\langle |f_k|^2 \rangle = {1\over 2a^2\omega_k}$ (in physical variables), mimicking the quantum vacuum fluctuations of the ground state of a scalar field in a FRW background. Let us recall that the Higgs is frozen in slow roll until the oscillation condition Eq.~(\ref{eq:OscCondition}) is attained at $z = z_{\rm osc}$; see the bottom panel of Fig.~\ref{fig:Figure1}. Hence, during the time $0 \leq z < z_{\rm osc}$, we only evolve in the lattice Eq.~(\ref{eq:eoms1a}), corresponding to the slow rolling of the Higgs field (the homogeneous mode of the $\chi_i$ fields is kept to zero). At $z = z_{\rm osc}$, we add the small inhomogeneous Fourier modes to all fields, and from then on, we evolve together Eqs.~(\ref{eq:eoms1a}) and (\ref{eq:eoms2}). The reader can find more details about our methodology for introducing the initial conditions in Appendix \ref{App-Init}. 

We have run simulations for different values of $\beta = \lambda^{1/4}\alpha$. Since we know that $10^{-3} \lesssim \alpha < 1$ and $10^{-5} < \lambda \lesssim 10^{-2}$, we find that $10^{-4} \lesssim \beta < 1$. We have thus run simulations for $\beta = 0.5, 0.1, 10^{-2}, 10^{-3}$ and $10^{-4}$. Note that the root mean square of $\beta$ is $\beta_{\rm rms} \simeq 0.115\lambda_{001}^{1/4}$ $\approx 0.115, 0.065, 0.037, 0.020$ for $\lambda_{001} = 1, 10^{-1}, 10^{-2}, 10^{-3}$, respectively. The probability distribution Eq.~(\ref{eq:ProbEQ}) for $\alpha$ (and hence $\beta$) is very non-Gaussian and, independently of $\lambda$, $\beta \gtrsim 0.5$ is exponentially suppressed. The range $10^{-4} \leq \beta \leq 0.1$ is obtained with more than $99\%$ ($99.8\%, 99.7\%, 99.4\%, 99.1\%$ for $\lambda_{001} = 1, 0.1, 0.01, 0.001$), while $\beta < 10^{-4}$ is attained with $< 1\%$ for all values of $\lambda$. Hence the values of $\beta$ that we have chosen, $\beta \in [10^{-4},0.5]$, sample fairly the range of random initial Higgs amplitudes dictated by $P_{\rm eq}(\varphi)$. 

The actual value of $\lambda$ is quite uncertain, since it depends on the energy scale of inflation. Besides, for a given Hubble rate $H_*$, it can still vary significantly given the uncertainties in $m_H, \alpha_s$ and $m_t$ (mostly in the latter). Due to this, for each value of $\beta$, we have chosen a set of 26 resonance parameters $q \equiv {g^2\over 4\lambda}$, logarithmically spaced between $q = 5$ and $q = 3000$. This corresponds to sampling the Higgs self-coupling from $\lambda \sim 10^{-5}$ to $\lambda \sim 10^{-2}$. Scanning this way $\beta$ and $q$ led us to characterize the behavior of the system, scrutinizing all possible different outcomes depending on $\lambda$ and $\varphi_*$. In Table~\ref{tab:I}, we list the values of all the resonance parameters $q$ that we have considered. We have guaranteed that by sampling different values, we include both the cases in which $q$ is within a resonance band of the Lam\'e equation, or in the middle of two bands (see Section \ref{sec:III}). 

Note that we have run simulations for three different expansion rates, corresponding to a MD universe, a RD universe, and a KD universe, given by $\omega=0,{1\over3}$ and $1$ in Eq.~(\ref{eq:ExpRate}), respectively. The following results in this section will be presented only for a RD background. The generalization to other expansion rates will be considered in Section~\ref{sec:vi}.

\begin{table}
    \begin{tabular}{| c | c | c | c | c | } \hline                       
        $q_W$ & $\lambda_{001} $ & $k_{\rm min} (q_W)$ & $k_{\rm max} (q_W)$  \\ \hline 
        $5$ & $1.5$ & $0.72$ & $ 1.09 $ \\ \hline 
        $6$ & $1.25$ & $0$ & $ 0.97 $ \\ \hline 
        $8$ & $0.938$ & $0$ & $ 0.69 $ \\ \hline 
        $9$ & $0.833$ & $0$ & $ 0.49 $ \\ \hline
        $11$ & $0.681$ & $1.33$ & $ 1.54 $ \\ \hline
        $14$ & $0.536$ & $0.67$ & $ 1.26 $ \\ \hline
        $18$ & $0.417$ & $0$ & $ 0.83 $ \\ \hline
        $23$ & $0.326$ & $1.43$ & $ 1.72 $ \\ \hline
        $29$ & $0.259$ & $0$ & $ 1.24 $ \\ \hline
        $37$ & $0.203$ & $1.75$ & $ 2.02 $ \\ \hline
        $48$ & $0.156$ & $0$ & $ 1.22 $ \\ \hline
        $61$ & $0.123$ & $1.36$ & $ 1.92 $ \\ \hline
        $79$ & $0.095$ & $2.06$ & $ 2.38 $ \\ \hline
        $101$ & $0.074$ & $0$ & $ 0.91 $ \\ \hline
        $130$ & $0.058$ & $0$ & $ 1.10 $ \\ \hline
        $167$ & $0.045$ & $0$ & $ 0.88 $ \\ \hline
        $214$ & $0.035$ & $2.34$ & $ 2.83 $ \\ \hline
        $275$ & $0.027$ & $0.56$ & $ 2.21 $ \\ \hline
        $354$ & $0.021$ & $2.71$ & $ 3.22 $ \\ \hline
        $454$ & $0.017$ & $0$ & $ 1.43 $ \\ \hline
        $584$ & $0.013$ & $0$ & $ 1.42 $ \\ \hline
        $750$ & $0.010$ & $2.93$ & $ 3.65 $ \\ \hline
        $1030$ & $0.0073$ & $1.18$ & $ 3.04 $ \\ \hline
        $1550$ & $0.0048$ & $0$ & $ 2.85 $ \\ \hline
        $2200$ & $0.0034$ & $0$ & $ 1.37 $ \\ \hline
        $3000$ & $0.0025$ & $0.87$ & $ 3.72 $ \\ \hline
    \end{tabular}
    \caption{Different resonance parameters $q$ used in the simulations, together with the corresponding values of the Higgs self-coupling derived for $g^2 = g^2_{W} \simeq 0.3$. For each case, we also provide the minimum and maximum momenta (in units of $H_*$), $k_{\rm min} \leq k \leq k_{\rm max}$, of the first resonance band. Half of the cases have a band down to $k_{\rm min} = 0$, while the others have $k_{\rm min} > 0$.}\label{tab:I}
    
\end{table}

Our simulations depend only on two parameters, $q$ and $\beta$. For each pair of values $(q,\beta)$, we have run simulations on a lattice with $N=128$ points per dimension, with periodic boundary conditions. The minimum momentum captured in each run is $k_{\rm m} = {2\pi\over Ndx}$, with $dx$ being the lattice spacing. The maximum momentum sampled in the lattice is $k_{\rm M} = {\sqrt{3}N\over 2} k_{\rm m}$. The length of the lattice box side is $L = N dx$. For each value of $\beta$ and $q$, we have made sure that our results are not sensitive to the lattice spacing $dx$ and/or the lattice size $L$. More details about these issues are given in Appendix \ref{App-Lattice}.

In Fig.~\ref{fig:means} we plot, as a function of time, the volume-average of the modulus of the (conformally transformed) Higgs field $|h|$. In this figure, we show the outcome corresponding to $\beta = 0.01$, and four different resonance parameters, $q = 8$, $14$, $101$ and $354$. The values $q = 8, 101$ are centered close to the middle of a resonance band of the Lam\'e equation, while $q = 14, 354$ are between adjacent bands. In this figure we also show the corresponding envelope curve of the Higgs oscillations. One conclusion is immediately clear: the time scale of the Higgs amplitude decay depends noticeably on $q$. By running simulations for each of the $q$ values displayed in Table~\ref{tab:I}, we have fully characterized the $q$ dependence of the Higgs decay. Note that in Table~\ref{tab:I} we have also indicated the range of momenta $k_{\rm min} \leq k \leq k_{\rm max}$ excited for each value of $q$, according to the Lam\'e equation. Such a range corresponds to the band with the largest Floquet index $\mu_{\rm max}$, which coincides in all cases with the most infrared band; see Fig.~\ref{fig:FloquetVarious}. The $\mu(k)$ index was obtained by solving the Lam\'e equation for a given $q$ parameter, and finding the range of momenta such that $\mu(k) > 0$. The band structure can be well appreciated in Fig.~\ref{fig:FloquetVarious}, where we plot $\mu(k)$ for each of the values of $q$ listed in Table I. As mentioned, we have sampled all possible cases, including when $q$ is within a resonant band (either close to the middle or to the extremes of the band), and hence $k_{\rm min} = 0$, or simply outside of any band (between adjacent bands), and then $k_{\rm min} > 0$.

\begin{figure}[t]
    \begin{center}
        \includegraphics[width=8.7cm]{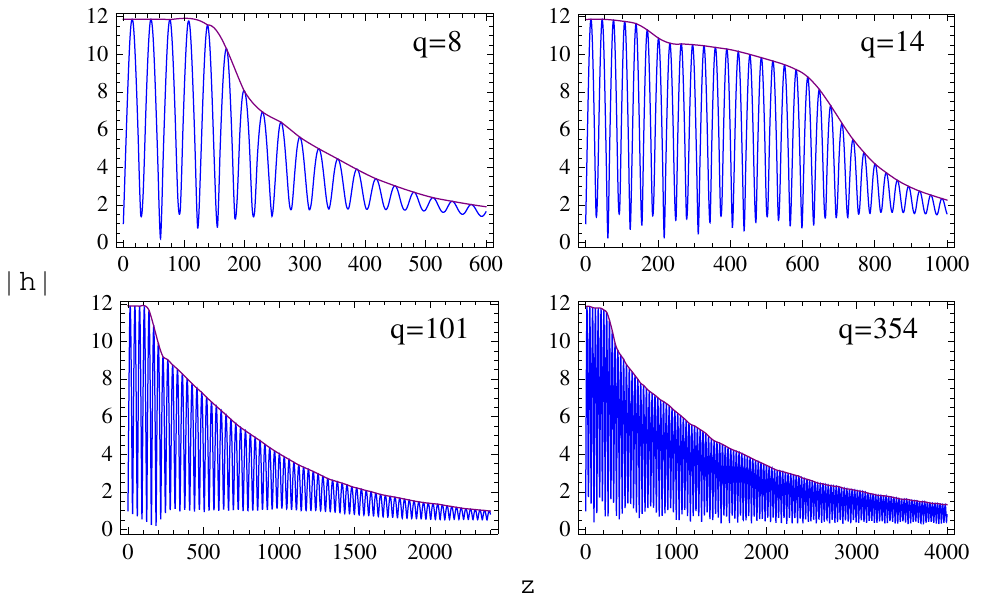}
    \end{center}
    \caption{Volume-averaged value of the Higgs field $|h|$ as a function of time, for four different resonance parameters,  $q=8,14,101$ and $354$. Also plotted, the corresponding envelope functions of the oscillations. All cases correspond to $\beta = 0.01$.}\label{fig:means}
\end{figure}

Before examining in more detail the general behavior of all the fields in the system, we can make some comments about the Higgs behavior. First of all, let us note that $h$ oscillates with a period $T$ which is, as expected, independent of the value of $q$. Even if it cannot be really appreciated in Fig.~$\ref{fig:means}$, we have checked that the period coincides initially with the analytical expression given by Eq.~(\ref{eq:Period}), until it becomes slightly modulated due to the interactions with the $\chi$ fields (though it does not change significantly). Looking at the different panels of Fig.~$\ref{fig:means}$, it seems that the Higgs decay is slower the greater the resonance parameter $q$ is. This is very opposite to the intuition gained by the study of the Lam\'e equation in Section \ref{sec:III}, which dictates that the larger the $q$, the shorter the decay time of the Higgs.\footnote{Contrary to 'popular wisdom' about parametric resonance, the time scale $z_{\rm eff}$, identified with the 'oscillatory field' decay time in the linear analytical approximation, is in practice mostly independent of $q$. It is true that the larger the $q$ the shorter the decay, but the dependence is only logarithmic [recall Eq.~(\ref{eq:EffEnergyTransferTimeScale})], and the number of oscillations does not change appreciably.} We thus see on this the first difference between the simplified study of the system of scalar fields in the linear regime (Section~\ref{subsec:III.a}), and the real outcome when nonlinearities are incorporated in lattice simulations. We will further comment on this issue later on.

\begin{figure}[t]
    \begin{center}
        \includegraphics[width=8cm]{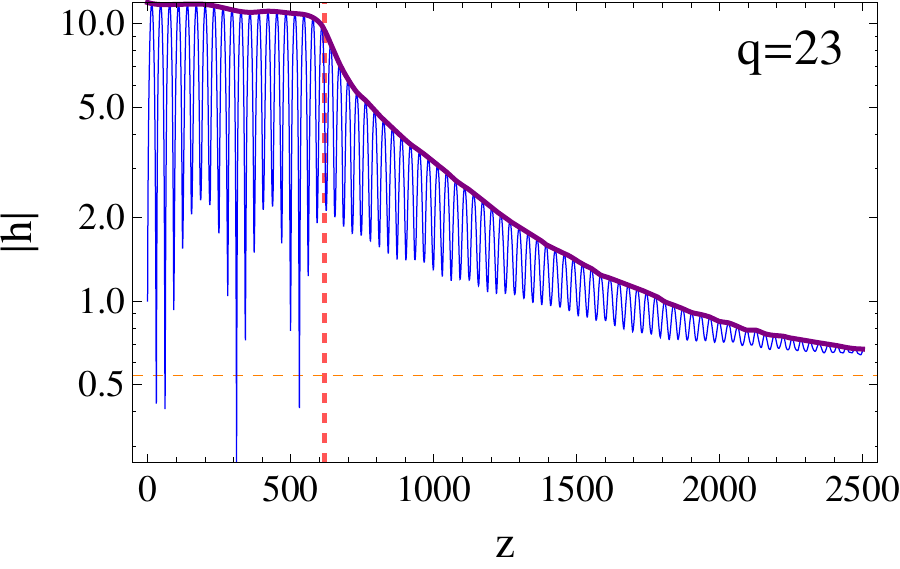}\vspace*{5mm}
        \includegraphics[width=8cm]{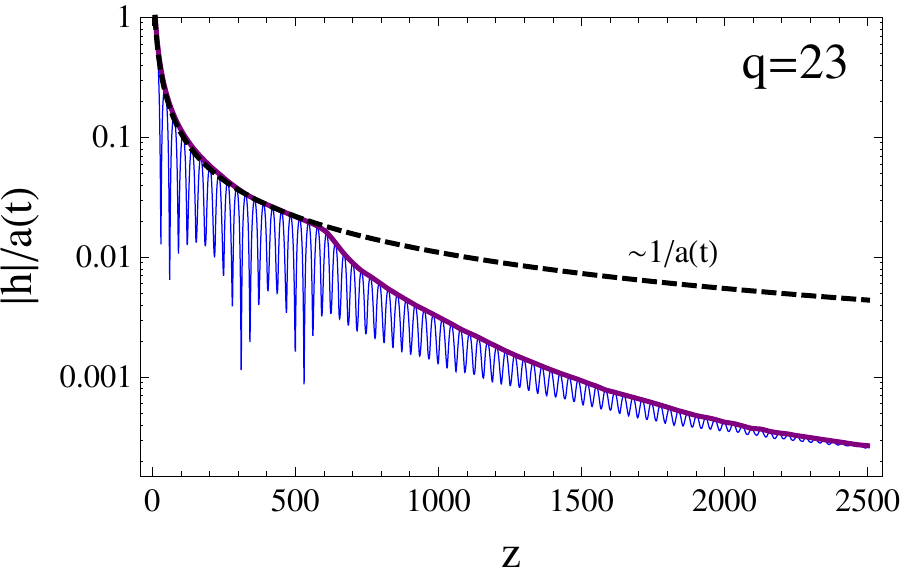}
    \end{center}
    \caption{Volume-averaged value of the Higgs modulus for $q=23$, $\beta = 0.01$ and RD. An initial plateau until $z = z_i$ can be clearly distinguished in the top panel, where we plot the conformally transformed Higgs. At later times $z > z_i$, the amplitude of the Higgs drops abruptly, due to its decay into the $\chi$ fields. In the lower panel we plot the physical Higgs $|\varphi|/\varphi_* = |h|/a$, where we can appreciate that the plateau for $h$ translates into a dilution $\propto 1/a$ for $\varphi$, due to the expansion of the universe. The decay of the Higgs into the other fields at later times is manifested by a significant decrement of $|\varphi|$ well below the $1/a$ decaying envelope.}\label{fig:means2}
\end{figure}

One can distinguish two different stages in each decay process. Let us look, for instance, at the upper panel of Fig.~\ref{fig:means2}, where the Higgs modulus is plotted for $q=23$, and where we also include the envelope curve of the oscillations. One can clearly appreciate that initially, and for some time, the envelope is approximately constant, reducing its amplitude only slightly. This is observed as a $plateau$ feature in the upper panel of Fig.~\ref{fig:means2}. The vertical dashed line in the figure indicates the end of this initial behavior, after which a second stage of rapid decay follows. Let us note that when we talk about the decay of the Higgs amplitude, we refer to the conformally transformed one $h$. The amplitude of the physical Higgs ${\varphi}/\varphi_* = h / a(t)$ is always decaying with the scale factor, no matter what its coupling to other species is. Before the second stage starts, the physical Higgs amplitude $\varphi$ decays mostly due to the expansion of the Universe, and not because of an efficient transfer of energy into the scalars. However, both effects are combined afterwards, producing an even more sharp decay of the physical amplitude. This is clearly seen in the lower panel of Fig.~\ref{fig:means2}. 

In order to understand better this two-stage behavior, we plot the different contributions to the total energy of the system as a function of time. The energy density can be conveniently written as
\bea \rho (z) &=& V_* {E_t (z)\over a(z)^4}\,,~~~ V_* \equiv \frac{\lambda \varphi_*^4}{4} \,,\\
E_t (z) &=& E_{\rm K}^{\varphi} + E_{\rm V} + E_{\rm G}^{\varphi} + E_{\rm K}^{\chi} +  E_{\rm G}^{\chi} + E_{\rm int} \,, \label{energy_sc} \eea
where, for our choice of variables, the Higgs and $\chi$ field contributions to the kinetic (K) energy are given by ($~\dot{} \equiv d/dt, ~' \equiv d/dz$)
\bea E_{\rm K}^{\varphi} &\equiv & {a^4\over V_*}{{\dot\varphi}^2\over 2a^2} = \frac{2}{\beta^2} \left(h' - h\frac{a'}{a} \right)^2\,,\\
E_{\rm K}^{\chi} &\equiv & {a^4\over V_*}{\dot\chi_i\dot\chi_i\over 2a^2} = \frac{2 \lambda}{\beta^4} \sum_{i=1}^3 \left( X_i' - X_i \frac{a'}{a} \right)^2 \ , \eea
the gradient (G) contributions by 
\bea  
E_{\rm G}^{\varphi} &\equiv & {a^4\over V_*}{{|\vec{\nabla}\varphi|}^2\over 2a^2} = \frac{2}{\beta^2} |\vec{\nabla} h|^2 \,, \\E_{\rm G}^{\chi} &\equiv & {a^4\over V_*}{\vec{\nabla}\chi_i\vec{\nabla}\chi_i\over 2a^2} = \frac{2\lambda}{\beta^4} \sum_{i=1}^3 |\vec{\nabla}X_i|^2 \,,
\eea
and finally, the Higgs potential (V) energy and the interaction (int) term, by
\bea 
E_{\rm V} &\equiv & {a^4\over V_*}{\lambda \varphi^4\over 4} = h^4 \,,\\
E_{\rm int} &\equiv & {a^4\over V_*}{e^2\over 2}\varphi^2\chi_i\chi_i = \frac{2 e^2}{\beta^2} h^2 \sum_i X_i^2\,.
\eea

\begin{figure}[t]
    \begin{center}
        \includegraphics[width=8.5cm]{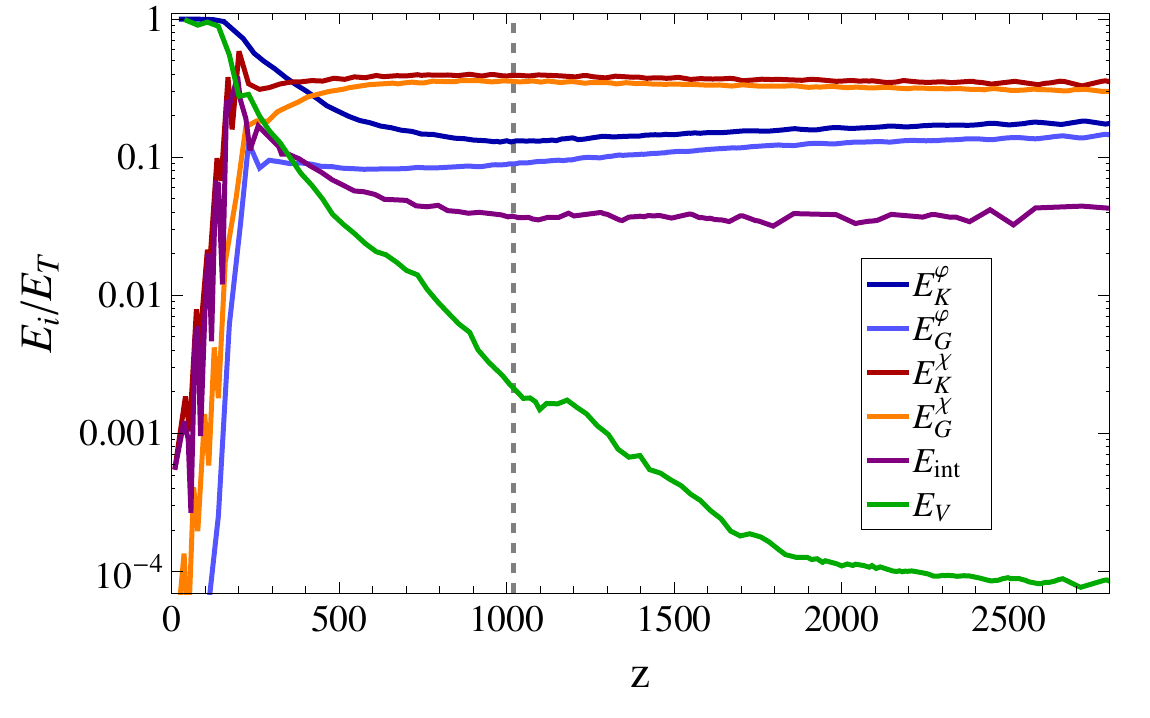}\vspace*{5mm}
        \includegraphics[width=8.5cm]{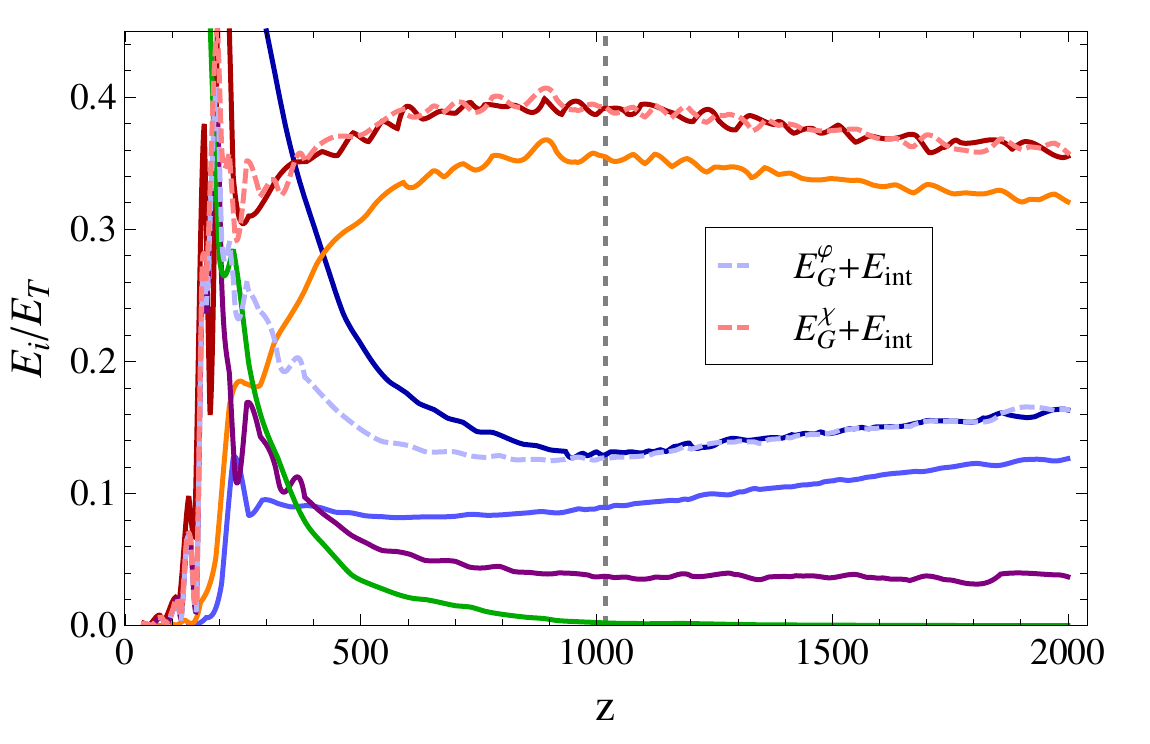}
    \end{center}
    \caption{Top: We show the envelope curves of the oscillations of the different contributions to the total energy $E_t(z)$, obtained for $q=8$, $\beta = 0.01$ and RD. The gray, vertical dashed line corresponds to the decay time $z_e$ for $q=8$. Bottom: Same quantities as in the upper figure (same color coding), but zooming in the area of interest. We also add two new lines, a pink one corresponding to the sum of the Higgs gradient energy and the interaction energy, and a light blue line, representing the sum of the $\chi$ fields' gradient energy plus the interaction energy. We see that the decay time indicates equally good the time when the Higgs kinetic energy stops decaying, and the time when equipartition is set.}\label{fig:energies1}
\end{figure}

In Fig.~\ref{fig:energies1} we have plotted the different contributions to $E_t(z)$ for the parameters $\beta = 0.01$ and $q=8$. Initially, the system is dominated by the kinetic and potential energy densities of the Higgs. This corresponds to the regime of anharmonic oscillations of the Higgs condensate described in Section~\ref{sec:II}, for when the coupling to other fields was ignored ($g^2 \rightarrow 0$). However, in reality, as soon as the Higgs starts to oscillate, there is an energy transfer into any species coupled to the Higgs. Each time the Higgs crosses zero, a fraction of its energy goes into the $\chi$ fields. Initially, the amount of energy transferred at each zero crossing is small relative to the total energy stored in the Higgs. Therefore, it takes some time until the transfer becomes noticeable. The Higgs energy components represent the dominant contribution to the total energy during the initial oscillations, so the Higgs evolves initially without really noticing the presence of the other fields. Eventually, at the time $z = z_i$, the energy transferred into the $\chi$ fields becomes significant enough, compared to the Higgs energy itself, say a fraction $\rho_\chi/\rho_{\varphi} = \delta < 1$. The Higgs condensate becomes affected by the transfer of energy into the other fields when $\delta \gtrsim \delta(z_i) \equiv 0.1$. From then onwards, the Higgs continues pumping energy into the other fields at $z > z_i$, but the amount of energy transferred at each zero crossing is no longer a small fraction of the energy available in the Higgs condensate itself. Therefore, soon after backreaction becomes noticeable at $z = z_i$, the previously exponential growth of the $\chi$ fields energy densities stops, eventually saturating to a fixed amplitude. This is clearly seen in Fig.~\ref{fig:energies1}, where the gradient and kinetic energy densities of the $\chi$ fields saturate to an almost constant amplitude. This happens because the Higgs has not enough energy anymore to accomplish transferring a sizable fraction of energy into the $\chi$ fields. At the same time, immediately after $z = z_i$, the Higgs energy density drops abruptly. This is so because the amount of energy transferred from the Higgs into the other fields, even if not significant anymore compared to the energy stored in the $\chi$ fields (hence the saturation of their growth), represents a significant fraction of the energy available in the Higgs at that moment. Therefore, the energy of the Higgs (mostly dominated by the kinetic contribution) drops abruptly, as can be clearly seen, for instance, from $z_i \approx 175$ to $z \approx 900$, for the case depicted in Fig.~\ref{fig:energies1}. 

\begin{figure}[t]
    \begin{center}
         \includegraphics[width=8.3cm]{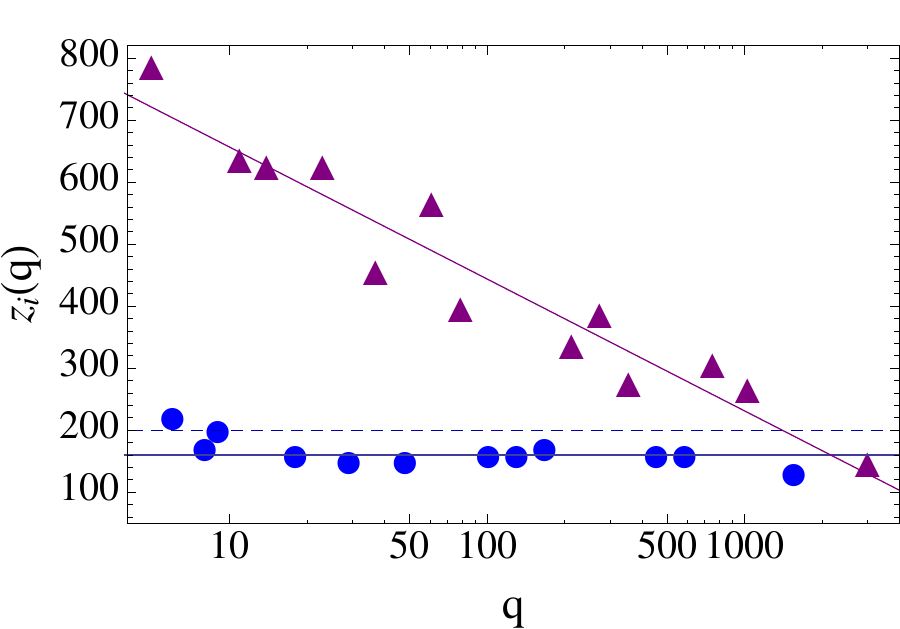}
        \includegraphics[width=8.5cm]{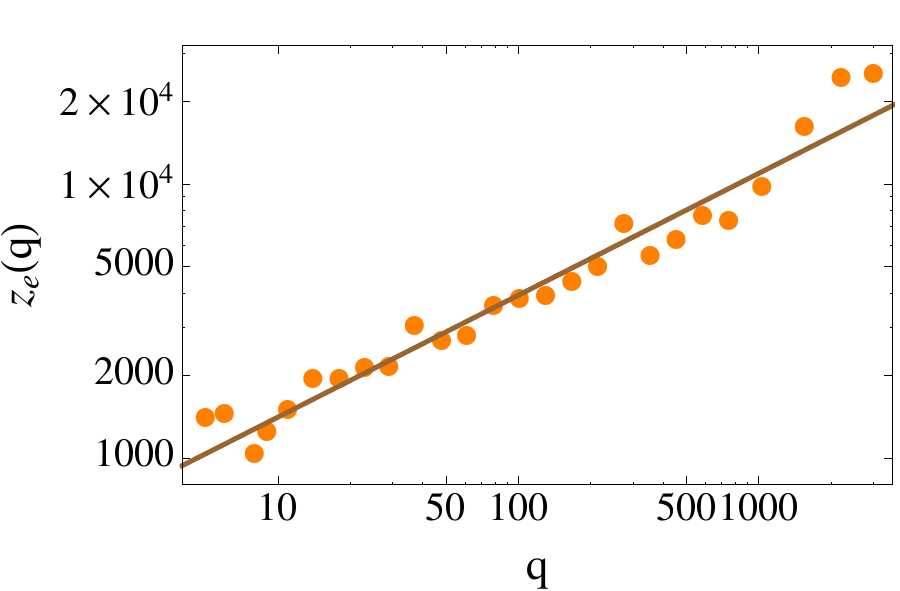}
    \end{center}
    \caption{Top: The different times $z_i (q)$ obtained from our simulations, for RD ($\omega = 1 / 3$) and $\beta = 0.01$.  Purple triangles and blue circles correspond to $q$ parameters inside or outside a resonance band of the Lam\'e equation respectively.  The blue and purple continuous lines correspond to the best fit to the circles and triangles respectively; Eq.~(\ref{eq:FitZsScalarsQnoResonance}). The dashed line corresponds to the analytical estimate $z_{\rm eff} \approx 200$, obtained from Eq.~(\ref{eq:EffEnergyTransferTimeScaleApprox}) with $\bar{\mu}_k = 0.2$. Bottom: The different points show the Higgs time decay $z_e (q)$ as a function of $q$ obtained from our simulations for the same $(\omega,\beta)$ values as the upper panel. The brown line corresponds to the best fit, Eq.~(\ref{eq:zs(q)}). }\label{fig:scalar-fit}
\end{figure}

Note that when the Higgs energy density starts decreasing significantly at $z \gtrsim z_i$, the Higgs amplitude also starts decreasing noticeably. However, while the Higgs energy density eventually stops decaying and saturates to an almost constant value, the amplitude $|h|$, instead, continues decreasing during a much longer time. The long-lasting decay of the Higgs amplitude induces the decrease of the potential energy of the Higgs, even long after the dominant energy components of the Higgs have saturated, as can be clearly appreciated in Fig.~\ref{fig:energies1}. This is simply due to the fact that the Higgs keeps on oscillating and hence transferring energy into the $\chi$ fields. Since soon after $z = z_i$ the energy in the Higgs becomes smaller than the energy in the $\chi$ fields, the continuous transfer of energy represents only a marginal fraction of the energy already accumulated in the latter. Hence, the amplitude reached by the gradient and kinetic energy terms $E_K^{\chi}, E_G^{\chi}$ is not affected anymore, whereas the amplitude of the Higgs potential energy continues decreasing. Eventually, the transfer of energy from the Higgs becomes inefficient and $E_V$ also saturates to an approximately constant value. By then, however, the Higgs potential energy is completely irrelevant compared to the gradient and kinetic counterparts. 

A very relevant aspect to note is that when all the energy contributions stop growing or decreasing abruptly (with the exception of the Higgs potential energy, which keeps on falling for a long time), the energy components reach equipartition. In particular, some time at $z > z_i$, the kinetic energy $E_K^{\varphi}$ of the Higgs becomes equal to the sum of the Higgs gradient energy plus the interaction energy, $E_G^{\varphi} + E_{\rm int}$; see the lower panel of Fig.~\ref{fig:energies1}. In other words, equipartition in the Higgs sector holds\footnote{In reality, it should be $E_K^{\varphi} = E_G^{\varphi} + E_{\rm int} + E_{V}$, but $E_{V}$ is so small by then, that it does not make a difference to add it or not.} as $E_K^{\varphi} = E_G^{\varphi} + E_{\rm int}$.
Similarly, in the $\chi$ fields, the sum of their gradient energy plus the interaction term, 'equipartitionates' with their kinetic energy, $E_K^{\chi} = E_G^{\chi} + E_{\rm int}$, as can also be well appreciated in the lower panel of Fig.~\ref{fig:energies1}.

All features described so far are, of course, not specific to the particular case $q = 8$, $\beta = 0.01$ and RD, shown in Fig.~\ref{fig:energies1}. A similar behavior is observed in the outcome of the field distribution for other choices of $\beta$, $q$ and $\omega$. That is, there is always initially a $plateau$-like stage during which the Higgs (conformal) amplitude remains almost constant (or changes only marginally) for a few oscillations. Then, the amplitude decreases fast when the backreaction from the $\chi$ fields becomes noticeable, which causes at the same time the ceasing of the exponential growth of the $\chi$ fields energy density. Eventually, all fields relax into a stationary distribution with exact equipartition $E_K^{\varphi} = E_G^{\varphi} + E_{\rm int}$ and $E_K^{\chi} = E_G^{\chi} + E_{\rm int}$. On the other hand, $E_V^{\varphi}$ becomes completely negligible as compared to any other energy term $E_i$, in correspondence with the decay of the Higgs amplitude, which carries on after equipartition is set. The duration of the different stages, for a given expansion rate, is directly related to the specific values of the parameters $\beta$ and $q$. In particular, the duration of the initial plateau is directly dependent on the band structure of the Lam\'e equation.

We have characterized the dependence of $z_i$ with the resonant parameter $q$; see Fig.~\ref{fig:scalar-fit}. Let us recall that $z_i$ corresponds to the moment when the energy transferred into the $\chi$ fields is sufficiently large so that the Higgs amplitude and energy density starts to decrease. Therefore, this is the moment that should be compared to the analytical estimate Eq.~(\ref{eq:EffEnergyTransferTimeScaleApprox}) of the Higgs decay time $z_{\rm eff}$, derived in Section~\ref{subsec:III.a}. 

The $z_i(q)$ behavior can be characterized by  
\begin{eqnarray}
\label{eq:FitZsScalarsQnoResonance}
z_i (q) \sim \left\lbrace\begin{array}{lcr}
160& ,& q \in {\rm Resonant~Band}\vspace*{1.5mm}\\
869 -92\log{q} & ,& q \notin {\rm Resonant~Band}\\
\end{array} \right.
\end{eqnarray}
If a given $q$ is within a resonant band, $z_i(q)$ is almost independent of $q$, as appreciated in Fig.~\ref{fig:scalar-fit}.
For RD and $\beta = 0.01$, our analytical estimate Eq.~(\ref{eq:EffEnergyTransferTimeScaleApprox}) predicts $z_{\rm eff} \simeq 200$, which is reasonably similar to the fit found from our numerical outcome, $z_i(q) \approx 160$. The analytical estimates are only an approximation to the real dynamics, and one cannot expect anything more than a reasonable order-of-magnitude prediction, as is indeed the case. More importantly, the analytical calculation predicts that $z_{\rm eff}$ should be only dependent on $q$ logarithmically, which in practice implies that for mildly broad resonance parameters as the ones we have, $q \sim \mathcal{O}(10)-\mathcal{O}(10^3)$, $z_{\rm eff}$ is essentially independent of $q$, as is indeed well appreciated in Fig.~\ref{fig:scalar-fit}. 

The dependence of $z_i(q)$ with $q$'s outside resonance bands is also logarithmic, though with a big coefficient. As it can be appreciated in the upper panel of Fig.~\ref{fig:scalar-fit}, for $q \lesssim 10^2$ it is a factor $\sim$2-4 larger than the analytical prediction Eq.~(\ref{eq:EffEnergyTransferTimeScaleApprox}), but becomes of the same order for $q \sim 10^2-10^3$, modulo a factor $\sim$1-2. Possibly, for $q \gg 10^3$, $z_i(q)$ will become smaller, but as said before, such regime is never valid in our case of study.

\begin{figure*}[t]
    \begin{center}
        \includegraphics[width=7.8cm]{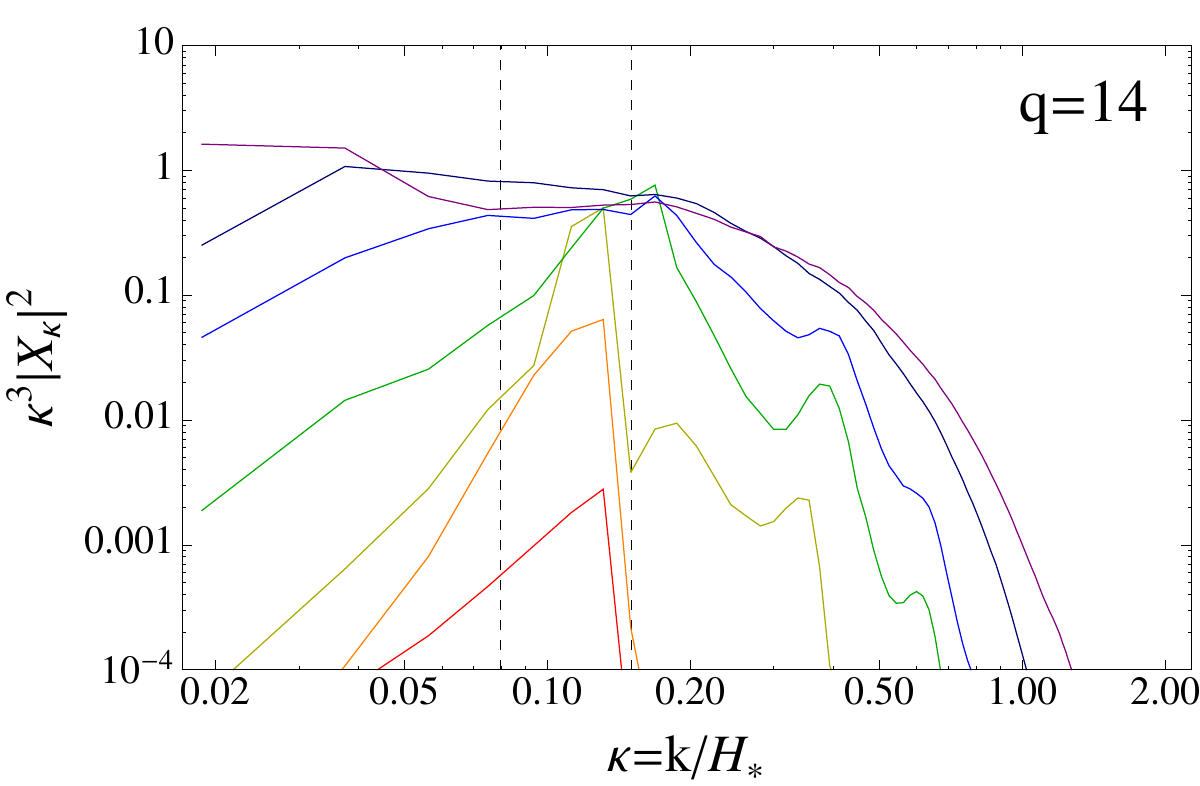} \hspace{0.2cm}
        \includegraphics[width=9.3cm]{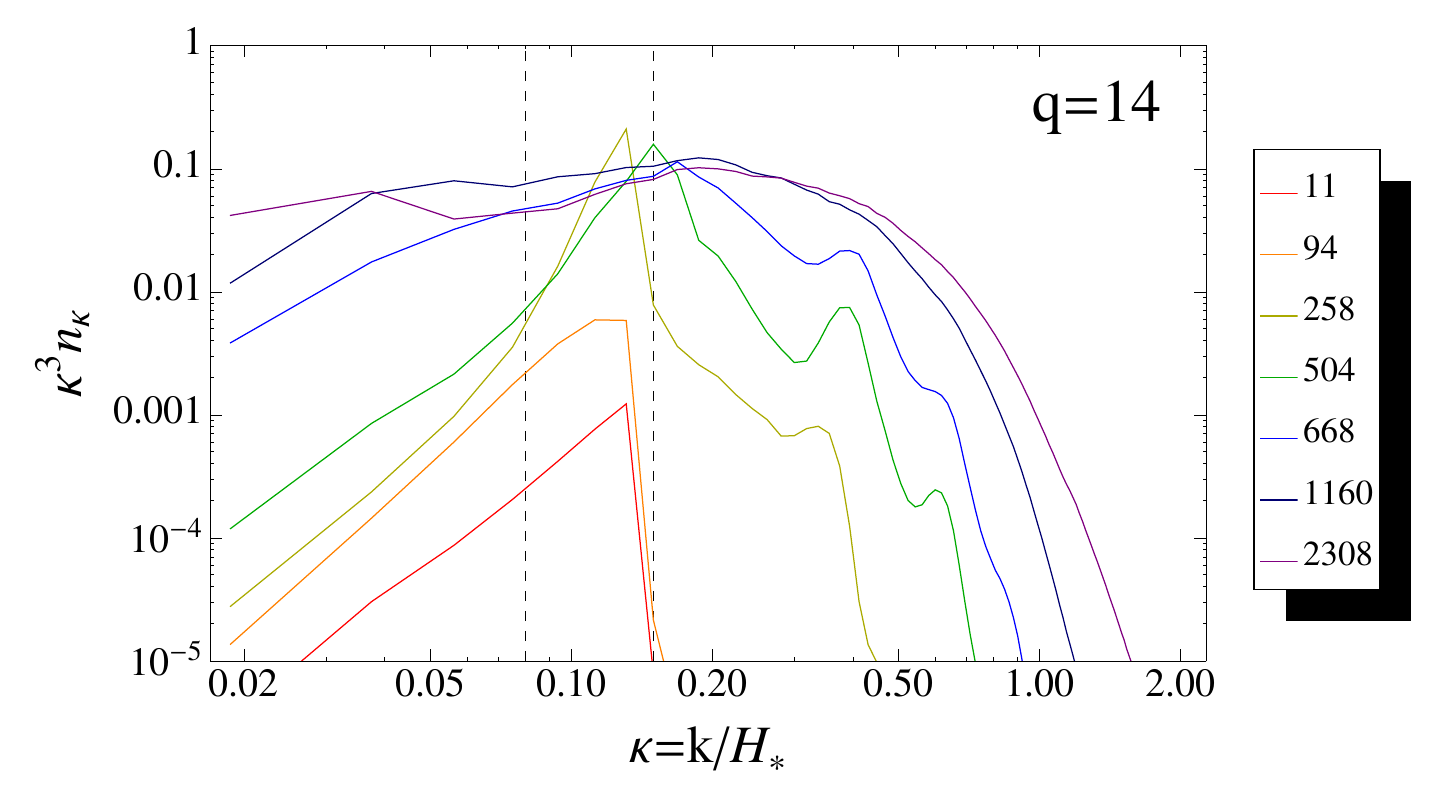}
        \includegraphics[width=7.8cm]{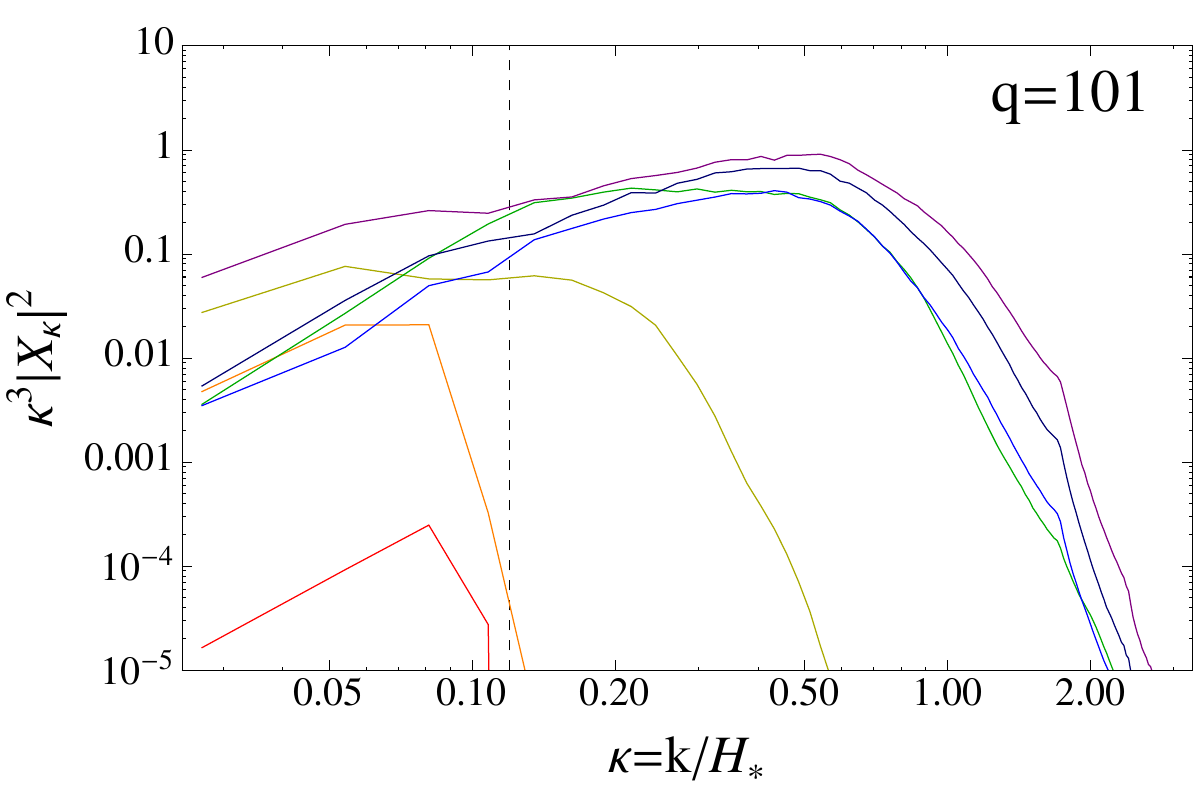} \hspace{0.2cm}
        \includegraphics[width=9.3cm]{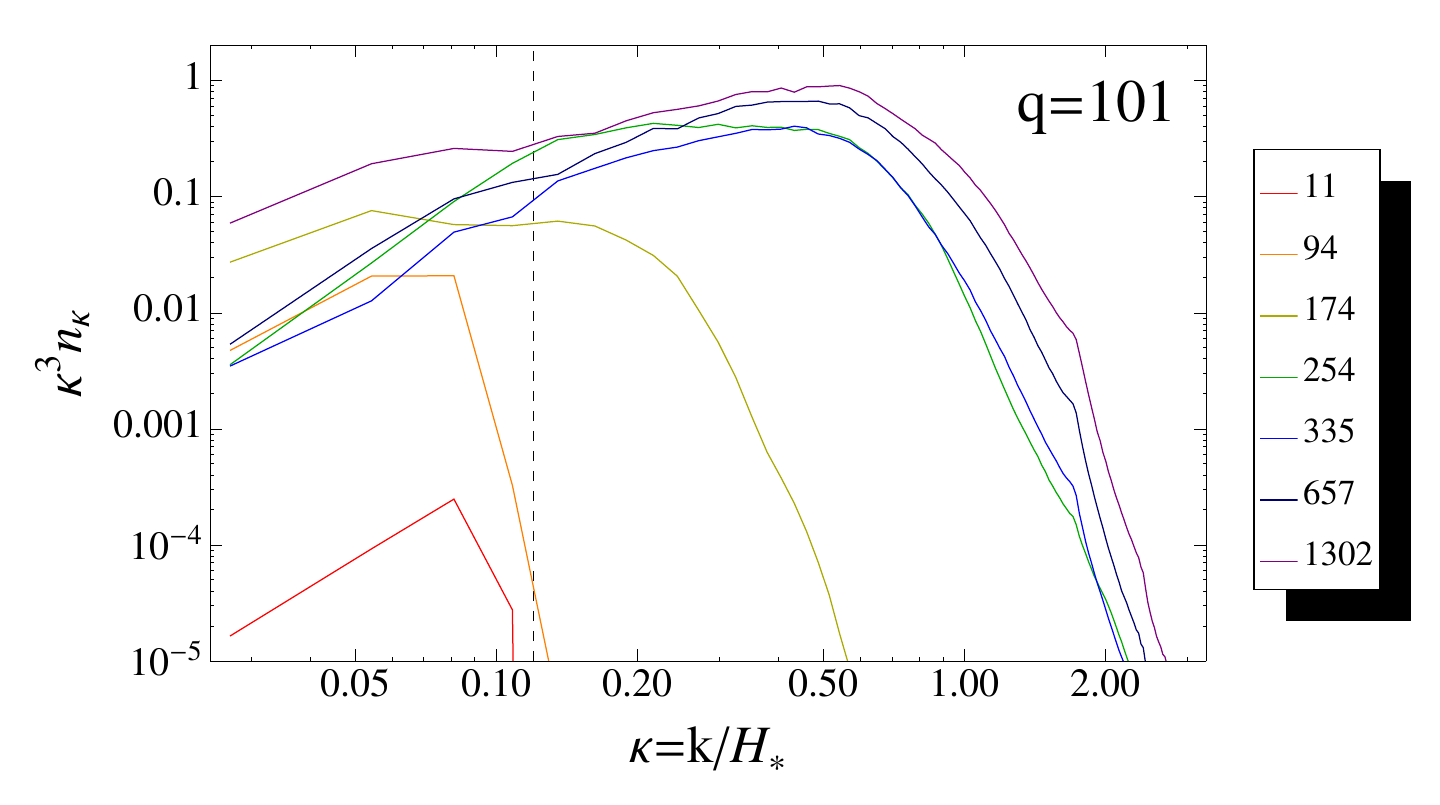}
    \end{center}
    \caption{Left: Spectra for $q=14, 101$ of one of the scalar fields, $\chi_1$. Right: Occupation number of $\chi_1$ field. The dashed vertical lines in the four figures indicate the position of the corresponding band of the Lam\'e equation. Note that the units used to express the momentum are different from the ones used in Fig.~\ref{fig:FloquetVarious}.} 
    \label{fig:spectraScalars}
\end{figure*}

In light of the results of this section, we see that the Higgs decay should be identified, rather than with $z_i$, with the abrupt drop of the Higgs energy density, some time afterwards at $z > z_i$. After the drop, the kinetic contribution $E_{K}^{\varphi}$ (which is the dominant energy component of the Higgs) enters into a stationary regime, equipartitioned with $E_G^{\varphi} + E_{\rm int}$. The onset of this regime signals the end of the decrease of the Higgs kinetic energy. We therefore provide a definition of the decay time of the Higgs, $z_e$, as the moment when equipartition (within the Higgs sector) holds better than a given percentage. In practice, we operationally determine $z_e$ as the moment when the equality $E_{K}^{\varphi} \simeq E_G^{\varphi} + E_{\rm int}$ holds to better than $1\%$. Defining like this the Higgs decay might seem arbitrary, but when looking carefully at the evolution of the energy components, we see that the end of the drop of the Higgs kinetic energy $E_{K}^{\varphi}$, coincides always with the onset of its equipartition with $E_G^{\varphi} + E_{\rm int}$, for all resonant parameters. From then onwards, i.e.~for $z > z_e$, all energy components (with the exception of the Higgs potential) enter into a stationary regime, evolving very slowly, preserving all the time the equipartition condition, $E_K^{\varphi} \simeq E_G^{\varphi} + E_{\rm int}$ and $E_K^{\chi} \simeq E_G^{\chi} + E_{\rm int}$. 

The dependence of the decay time scale $z_e$ versus $q$ is shown in the lower panel of Fig.~\ref{fig:scalar-fit}. A fit to this relation is given by
\be z_e (q) = 507 q^{0.44} \ . \label{eq:zs(q)} \ee
This is valid for $\beta=0.01$ and for a RD ($w$) background. As we shall explain in Section \ref{sec:vi}, this fit can be generalized to other $\beta$ values within our range of interest, and to other expansion rates (characterized by the equation of state $w$), as
\be z_e (q) \approx 50.7 \beta^{\frac{-(1+ 3 \omega)}{3 (1 + \omega)}} q^{0.44} \ . \label{eq:zs(q)-extrap} \ee

As we can see, the behavior of the Higgs decay time is actually independent of whether $q$ is within or outside a resonance band. More remarkably, the growth of $z_e(q)$ with $q$ is actually quite contrary to the intuition obtained from solving the Lam\'e equation. In the linear regime $z < z_i$, when the Lam\'e equation is valid, we expect that the larger the resonance parameter, the faster the transfer of energy from the Higgs to its decay products. Such trend is clearly observed (see upper panel of Fig.~\ref{fig:scalar-fit}), where $z_i(q)$ either changes only logarithmically or decreases with $q$, for parameters within or outside resonance bands, respectively. It is, however, $z_e$, as explained, that should be interpreted as the decay time of the Higgs. The behavior of $z_e$ is set by the nonlinearities of the problem, as opposed to $z_i$, which is determined by the linear regime. This results in a completely opposite trend to $z_i$, given the growth of $z_e$ with $q$. This remarkable fact, due to the nonlinear behavior of the system, represents one of the most relevant results of the paper. 

To conclude the section, we will briefly describe the dynamics of the system in the spectral domain. During the initial stages, the modes that are excited correspond to those in the band structure of the Lam\'e equation. We clearly see this for $z < z_i (q)$ in Fig.~\ref{fig:spectraScalars}, where we plot both the field spectra $k^3|X_k|^2$ and its occupation number $k^3n_k$. We also indicate with dashed lines the resonance bands. As the amplitude of the modes within the resonance bands grows, the system becomes more and more nonlinear. Rescattering among modes occurs, and the bands become wider. Due to the coupling of the modes through Eqs.~(\ref{eq:eoms1a}) and (\ref{eq:eoms2}), the initial parametric resonance of the $\chi_k$ modes within the resonance bands, excite at the same time Higgs modes $\varphi_{k'}$, which then rescatter off other modes $\chi_{k''}$, and so on. As a consequence, the power spectrum of the fields grows exponentially and widens, with a typical width $0 \leq k \lesssim \mathcal{O}(10)k_*$. As we have discussed in detail, at late times $z \gtrsim z_e$ the fields enter into a stationary stage, characterized by equipartition and a very slow evolution of the energy densities. This stage is indeed associated with a turbulent regime, typically expected to be developed due to the nonlinear character of a multifield interacting system~\cite{Micha:2004bv,Micha:2002ey} (see also~\cite{DiazGil:2005qp,DiazGil:2007qx}). The onset of this regime translates into the field distributions entering in a self-similar evolution, with the occupation numbers verifying a scaling law of the type,
\begin{eqnarray}
n(k,t) \simeq t^{-q}n_o(kt^{-p})\,,
\label{eq:SelfSimTurb}
\end{eqnarray}
with $p < 1$ and $q/p \gtrsim 1$ typically, and $n_o(k)$ a universal function specific to each species. We have checked that at late times $z \gg z_e$, the evolution of the Higgs occupation number follows quite accurately Eq.~(\ref{eq:SelfSimTurb}), with $p \approx 1/4$ and $q/p \approx 2.7$. The late-time evolution of the occupation number of the $\chi$ fields, however, can be fitted into the form of Eq.~(\ref{eq:SelfSimTurb}) only to some extent, since any value between $p = 1/7$ and $p = 1/12$ does an equally good job (as long as $p/q$ changes accordingly between $3$ and $4$), and the high-momentum tails are always somewhat offset with respect to the $n_o(k)$ tails. Eventually the system is expected to relax into a thermal distribution. The turbulent regime is, however, not very efficient in transferring energy from the long-wave modes to the high-momentum region, so an eventual total thermalization is indeed a long way off from the onset of the stationary regime (also from our typical running times in the simulations).

In the next section, we will present a similar analysis of the properties of the Higgs decay process, but finally introducing the gauge nature of the interactions. Before we move on, let us recall again that all our results of Section~\ref{sec:iv}, correspond to RD and were obtained for a fixed value $\beta = 0.01$. We will devote Section~\ref{sec:vi} to an analysis of how the results change when varying the Higgs initial amplitude (i.e.~$\beta$) and the background expansion rate (i.e.~$w$).

\section{Lattice Simulations, Part 2: Abelian-Higgs Modeling}
\label{sec:v}

In this section, we study the properties of the Higgs decay modeling the system with an Abelian-Higgs framework. In this approach, and in contrast with the global scenario, we introduce for the first time a gauge structure in the interactions. The differences and similarities in the results of these two scenarios will be scrutinized. We will approximate the action of the electroweak sector of the standard model, invariant under the local SU(2)$\times$U(1) symmetry group, by a local U(1) gauge theory. This is justified in principle because, as we will show explicitly in Section~\ref{subsec:BeyondAH}, the corrections due to the non-Abelian nature of the SM interactions, are not expected to play any significant role.

Let us note that for practical reasons, we will continue considering a system where the Higgs is only coupled to a single gauge boson, with resonance parameter $q = g^2/4\lambda$. This way we will be able to compare directly the results from the gauge theory, with those from the previously studied global scenario. Towards the end of this section, however, we will consider the real case of the Higgs decaying simultaneously into the three gauge bosons $W^+$, $W^-$ and $Z$. Only in that way we will be really approaching realistically the dynamics of the SM. Remarkably, as we will demonstrate in Section~\ref{sec:vB}, the system of three (Abelian) gauge bosons can be effectively mapped into a system with only one gauge field, with effective resonance parameter $q = q_Z + 2 q_W$. Thanks to this, we will show explicitly that any analysis carried out with only one gauge boson, can be used directly, after applying an appropriate mapping, to fully understand the dynamics of the system with three gauge bosons. We will justify $a~posteriori$ this way, the ability and utility of modeling the system with a single gauge boson, as considered so far.

The Abelian-Higgs model with one gauge boson is described by the action $S = \int \mathcal{L}\, d^4x$, with Lagrangian
\be - \mathcal{L} =   (D_{\mu} \varphi)^* (D^{\mu} \varphi ) + \frac{1}{4 e^2} F_{\mu \nu} F^{\mu \nu} + \lambda (\varphi^* \varphi )^2  \ , \label{action}\ee
where the covariant derivative and field strength are defined as usual,
\be D_{\mu} = \partial_{\mu} - i A_{\mu}  \ ,  \hspace{0.5cm} F_{\mu \nu} \equiv \partial_{\mu} A_{\nu} - \partial_{\nu} A_{\mu} \ . \ee
Here, $e$ is the Abelian coupling strength representing the coupling of either one of the $W^\pm$ or $Z$ gauge fields. As before, in order to mimic correctly the Higgs-gauge interactions, we need to take $e^2 = g^2/4$, with $g^2 = g_W^2$ or $g_Z^2$, respectively for $W$ or $Z$ bosons.

Since we are working with a system invariant under a local $U(1)$ transformation, we must take consequently the Higgs as a complex field. In terms of its components we shall write it as
\be \varphi = \frac{1}{\sqrt{2}} ( \varphi_1 + i \varphi_2 ) \ ,  \hspace{0.6cm} \varphi_i \in \mathfrak{R} \ . \ee
From Eq.~(\ref{action}) we derive the following equations of motion
\bea \ddot{\varphi} - D_i D_i \varphi + 2 \mathcal{H} \dot{\varphi} + 2 \lambda a^2 (t) |\varphi|^2\varphi  &=& 0 \ , \label{eom1} \\
\partial_0 F_{\mu 0} - \partial_i F_{\mu i} + 2 e^2 a^2 (t) \mathfrak{Im}[ \varphi^* D_{\mu} \varphi ]  &=& 0 \ .  \label{eom2} \eea

As we are dealing with a gauge theory, we have a gauge freedom in the choice of the field components. This allows us to set, from now on, the condition $A_0=0$. In this case, the EOM of the gauge fields, Eq.~(\ref{eom2}), can be written in terms of its components as
\bea
\ddot{A}_j + \partial_j \partial_i A_i - \partial_i \partial_i A_j &=& 2 e^2 a^2 (t) \mathfrak{Im} [ \varphi^* D_j \varphi ] \ ,\label{eomb} \\
\partial_i \dot{A}_i &=& 2 e^2 a^2(t)  \mathfrak{Im}[\varphi^* \varphi] \ . \label{eomc} \eea
Eq.~(\ref{eomc}) is the Gauss law, which represents a constraint that the solution to Eqs.~(\ref{eom1}) and (\ref{eomb}) must preserve at all times. When solving these equations in a three-dimensional lattice, one must of course check that the Gauss constraint Eq.~(\ref{eomc}) (or more specifically, its equivalent discretized version) is indeed preserved during the whole evolution of the system. We also define the gauge-invariant electric and magnetic fields as usual
\be E_i \equiv \dot{A}_i \ , \hspace{0.6cm} B_i = \frac{1}{2} \epsilon_{i j k} ( \partial_j A_k - \partial_k A_j )  \ .\ee

As in the global scenario, it is really useful to redefine the spacetime and field variables. On the one hand, we change to the same set of dimensionless spacetime coordinates $z^{\mu} = (z^0, z^i)$ introduced in Section \ref{sec:iv}, 
\be z \equiv z^0 = H_* t \,,~~~ z^i = H_* x^i \label{def-st2} \ . \ee
On the other hand, it is also convenient to define new Higgs and gauge field dimensionless variables as
\be h_{j} \equiv  \frac{a(z)}{a_*} \frac{\varphi_j}{\sqrt{2} \varphi_*} \ ,  \hspace{1cm} V_{i} \equiv \frac{1}{H_*} A_{i}  \ . \label{defh} \ee
(with $j=1,2$; $i=1,2,3$) where $\varphi_* \equiv |\varphi (t_*)|$ is the initial modulus of the complex Higgs field at the end of inflation. To distinguish between different variables, we use a dot or a prime to denote differentiation with respect conformal or natural variables ($~\dot{} \equiv d/dt, ~' \equiv d/dz$), respectively. From now on, all spatial derivatives will also be with respect the new variables, unless otherwise stated. We also define a dimensionless covariant derivative as 
$$\mathcal{D}_i \equiv \frac{ \partial}{\partial z_i} - i V_i\,.$$
With these changes, Eqs.~(\ref{eom1})-(\ref{eomc}) can be written as
\bea 
h_1''  - \mathfrak{Re}[\mathcal{D}_{i} \mathcal{D}_{i} (h_1 + i h_2) ] + \beta^2 (h_1^2 + h_2^2) h_1  &=& h_1 \frac{a''}{a} \ ,   \label{eomb1}\\
h_2''  - \mathfrak{Im}[ \mathcal{D}_{i} \mathcal{D}_{i} (h_1 + i h_2) ] +  \beta^2 (h_1^2 + h_2^2) h_2  &=& h_2 \frac{a''}{a} \ , \hspace{0.8cm} \label{eomb2}\\
V_j'' + \partial_{j} \partial_{i} V_i - \partial_{i} \partial_{i} V_j &=&  j_i (z) \ ,  \label{eomb3}\\
\partial_i V_{i}' &=&  j_0 (z) \ , \label{eomb4}  
\eea
where the current $j_{\mu} (x)$ is defined by
\be j_{\mu} (x) \equiv q \beta^2  \mathfrak{Im} [ (h_1 - i h_2) \mathcal{D}_{\mu} (h_1 + i h_2) ] \ .\ee
Finally, we also define dimensionless electric and magnetic fields as
\bea \mathcal{E}_i \equiv V'_i = \frac{E_i}{H_*^2}\,,~ ~ \mathcal{B}_i \equiv \frac{1}{2} \epsilon_{i j k} ( \partial_j V_k - \partial_k V_j ) = \frac{B_i}{H_*^2}\,.
\eea
In this work, we have solved the system of Eqs.~(\ref{eomb1})-(\ref{eomb4}) in three-dimensional lattices. More specifically, we have solved a gauge-invariant set of analogous equations in a discrete spacetime. In all simulations, we have ensured that the lattice analogue of the Gauss conservation law Eq.~(\ref{eomb4}) is preserved by the time evolution of the system to the machine precision. The reader can find more details of our lattice formulation in Appendix \ref{App-Lattice}.

We have considered the following initial condition for the homogeneous modes of the fields. From Eq.~(\ref{defh}), we have by construction $|h_*| \equiv |h (t_*) | = \sqrt{ h_{1*}^2 + h_{2*}^2 } = 1$ at the end of inflation. As long as this condition is satisfied, we can freely distribute this initial value between the components $h_{i*} \equiv h_i (t_*)$, thanks to the symmetries of the model. A convenient choice is
\be h_{1*} = 1 \ ,  \hspace{1cm} h_{2*} = 0 \label{h1h2initial} \ . \ee
As we are evolving the system of equations from the end of inflation, the Higgs initial velocity must obey the slow-roll condition $\dot\varphi_i(t_*) = -{\lambda a_*^2\varphi_*^2\varphi_i/ 2H_*}$. With the choice of Eq.~(\ref{h1h2initial}), the slow-roll condition reads
\be h_{1*}' =  1 - \frac{\beta^2}{2} \ , \hspace{1cm} h_{2*}' = 0 \ . \label{h-init}\ee
We also set the homogeneous mode of the gauge bosons to zero, $V_{i*} = V'_{i*} = 0$, until the onset of the oscillations at $z = z_{\rm osc}$. 

The system is solved in the following way. First, for the times $0< z < z_{\rm osc}$, we only evolve the homogeneous Higgs field with Eqs.~(\ref{eomb1}) and (\ref{eomb2}), while the homogeneous gauge fields are kept to zero. At $z = z_{\rm osc}$, we add fluctuations on top of the homogeneous modes of the different fields, allowing the gauge boson production to take place. Over the homogeneous mode of each Higgs component, we add Fourier modes with a spectrum $\langle |f_k|^2 \rangle = {1\over 2a^2\omega_k}$ (in physical variables), which mimics again the vacuum fluctuations of the ground state of a scalar field in a FRW background. Let us note that the initialization of the Higgs field given by Eqs.~(\ref{h1h2initial}) is indeed crucial for justifying the fact that we ignore cross terms in the initial spectra of fluctuations. Thanks to the gauge rotation Eq.~(\ref{h1h2initial}) and the slow-roll condition Eq.~(\ref{h-init}), we see that the two components of the Higgs are not mixed in the initial trajectory in the $(h_1,h_2)$ plane, and hence only the diagonal terms of the spectra of initial fluctuations are needed. See~\cite{Lozanov:2014zfa,Amin:2014eta} for more details about this and other issues on the initialization of multifield systems.

Due to the gauge nature of the system, the initialization of the gauge fields is more subtle and delicate than in the case of scalar fields. In this case, the fluctuations we add to the gauge fields must preserve the Gauss constraint Eq.~(\ref{eomb4}) initially at every lattice point. Thus, given the spectrum of Higgs fluctuations, we fix the gauge fluctuations as given by the right-hand side of Eq.~(\ref{eomb4}). More specifically, we fix the gauge fields' amplitude in momentum space as
\bea 
V'_{i} (\vec{k}, z_{\rm osc} ) &=& i \frac{k_i}{k^2} j_0 (\vec{k}, z_{osc})\ , \label{eq:gauss-k}  
\eea
where in the lattice this is done with the corresponding lattice momenta (see~\cite{Figueroa:2011ye} for a discussion), corresponding to the choice of lattice finite difference operators that mimic continuous derivatives. The implementation of these initial conditions is described in more detail in Appendix \ref{App-Init}. In particular, we discuss the importance of setting appropriately the Higgs fluctuations so that we can impose correctly Eq.~(\ref{eq:gauss-k}). From $z \geq z_{\rm osc}$ onwards, the Gauss law is then preserved to machine precision by the gauge-invariant evolution of the system. How this is checked is discussed in Appendix~\ref{App-Lattice}.

\begin{figure}[t]
    \begin{center}
        \includegraphics[width=8.5cm]{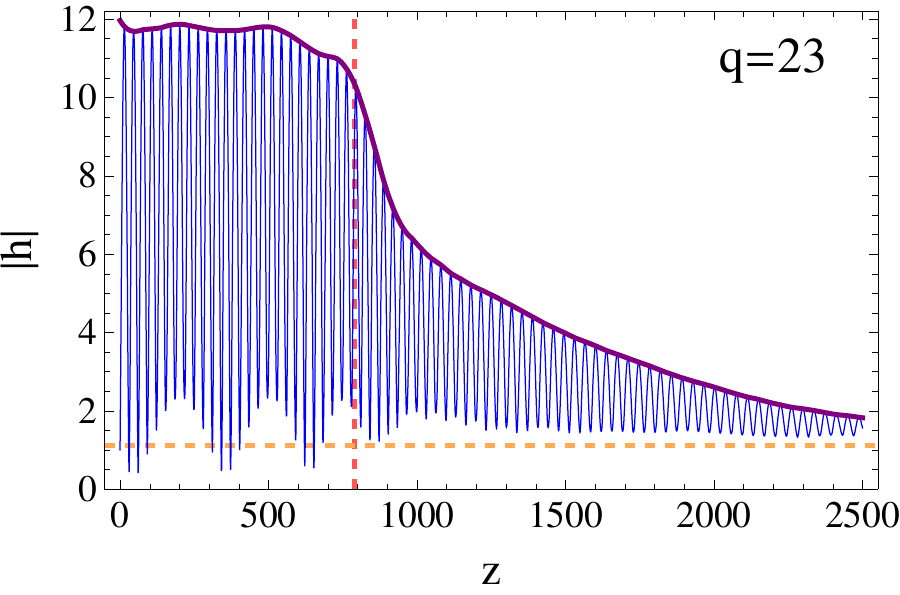}
        \includegraphics[width=8.5cm]{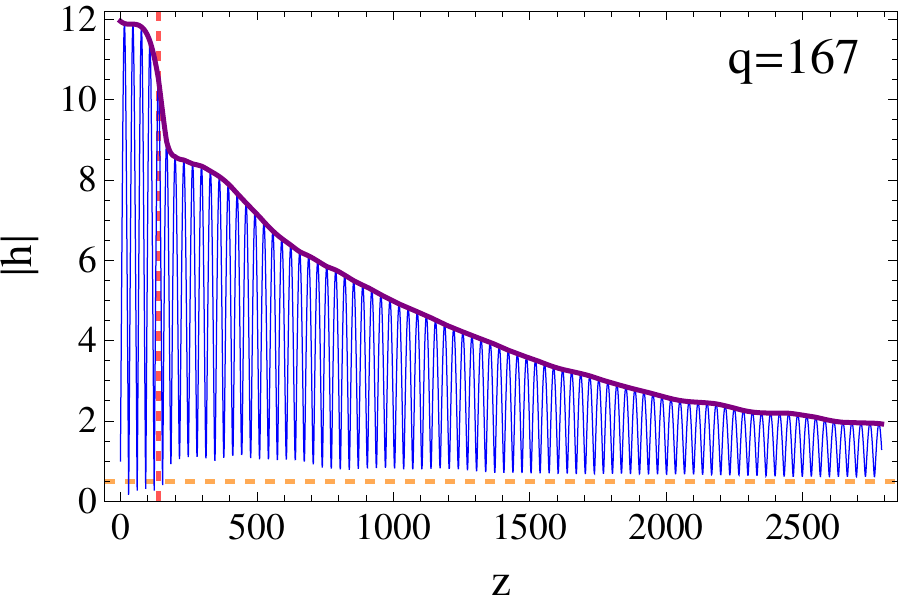}
    \end{center}
    \caption{We show in blue the volume-average value of the conformal Higgs field $ |h| $ as a function of time for the resonance parameters $q=23$ and $q=167$, and in purple the maximum amplitude of the oscillations. The red line indicates the approximate time at which the initial plateau finishes and the Higgs decay starts. The orange line indicates the value of $| h | $ at which this function stabilizes at long times, $h_f$.} \label{fig:means-gauge}
\end{figure}

We now present the main results of the lattice simulations carried out for the Abelian-Higgs model. Like in the global scenario of Section~\ref{sec:iv}, we have run simulations for the resonance parameters given in Table~\ref{tab:I}, ranging from $q=5$ to $q=3000$. These values correspond to $\lambda$ values between $2.5 \times 10^{-5}$ and $1.5 \times 10^{-2}$ for the $W$ boson, and $5 \times 10^{-5}$ and $3 \times 10^{-2}$ for the $Z$ boson. We have also run different simulations for $\beta=10^{-4},10^{-3},10^{-2},10^{-1}$ and $0.5$. The justification of this choice of parameters has been explained in detail in Section~\ref{sec:iv}. Again, all results presented in this section will be obtained for a RD background ($w = 1/3$) and for the $\beta=0.01$ value. In Section~\ref{sec:vi} we will explain how these results can be extrapolated to other values of $\omega$ and $\beta$.

One of the main differences of the Abelian-Higgs model with respect to the global scenario is that now the Higgs field is described by a set of two components $h_1, h_2$, combined in a complex variable $h = h_1+ih_2$. This means that the quantity of interest that we must study is the average value of the Higgs modulus, $|h| \equiv \sqrt{h_1^2 + h_2^2} $. Note that in the global case of Section~\ref{sec:iv}, we analyzed the analogous quantity to $|h|$, corresponding to the absolute value of the real Higgs. This way, the results presented here about the decay of the Higgs amplitude can be easily compared with those of the global scenario.

We have plotted in Fig.~\ref{fig:means-gauge} the volume-average of the Higgs modulus $|h|$ as a function of time, for the two resonance parameters $q=23$ and $q=167$. We have plotted the corresponding oscillations' envelope curve by joining all local maxima with a smooth line. Remember that, according to what we discussed in Section \ref{sec:III}, all resonance parameters can be divided into two groups: those placed within a resonance band of the Lam\'e equation, with an interval of excited momenta of the type $0 \leq k \lesssim k_{\rm max}$, and those which have a smaller band of the type $0 < k_{\rm min} \lesssim k \lesssim k_{\rm max}$. We recall that examples of these two groups are shown with continuous and dashed lines, respectively, in Fig.~\ref{fig:FloquetVarious}. In this regard, $q=23$ belongs to the second group, and $q= 167$ to the first. The initial period of oscillations, before the amplitude of the Higgs drops significantly, fits well the analytical estimate of Eq.~(\ref{eq:Period}). This is expected even in the present case with a complex field, since before the Higgs notices the presence of the gauge fields, the dynamics of the Higgs radius is still effectively equivalent to the absolute value of a real degree of freedom. When the Higgs amplitude starts decreasing due to its transfer of energy into the gauge bosons, the period of oscillations is slightly modulated, but never significantly.

\begin{figure}[t]
    \begin{center}
        \includegraphics[width=8.5cm]{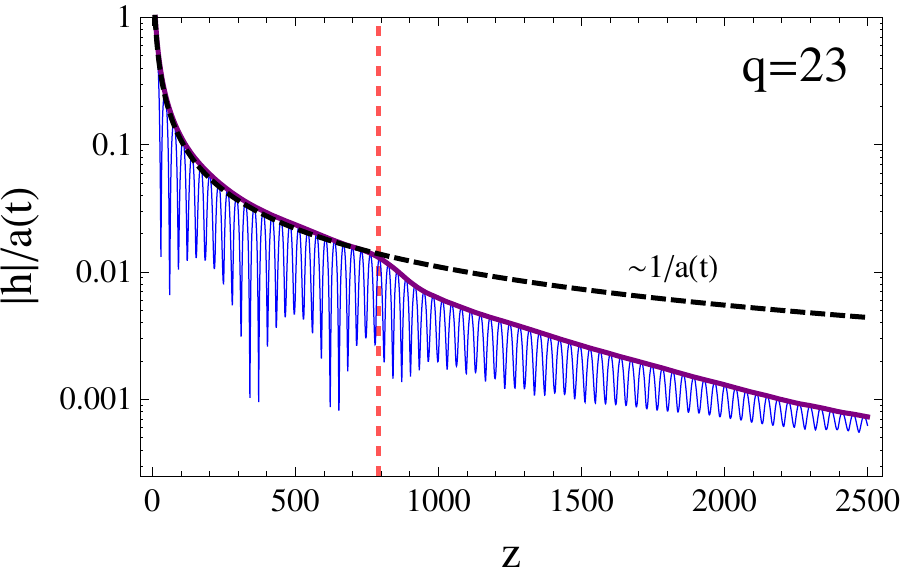}
    \end{center}
    \caption{We show the volume-average quantity $ |h| / a(t) \propto \varphi$ for a RD universe. The dashed, red vertical line shows the time $z_i$, and the dashed, black line shows $\sim 1 / a(t)$.} \label{fig:means-gauge2}
\end{figure}

We find that the Higgs amplitude behaves qualitatively in a similar way as in the global scenario. This can be rapidly seen by comparing Fig.~\ref{fig:means-gauge} to the equivalent Fig.~\ref{fig:means2} of the global scenario. In both scenarios, there is first a stage of few oscillations during which the (conformal) Higgs amplitude does not decay, corresponding to a {\it plateau} in the envelope function. After that, at times $z \gtrsim z_i (q)$, the Higgs amplitude starts decaying strongly. This time is indicated in both panels of Fig.~\ref{fig:means-gauge} with a red dashed vertical line. Finally, the rescaled Higgs amplitude approaches a constant value at late times, $|h| \rightarrow h_f$, which is indicated in both figures with an orange dashed horizontal line. It is important to emphasize again that the plateau is only manifest for the {\it conformal} Higgs field $|h|$, since the physical Higgs field $|\varphi|/\varphi_*$ decays as $\propto 1 / a (t)$, due to the expansion of the Universe. The key observation is that, for $z \lesssim z_i (q)$, it decays as the inverse of the scale factor $\propto 1/a$, while for $z \gtrsim z_i (q)$ the decay is much faster due to the energy transfer to the gauge fields. This can be clearly seen in Fig.~\ref{fig:means-gauge2}, shown for $q = 23$. Therefore, we conclude that the qualitative behavior of the system is very similar, almost identical, to the global scenario.

The time scale $z_i (q)$ signals, as in the global modeling, the moment at which the decay products (in this case, gauge bosons) have accumulated sufficient energy to start affecting the dynamics of the Higgs condensate. As before, this is understood better if we plot the different contributions to the energy, as a function of time. The energy density of the Abelian-Higgs model Eq.~(\ref{action}) is found to be
\be \rho (z) =  \frac{V_*}{a^4 (z)} E_t (z)\,,~~~~V_* \equiv {\lambda\over 4} |\varphi_*|^4\,, \label{rhoz_gauge}\ee
where $V_*$ is the value of the Higgs potential at the end of inflation. The function $E_t (z)$ is formed by the sum of the following contributions:
\be E_t (z) = E_{\rm K} + E_{\rm GD} + E_{\rm E} + E_{\rm M} + E_{\rm V} \ . \label{rhot-gauge} \ee
Here $E_{\rm K}$ and $E_V$ are the kinetic and potential energies of the Higgs field
\bea E_{\rm K}^{\varphi} &\equiv& \frac{a^4}{V_*} \frac{\sum_i \dot{\varphi}_i^2}{2 a^2} = \frac{2}{\beta^2} \sum_i^{2}  \left( h'_i - h_i \frac{a'}{a} \right)^2 \ , \nonumber \\
E_{\rm V} &\equiv& \frac{a^4}{V_*} \frac{\lambda ( \varphi_1^2 + \varphi_2^2 )^2}{4}= (h_1^2 + h_2^2)^2 \nonumber \ , \eea
$E_{\rm GD}$ is a gauge-invariant term formed by the product of two covariant derivatives of the Higgs field (hence containing the spatial Higgs gradients plus the interaction terms)
\bea  E_{\rm GD} &\equiv& \frac{a^4}{V_*} \frac{1}{2 a^2} \sum_i \mathfrak{Re}  [ (D_i (\varphi_1 + i \varphi_2))^* D_i (\varphi_1 + i \varphi_2) ] \nonumber \\
&=& \frac{2}{\beta^2} \sum_i \mathfrak{Re}  [ (\mathcal{D}_i (h_1 + i h_2))^* \mathcal{D}_i (h_1 + i h_2) ] \ , \eea
and $E_{\rm E}$ and $E_{\rm M}$ are the electric and magnetic energy densities
\bea E_{\rm E} \equiv \frac{a^4}{V_*} \frac{1}{2 e^2 a^4 } \sum_i E_i^2 = \frac{2 }{q \beta^4} \sum_i \mathcal{E}_i^2 \label{eq:electric-energy} \ , \\
E_{\rm M} \equiv \frac{a^4}{V_*} \frac{1}{2 e^2 a^4 } \sum_i B_i^2 =  \frac{2}{q \beta^4} \sum_i \mathcal{B}_i^2 \label{eq:magnetic-energy} \ .\eea

\begin{figure}[t]
    \begin{center}
        \includegraphics[width=8.1cm]{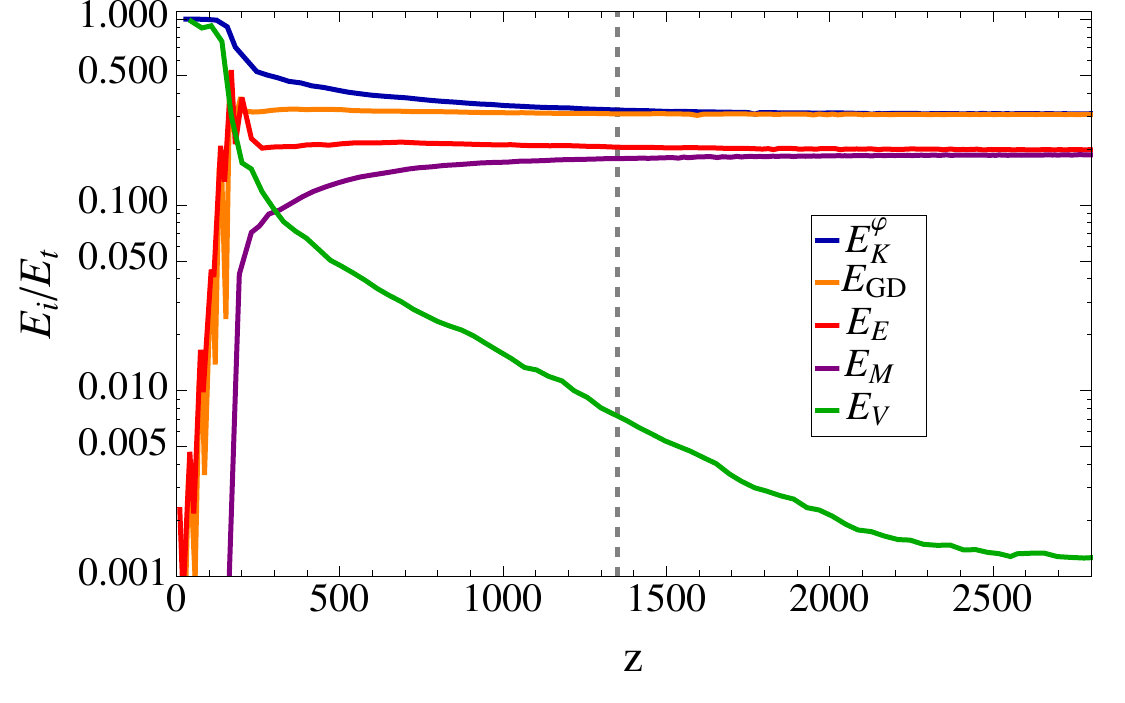}\vspace*{5mm}
        \includegraphics[width=8.5cm]{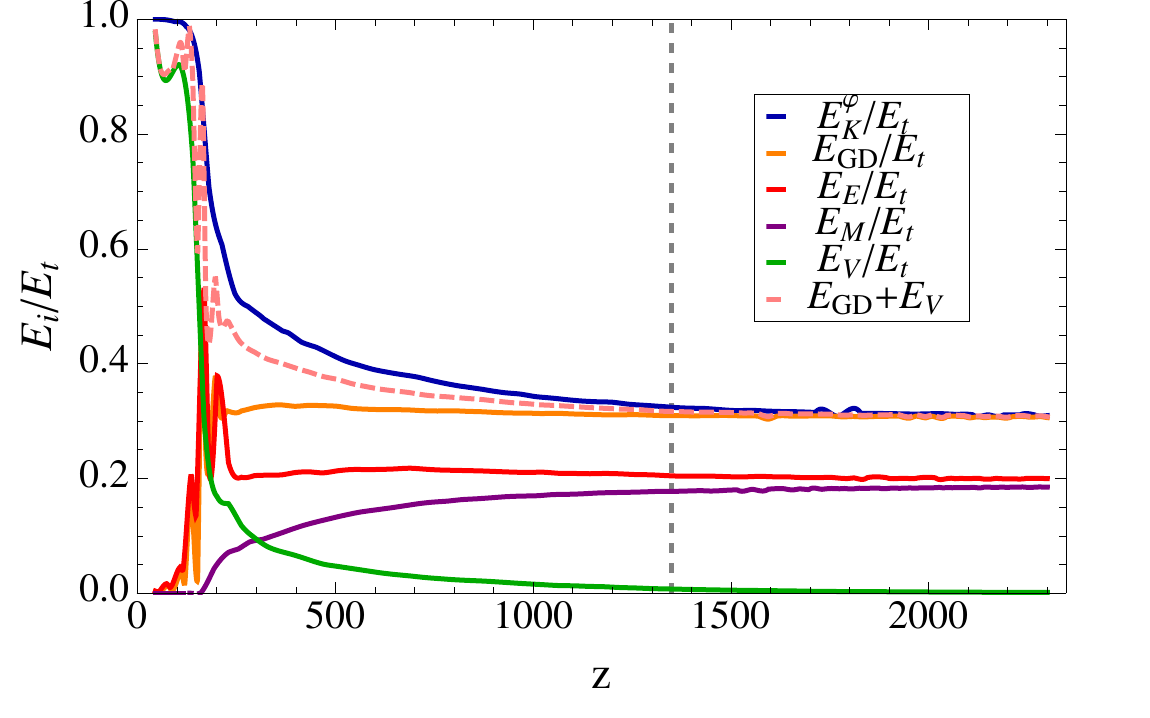}
    \end{center}
    \caption{Upper panel: We plot the different contributions to the total energy of the system as a function of time, $E_i / E_t$ [see Eq.~(\ref{rhot-gauge})], for $q=9$. All functions are oscillating, so we take the envelope of the corresponding oscillations for clarity. The dashed  vertical line signals the Higgs decay time $z_i (q)$. Lower panel: We plot the same quantities with the same color code as in the upper panel, but now $E_{\rm GD}$ and $E_{\rm V}$ appear dashed, and we have added a new pink line corresponding to $E_{\rm GD} + E_{\rm V}$, which is the quantity that equipartitionates with $E_{\rm K}$. Let us note that equipartition in the gauge sector, between the electric and magnetic contributions, is achieved later than in the scalar sector, at some time $z > z_e(q)$.}\label{fig:gauge-energy}
\end{figure}

We have plotted in the upper panel of Fig.~\ref{fig:gauge-energy} these quantities as a function of time for the resonance parameter $q=9$, which corresponds to a case within a resonance band. We also show in the lower panel of Fig.~\ref{fig:gauge-energy} the contribution of each energy component to the total, $E_i / E_t$, removing the oscillations of each component, and hence showing only the corresponding envelope functions. We see that initially the dominant contributions come from the kinetic and potential energies of the Higgs field. This corresponds to the oscillations of the condensate around the minimum of its potential, before it `feels' the gauge fields. Meanwhile, the other components of the energy, $E_{\rm E}$, $E_{\rm M}$ and $E_{\rm GD}$, grow really fast, due to the energy transfer from the Higgs into the gauge fields. Note that for the whole evolution of the system (until equipartition is reached), the electric energy clearly dominates over the magnetic energy. 

As in the global analogue, although gauge bosons are being strongly created, the Higgs condensate is at first unaffected. At $z \approx z_i (q)$ (indicated by a dashed red vertical line in the figures) the gauge energy has grown enough to start affecting significantly the Higgs condensate, and a sharp decrease of both the Higgs potential and kinetic energy start from then on. Physically, this happens when the fraction $\delta  \equiv E_{\rm E} / E_t < 1$ becomes sizeable. In particular, like in the global scenario, when $\delta \gtrsim 0.1$, we can clearly see the correspondence between the backreaction of the gauge fields over the Higgs field and the decrease in the Higgs amplitude.

As in the global scenario, for $z \gtrsim z_i (q)$ the Higgs kinetic and potential energies decrease sharply. The potential energy very soon becomes irrelevant compared to the other energy contributions, while the kinetic energy approaches an almost constant amplitude. Simultaneously, $E_{\rm GD}$ and $E_{\rm E}$ stop their growth, and also saturate to almost constant values. However, the magnetic energy continues to grow even after $E_{\rm GT}$ and $E_E$ have been stabilized. Finally, at $z = z_e$, the system arrives again at a stationary regime, in which equipartition between different components is clearly achieved. In this regime, 30\% of the total energy goes to the Higgs kinetic part, 30\% to $E_{\rm GD}$, 20\% to electric energy $E_{\rm E}$, and 20\% to magnetic energy $E_{\rm M}$. The potential energy $E_V$ also saturates to a constant, but it is very subdominant with respect to the other contributions. Quite remarkably, these numerical percentages are independent of the values $q$ and $\beta$ taken in our simulations. In other words, the final fractions of energies are universal within the Abelian-Higgs formulation\footnote{We expect this to be also the case if the non-Abelian nature of the interactions was considered, but only simulations of the full $SU(2)\times U(1)$ gauge group of the SM sector, can really prove it.}.

Let us analyze the equipartition regime in the gauge scenario in more detail. We observe that the kinetic energy of the Higgs field $E_{\rm K}$ eventually becomes equal to $E_{\rm V} + E_{\rm GD}$. Since $E_{\rm GD}$ is gauge invariant, it contains both the Higgs gradient terms plus the Higgs interactions with the gauge fields. One can then naturally identify this quantity with the analogous combination in the global scenario, given by the sum of the interaction term plus the Higgs gradients, $E_{\rm int} + E_{\rm G}^{\varphi}$. 

Similarly also to the global scenario, the potential energy keeps decaying even after equipartition has been established. In principle, we could then think of using just the equipartition relation $E_{\rm K} \simeq E_{GD}$, neglecting the contribution from the potential energy, as we did in the global case. However, in the moment when the rest of energy contributions are stabilized, the potential energy still represents $\sim 1\%$ of the total. So, although the potential energy becomes eventually subdominant, its rate of decay is slower than in the global scenario. This percentage, although small, is still significant at the moment in which equipartition is achieved. Therefore, it is better to follow the equipartition condition $E_K \simeq E_V + E_{\rm GD}$. The evolution of the different energy components and the achievement of equipartition can be well appreciated in Fig.~\ref{fig:gauge-energy}.

It is useful to define the Higgs decay time as the moment when the Higgs kinetic energy (which dominates over the potential energy) results stabilized at the onset of the stationary regime. As in the global scenario, we will call this quantity $z_e (q)$. Naturally, there is again some degree of arbitrariness in this definition. In the global scenario, we observed that a good operative criterion for defining $z_e$ was based on the degree of equipartition achieved. In our present gauge context, we have observed that an appropriate criterion is to take the moment when the relative difference between $E_{\rm K}$ and the sum $E_{\rm GD} + E_{\rm V}$ becomes less than $1\%$. We have indicated this time in Fig.~\ref{fig:gauge-energy}, with a dashed vertical line. As we just mentioned, in the global scenario we did not consider the contribution $E_{\rm V}$ of the Higgs potential energy into the equipartition equalities, simply because its contribution was already irrelevant when equipartition was reached. However, in the Abelian-Higgs scenario, the addition of this contribution to the covariant-gradient one $E_{\rm GD}$ is crucial. Even though $E_{\rm V}$ is also marginal in this case, if we were to consider just $E_{\rm GD}$ in the equipartition analysis, it would achieve equipartition with $E_{\rm K}$ (say to better than $1\%$) long after the Higgs kinetic energy density has started to saturate. As we can observe in Fig.~\ref{fig:gauge-energy}, our criterion $E_{\rm K} \simeq E_{\rm GD} + E_{\rm V}$ holding better than $1\%$, coincides very well with the moment when all relevant energy densities have just stopped either growing or decreasing. Hence it defines very well what we mean by the end of the Higgs decay.

\begin{figure}[t]
    \begin{center} \includegraphics[width=8.5cm]{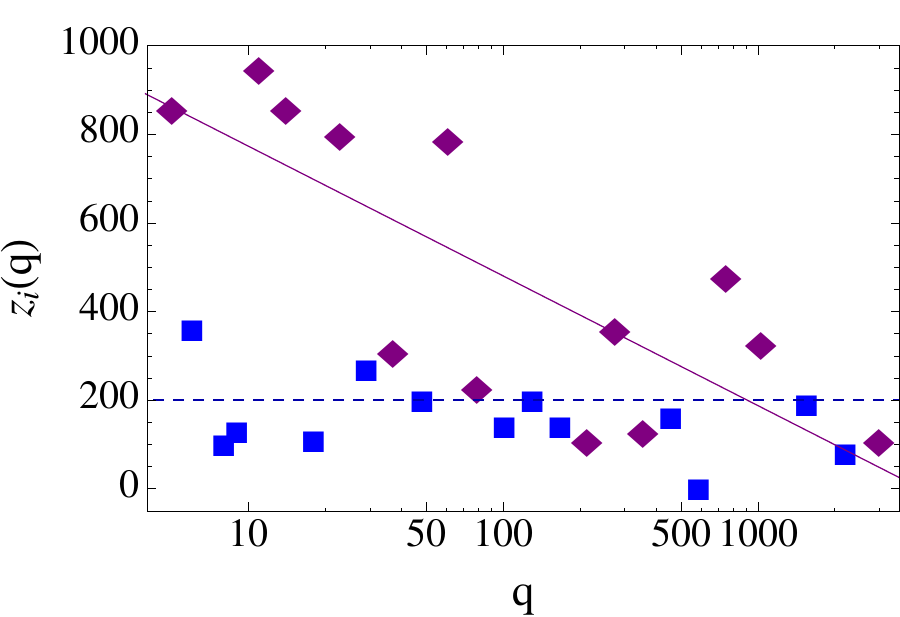}
    \includegraphics[width=8.2cm]{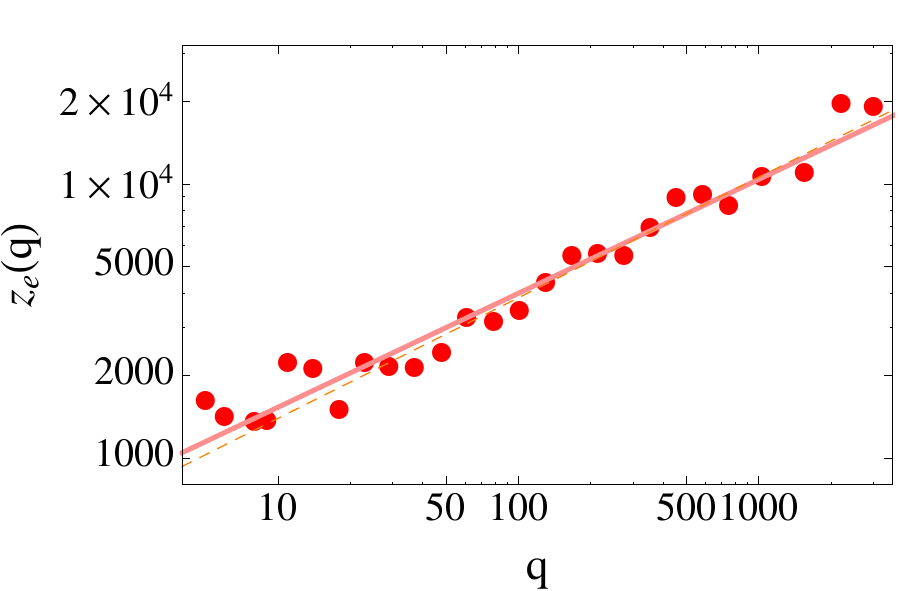}  \end{center}
    \caption{Top: Different values of $z_i (q)$ obtained for different resonance parameters $q$, for a RD universe and for $\beta=0.01$.  Blue squares correspond to $q$ values that are within a resonance band of the Lam\'e equation, while purple diamonds are points which are not. The purple line corresponds to the best fit (\ref{eq:FitZsGaugeQnoResonance}), while the dashed blue line corresponds to the analytical estimate $z_{\rm eff} \approx 200$, obtained from Eq.~(\ref{eq:EffEnergyTransferTimeScaleApprox}) ($\bar{\mu}_k = 0.2$). Bottom: Red points indicate the obtained Higgs decay times $z_e (q)$ as a function of $q$, for the same Abelian-Higgs simulations, while the red thick line shows the best fit (\ref{fit-gaugedecay}). The dashed yellow line shows the best fit of this same quantity obtained from the global simulations in Eq.~(\ref{eq:zs(q)}).} \label{fig:gauge-zgq}
\end{figure}

\begin{figure*}[t]
    \begin{center}
        \includegraphics[width=7.8cm]{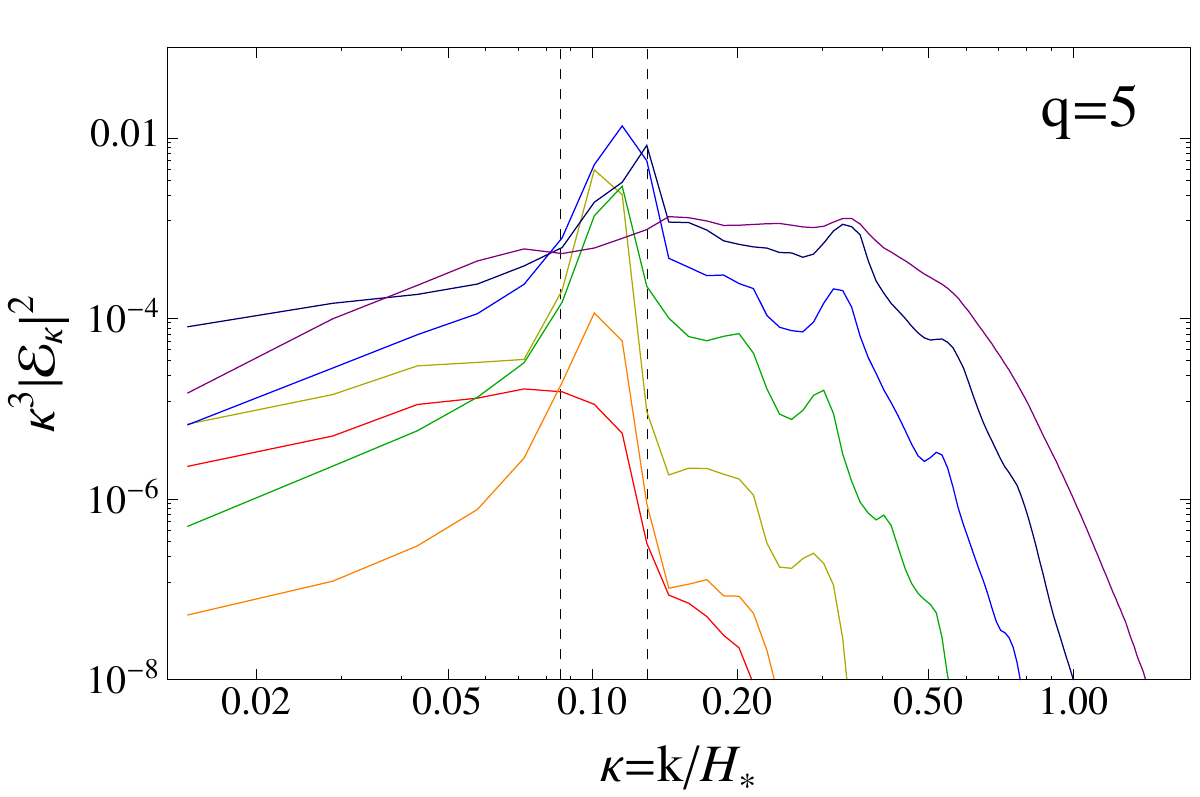} \hspace{0.2cm}
        \includegraphics[width=9.3cm]{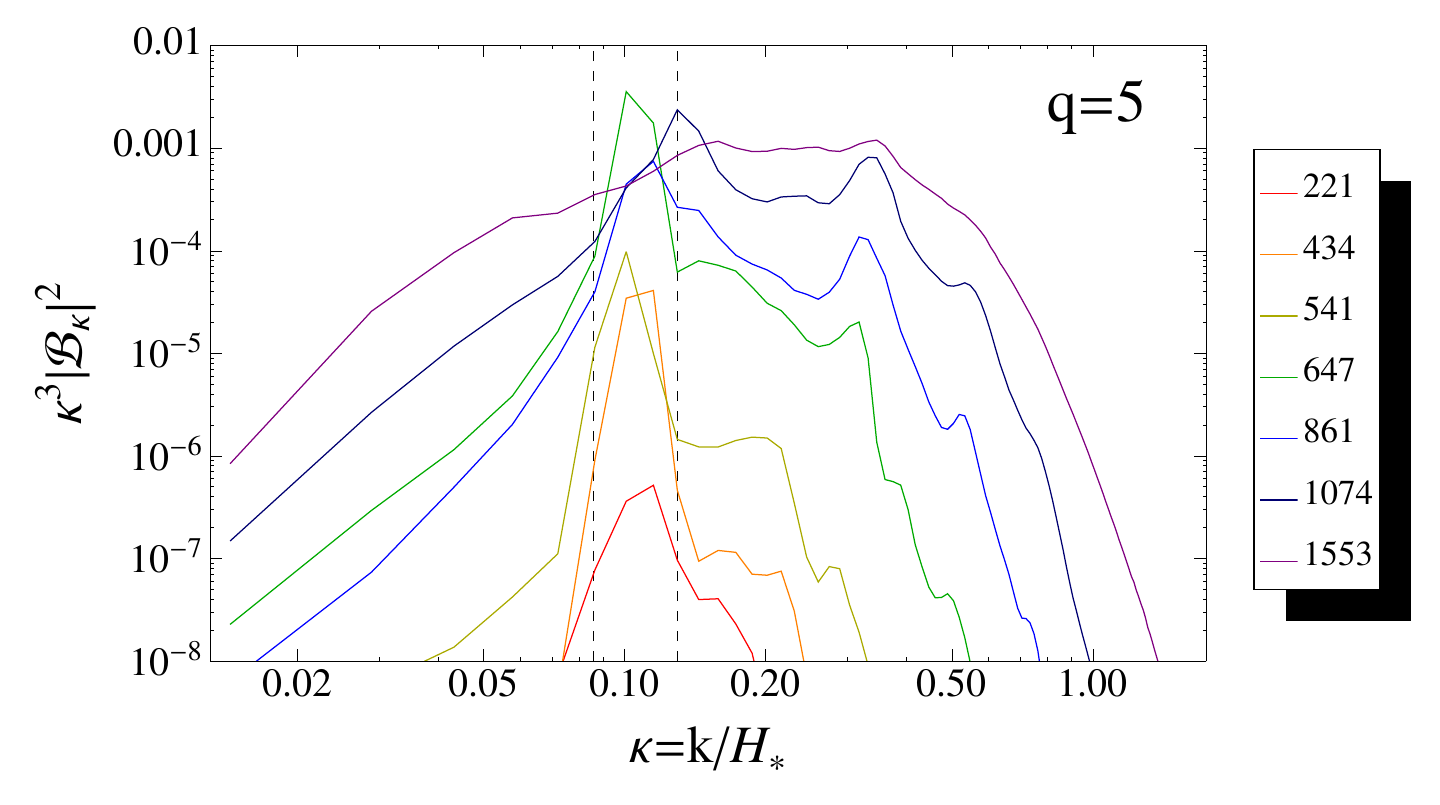}
        \includegraphics[width=7.8cm]{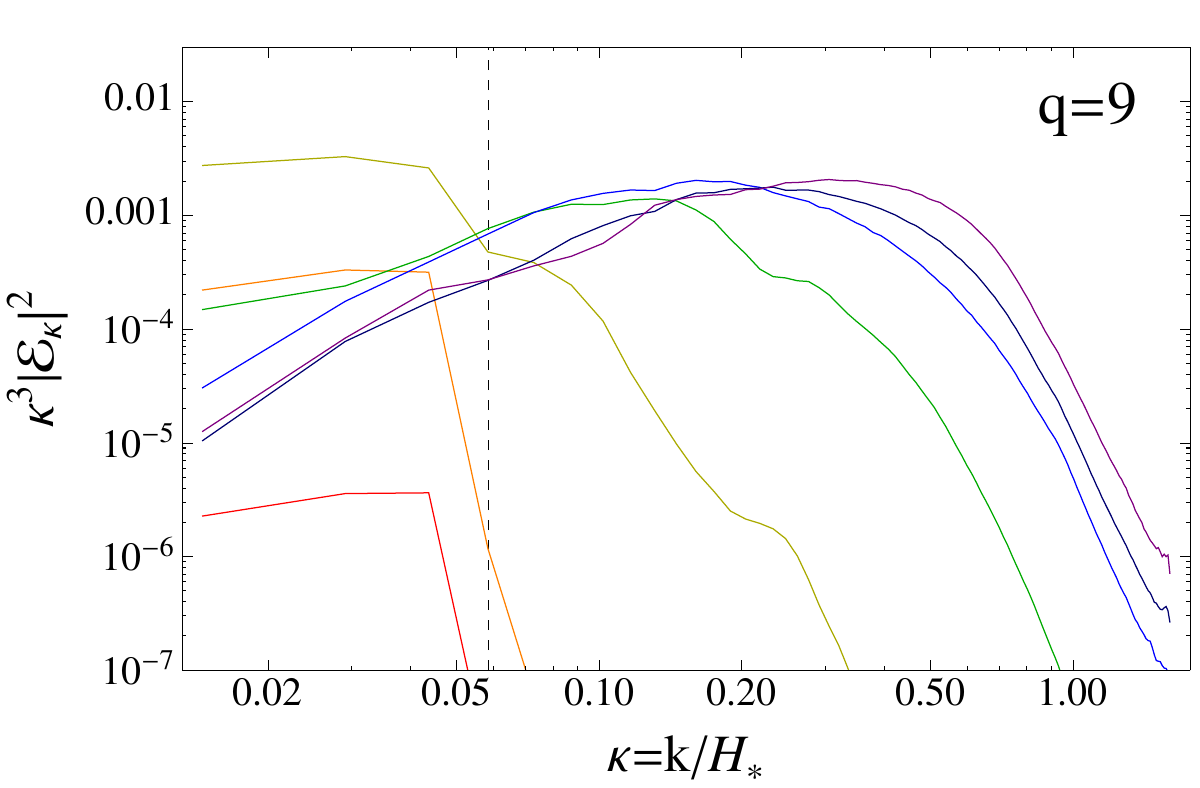} \hspace{0.2cm}
        \includegraphics[width=9.3cm]{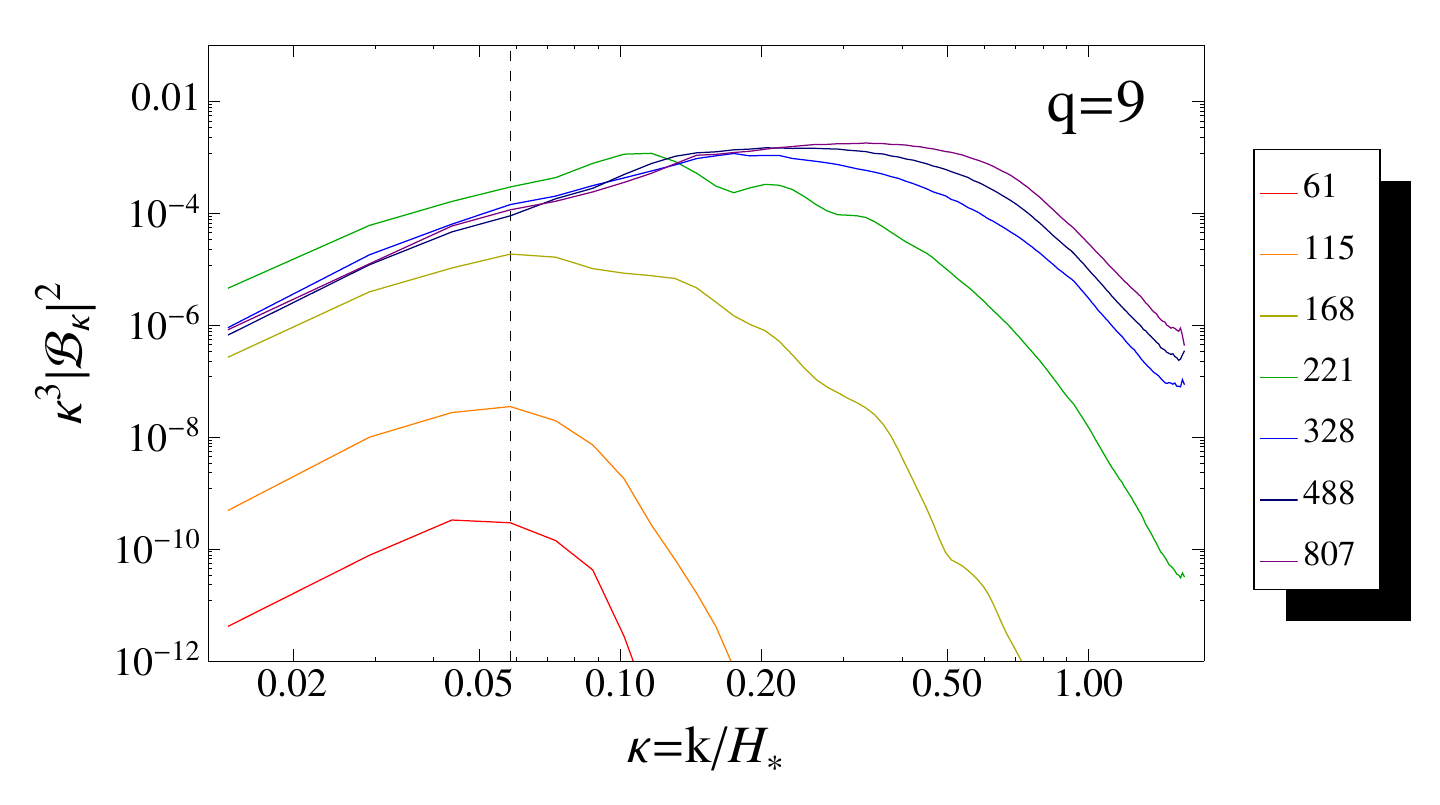}
    \end{center}
    \caption{Electric spectra $k^3 |\mathcal{E}_k|^2$ and magnetic spectra $k^3 |\mathcal{B}_k|^2$ for different times and for $q=5$ (upper figure) and $q=9$ (lower figure). The dashed, vertical lines indicate the corresponding position of the resonance band. The corresponding times at which the spectra are plotted are written at the right. Note that the units used to express the momentum are different to the ones used in Fig.~(\ref{fig:FloquetVarious}). } \label{fig:elecmag-spectra}
\end{figure*}

We have characterized again the dependence of $z_i$ and $z_e$ with the different $q$'s considered. We show in the upper panel of Fig.~\ref{fig:gauge-zgq} the behavior of $z_i (q)$. In the figure, blue squares correspond to $q$ values within a resonance band, and purple circles correspond to values outside bands. We see a clear trend, such that simulations with $q$ within resonance bands have a smaller $z_i (q)$ than those with $q$ between adjacent bands. %This is, however not a robust fact as it was in the global scenario (recall Fig.~\ref{fig:scalar-fit}), as purple points at relatively higher $q$ do not fulfill this property. 
Like in the global scenario, the order of magnitude of $z_i$ for blue squares is approximated quite well with the analytical estimate $z_{\rm eff} \approx 200$, obtained from Eq.~(\ref{eq:EffEnergyTransferTimeScaleApprox}), with $\bar{\mu}_k = 0.2$. At the same time, the  purple circles can be fitted as
\be z_i (q) \sim 1066 -127\log{q} \ , \hspace{0.3cm} q \notin {\rm Resonant~Band}\,,     \label{eq:FitZsGaugeQnoResonance}
\ee
but their dispersion is much worse than in the global case (recall the top panel of Fig.~\ref{fig:scalar-fit}).

In the lower panel of Fig.~\ref{fig:gauge-zgq}, we also plot $z_e$ as a function of the resonance parameter $q$. We have obtained the following phenomenological fit
\be z_e (q) = 588 q^{0.42}\,,\label{fit-gaugedecay} \ee
indicated in the figure with a red continuous line. Note that we have plotted as well the corresponding fit obtained from the global simulations, Eq.~(\ref{eq:zs(q)}), with a dashed line. Both fits coincide pretty well, indicating that the Higgs decay time $z_e(q)$ obtained in the global scenario constitutes already a very good estimation. To some extent this is surprising, since one could expect that the extra terms in the gauge field's EOM could play some role, like for example modulating the decay time $z_e(q)$ differently than in the case of only scalar fields. However, our results prove that this is not the case. In fact, they imply that the interaction term $g^2A_\mu A^\mu\varphi^2$ (which is the only one kept in the global scenario) is the most relevant one when determining the Higgs decay time scale and the onset of the stationary regime. 

Let us note again that the fit Eq.~(\ref{fit-gaugedecay}) is only valid for $\beta=0.01$ and for a RD background. Using the theoretical extrapolation that we will present in Section \ref{sec:vi}, this can be generalized to other $\beta$ and $w$ values as  
\be z_e (q) \approx 58.8 \beta^{\frac{-(1+ 3 \omega)}{3 (1 + \omega)}} q^{0.42} \ . \label{fit-gaugedecay-extrap} \ee

An alternative source of information about the Abelian-Higgs system comes from the spectra of the different fields. Since we are dealing with a gauge theory, all quantities of physical interest must be gauge invariant. We then plot in Fig.~\ref{fig:elecmag-spectra} the spectra of the electric and magnetic fields $k^3 |\mathcal{E}_k|^2$ and $k^3 |\mathcal{B}_k|^2$ at different times. In order to see the dependence of the spectra evolution on the analytical properties of the Lam\'e equation, we plot both spectra for two different resonance parameters, $q=5$ and $q=9$. The latter is placed in the middle of a resonance band, while the former is between the first and second resonance bands. The dashed vertical lines in the figures indicate the location of the respective resonance bands. In the case $q = 5$, one can clearly see that both spectra grow with time, as a consequence of the resonant excitation of gauge bosons. At initial times, there clearly appears a peak in both spectra, centered in the corresponding main resonance band. This confirms that the behavior derived from the Lam\'e equation describes well enough the real dynamics during the initial stages, even for the gauge theory. When the gauge bosons start to affect significantly the Higgs condensate, i.e. for $z \gtrsim z_i (q)$, both spectra start to displace to the right, populating modes of higher momenta. In this process, new subdominant peaks appear. As time goes on, the peaks disappear, and when the Higgs condensate has decayed [i.e for $z \gtrsim z_e (q)$], the stationary state is established. For the case $q=8$, the time scale $z_i (q) \approx 150$ is much smaller than for $q = 5$, and the resonance band is much wider. This is expected, as we include modes down to $k=0$. In this case, we see that the population of higher modes is much faster than for $q = 5$, and we do not observe additional subdominant peaks in the spectra. 

As a final remark, let us note that in the gauge scenario, none of the field spectra could be well fitted with similar scaling laws to those of the turbulent regime in the global case. After equipartition is reached in the gauge scenario, the field distributions evolve smoothly, slowing transferring power into higher modes, pretty much like in the global case. However, the evolution towards equilibrium cannot be really grasped by simple fitting formulas like Eq.~(\ref{eq:SelfSimTurb}).

\subsection{Beyond the Abelian-Higgs}
\label{subsec:BeyondAH}

The real nature of the SM interactions is non-Abelian, since the EW sector of the SM is $SU(2)\times U(1)$ gauge invariant. In the EOM of the gauge bosons there are therefore nonlinear terms\footnote{For the sake of clarity of the physics, we switch back to physical variables in the discussion of this subsection.} of the form $\sim g^2A^3, gA\partial A, g\partial A^2$, where we omit charge and Lorentz indices for simplicity. Following~\cite{Enqvist:2014tta}, one obtains that within the Hartree approximation, the terms $\sim g A\partial A, g\partial A^2$ vanish, so that in principle only the terms $\sim g^2A^3$ contribute effectively to the dynamics of the gauge fields. We can write the effective mass entering into the gauge fields' EOM, as given by their interactions with the Higgs, plus a contribution from their own non-Abelian self-interactions. Symbolically, we will write this as
\begin{eqnarray}
    m_A^2 = g^2\varphi^2 + \left\langle A^2 \right\rangle\,.
\end{eqnarray}
The Abelian-Higgs simulations capture the first term $g^2\varphi^2$, which is due to the interaction with the Higgs, and is responsible for the resonant excitation of the gauge fields. The self-induced mass due to the gauge-field self-interactions is, of course, not present in the Abelian approach. This second term describes the nonlinearities of the non-Abelian nature of the SM interactions. Hence, only when the gauge fields have been excited with a sufficiently high amplitude $\left\langle A^2 \right\rangle \gtrsim g^2\varphi^2$ may their presence have any relevance. The question, then, is when do the gauge fields reach the critical amplitude $A \sim A_{\rm c} \equiv g\varphi$? 

The answer can be easily found by analyzing the effective mass of the Higgs. The non-Abelian nature of the interactions does not add any extra contribution into the effective mass of the Higgs field, given by
\begin{eqnarray}
    m_\varphi^2 = \lambda\varphi^2 + g^2\left\langle A^2 \right\rangle\,.
\end{eqnarray}
These terms are already captured in our simulations, so the only difference in a non-Abelian simulation would come from the fact that $A_\mu$ is affected by the nonlinearities of its own EOM. The gauge fields backreact into the Higgs dynamics at the time $z = z_i(q)$, which corresponds physically with the moment when the amplitude of the gauge fields has grown -- due to parametric resonance -- up to $\left\langle A^2 \right\rangle \gtrsim \lambda\varphi^2 /g^2$. This condition corresponds, however, to a typical amplitude of the gauge fields $A \sim A(z_i) \equiv \sqrt{\lambda} \varphi / g$, which is much smaller than $A_{\rm c}$. In particular, $\frac{A(z_i)}{ A_{\rm c}} \sim \frac{1}{g \sqrt{q}} < 1$, for the typical broad resonant parameters $q \sim \mathcal{O}(10)-\mathcal{O}(10^3)$. The effective mass of the gauge bosons at $z \approx z_i$ is
\begin{eqnarray}
    m_A^2(z_i) &=& g^2\varphi^2 + \left\langle A^2 \right\rangle_{z_i} \approx g^2\varphi^2\left(1+{1\over g^2q}\right) \label{eq:nonab} \,,
\end{eqnarray}
where ${1\over g^2q}  \ll 1$. It is then clear that $m_A^2(z_i) \approx g^2\varphi^2$, as if there were no effect from the gauge-field self-interactions. By the time the gauge-field resonant production backreacts on the Higgs dynamics, the gauge fields stop growing, as explained in detail in Section~\ref{sec:v}. Therefore, the non-Abelian terms (neglected in the Abelian-Higgs approach), are not expected to play any significant role in the dynamics of the system.\footnote{It is possible though that for the mildest broad resonance parameters, such as $q \sim \mathcal{O}(10)$, there might be some effect from the non-Abelian terms, since in this case the $\frac{1}{g^2q}$ correction in Eq.~(\ref{eq:nonab}) is only marginally smaller than unity.} It is, however, likely that the presence of the non-Abelian terms will possibly change the details of the achievement of the equipartition regime. Therefore, although we do not expect the time scale $z_i(q)$ to change, the time scale $z_e(q)$ might perhaps change moderately in the presence of non-Abelian corrections. However, only non-Abelian lattice simulations, beyond our present work, can really quantify these questions. 

In light of this analysis, we see $a~posteriori$ that neglecting the nonlinearities due to the non-Abelian nature of the SM interactions was well justified.

\subsection{Abelian-Higgs model with three gauge fields} \label{sec:vB}

So far, we have studied the postinflationary Higgs dynamics in the lattice, mimicking its interaction with a single gauge boson using an Abelian-Higgs modeling. This has allowed us to obtain a bunch of interesting results, which depend greatly on the choice of the gauge boson resonance parameter, $q \equiv g^2 / (4 \lambda)$, with $g^2$ being the corresponding standard model coupling of either $W$ or $Z$ bosons. Naturally, we should include the three massive gauge bosons in our simulations (i.e.~the $W^{+}$, $W^{-}$ and Z), as in the EW sector of the standard model. Remarkably, the results presented so far for a single gauge field can be easily translated into the three-boson case, with an appropriate field redefinition. We explain this in what follows.

In the case of a Higgs decaying into three Abelian gauge fields, the Higgs equation can be written as 
\be h'' - \mathcal{D}_i \mathcal{D}_i h + \beta^2 |h|^2 h = h \frac{a''}{a} \label{eq:higgs-3gauge} \ee
where $h \equiv h_1 + i h_2$, and the covariant derivative is now
\be \mathcal{D}_i \equiv \frac{\partial}{\partial z^i} - i ( W^+_{i} + W^-_{i} + Z_i ) \ . \label{eq:covariant-3gauge}\ee
Here, $W^+_{\mu}$, $W^-_{\mu}$, and $Z_{\mu}$ are the  corresponding fields of the $W^+$, $W^-$, and Z bosons, respectively. We describe the three fields in the temporal gauge, so that their 0 components are null. The EOMs of either of the $W$ bosons are then
\bea
W''_j + \partial_{j} \partial_{i} W_i - \partial_{i} \partial_{i} W_j &=& q_W \beta^2  \mathfrak{Im} [ h^* \mathcal{D}_i h ] \ , \label{gaugeeom-viii2} \\
\partial_i W'_i &=&  q_W \beta^2  \mathfrak{Im} [ h^* h' ] \ , \label{gauss-viii2} \eea
with $q_W \equiv g_W^2 / (4 \lambda)$. Equivalently, the EOMs of the $Z$ boson are
\bea
Z''_j + \partial_{j} \partial_{i} Z_i - \partial_{i} \partial_{i} Z_j &=& q_Z \beta^2  \mathfrak{Im} [ h^* \mathcal{D}_i h ] \ , \label{gaugeeom-viii1} \\
\partial_i Z'_i &=&  q_Z \beta^2  \mathfrak{Im} [ h^* h' ] \label{gauss-viii1} \ , \eea
with $q_Z \equiv g_Z^2 / (4 \lambda)$. Note that there is a Gauss law for each gauge field, representing as before, dynamical constraints of the system. Interestingly, this system can be reduced, with an appropriate redefinition of the gauge fields, to the case of a Higgs decaying into a single gauge field studied above. To see this, let us define the following effective gauge field 
\begin{eqnarray}\label{eq:SuperGaugeField}
S_\mu \equiv W_\mu^+ + W_\mu^- + Z_\mu\,,
\end{eqnarray}
and the resonance parameter
\be q \equiv q_{\rm Z} + 2 q_{\rm W} = \frac{g_Z^2 + 2 g_W^2}{4 \lambda} \label{eq:qeff} \ . \ee
If we consider the mapping
\be W^{\pm}_\mu  \equiv \frac{q_W}{q} S_\mu \ , \hspace{0.3cm} Z_\mu \equiv \frac{q_Z}{q} S_\mu \label{sgaugefield} \ , \ee
automatically $S_0 = 0$, and we can then reduce both the $W$ EOM (\ref{gaugeeom-viii2})-(\ref{gauss-viii2}) and the $Z$ EOM (\ref{gaugeeom-viii1})-(\ref{gauss-viii1}) to just
\bea
S''_j + \partial_{j} \partial_{i} S_i - \partial_{i} \partial_{i} S_j &=& q \beta^2  \mathfrak{Im} [ h^* \mathcal{D}_i h ] \ ,  \label{eq:gaugeS1} \\
\partial_i S'_i &=&  q \beta^2  \mathfrak{Im} [ h^* h' ] \label{eq:gaugeS2} \ , \eea
where the covariant derivative of Eq.~(\ref{eq:covariant-3gauge}) is now simply 
\be \mathcal{D}_\mu \equiv \partial_\mu - iS_\mu \ . \label{covder-S}\ee

Therefore, the three gauge bosons can be described\footnote{Note that this property can be generalized to a Higgs coupled to $N$ Abelian gauge bosons, $V_i^c$ ($c = 1, 2, \cdots N$), with different resonance parameters $q_c$. If we define $S_i$ so that $V_i^c = (q_c/q) S_i$, all fields can be described with the same $eom$ of a single field with effective resonance parameter $q = \sum_c q_c$.} by a single \emph{effective} gauge boson $S_i$, coupled to the Higgs with the resonance parameter $q$ of Eq.~(\ref{eq:qeff}). This property is very useful, since we just need to introduce only one effective gauge field, Eq.~(\ref{eq:SuperGaugeField}), and the system is then fully described by Eqs.~(\ref{eq:higgs-3gauge}), (\ref{eq:gaugeS1}), (\ref{eq:gaugeS2}), and the covariant derivative (\ref{covder-S}). As an example, if we have $q_{W} = 14$ and $q_{Z}  \simeq 2 q_{W} = 28$, all three gauge bosons can be described by the EOM of a single gauge field with resonance parameter $q = 28 + 14 + 14 = 56$. In other words, the system behaves in such a way that the three gauge bosons have the same effective resonance parameter. From Eq.~(\ref{sgaugefield}), we find the following relation between the $W$ and $Z$ amplitudes
\be Z_i (z) = \frac{q_Z}{q_W} W_i^{+} (z) = \frac{q_Z}{q_W} W_i^{-} (z)   \ , \label{Eq:WZrelation}\ee
which at very high energies, when $q_{Z} \approx 2 q_{W}$, reduces simply to $Z_i (z) \approx 2 W^{+}_i (z)  \approx 2 W^{-}_i (z)$. Eq.~(\ref{Eq:WZrelation}) follows in all spacetime (and in the lattice, in all sites at all times). 

We have just seen that the dynamical equations of the Higgs coupled to three gauge bosons can be reduced to a system with the Higgs coupled to only one gauge boson, with resonance parameter $q = q_Z + 2 q_W$. The equivalence between these two systems is actually a mathematical identity. For the sake of verification, we have checked that the results in terms of $z_i$ and $z_e$, are indeed identical when comparing the simulations of one effective gauge boson $S_\mu$ with the resonance parameter $q = q_Z + 2 q_W$, and the simulations of two $W$ bosons with the resonance parameter $q_W$ each, plus a $Z$ boson with resonance parameter $q_Z$. 

Given the above equivalence, in principle, we should then be able to translate the analysis from the simulations discussed so far, for only one gauge field, into the real scenario with the three gauge fields $W^\pm, Z$. Strictly speaking, however, both scenarios are not really identical, if we compare them for the same $q$ and $\beta$. To understand this, let us identify the gauge boson of the single gauge field simulations presented so far, e.g.~with one of the $W$ bosons. For a fixed value of its resonance parameter $q = g_W^2/4\lambda$, we conclude that the Higgs self-coupling is $\lambda = \lambda_{1B} \equiv g_W^2/4q$. In the case of the three gauge bosons, however, the effective field $S_\mu$ (exactly equivalent to the three-gauge-boson system), has a resonance parameter $q = (2g_W^2+g_Z^2)/4\lambda$. From there we deduce that for the same $q$, the Higgs self-coupling in this case is $\lambda = \lambda_{3B} \equiv (2g_W^2+g_Z^2)/4q$, which differs in a factor $(2+(g_Z/g_W)^2)$ with respect to $\lambda_{1B}$. In other words, we would be comparing systems with different Higgs self-couplings, and hence not equivalent. Since we want to compare the two systems for the same $\beta \equiv \sqrt{\lambda}\alpha$, the difference in $\lambda$ translates into a different $\alpha$, and hence into different initial Higgs modes. According to the initial spectra given by Eqs.~(\ref{fluct2}), (\ref{rms-hj}), the fluctuations depend explicitly on $\lambda$, as a reflection of their dependence on $\alpha$ (after having fixed $\beta$). Given our choice of variables, the fluctuations added at the time $z_{\rm osc}$, are then slightly different in the two scenarios. As a consequence, there are also differences in the gauge initial fluctuations, as we impose, following the procedure of Eq.~(\ref{eq:gauss-k}),
\bea 
S'_{i} (\vec{k}, z_{\rm osc} ) &=& i \frac{k_i}{k^2} j_0 (\vec{k}, z_{osc})\ , \label{eq:effgauss-S}   
\eea
with $j_0 (\vec{k}, z_{\rm osc})$ the Fourier transform of $j_0 (\vec{z}, z) \equiv q_{\rm eff} \beta^2  \mathfrak{Im} [ h^* h' ]$, evaluated at $z = z_{\rm osc}$. 

\begin{figure}[t]
    \begin{center}
        \includegraphics[width=8.5cm]{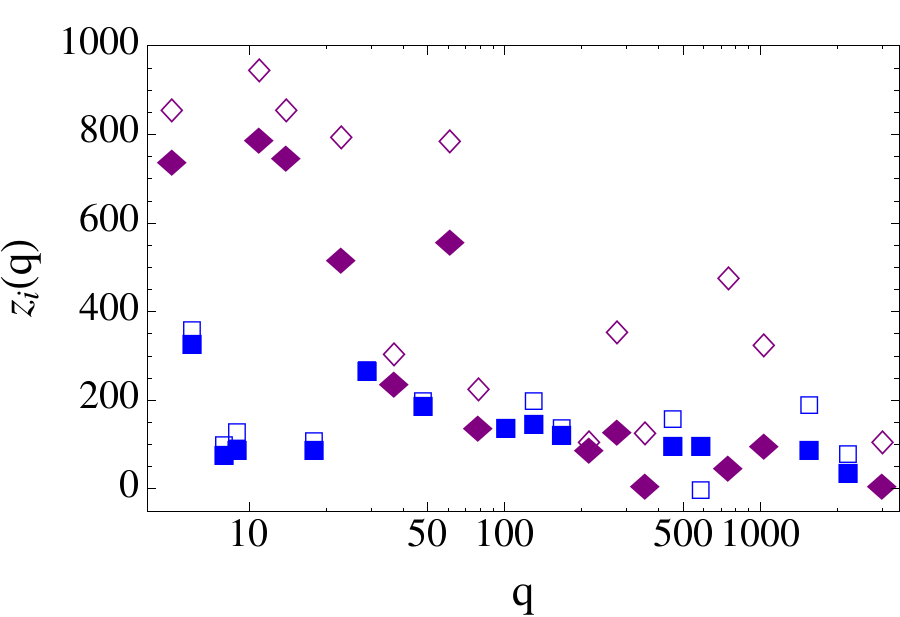}\\
       \includegraphics[width=8.2cm]{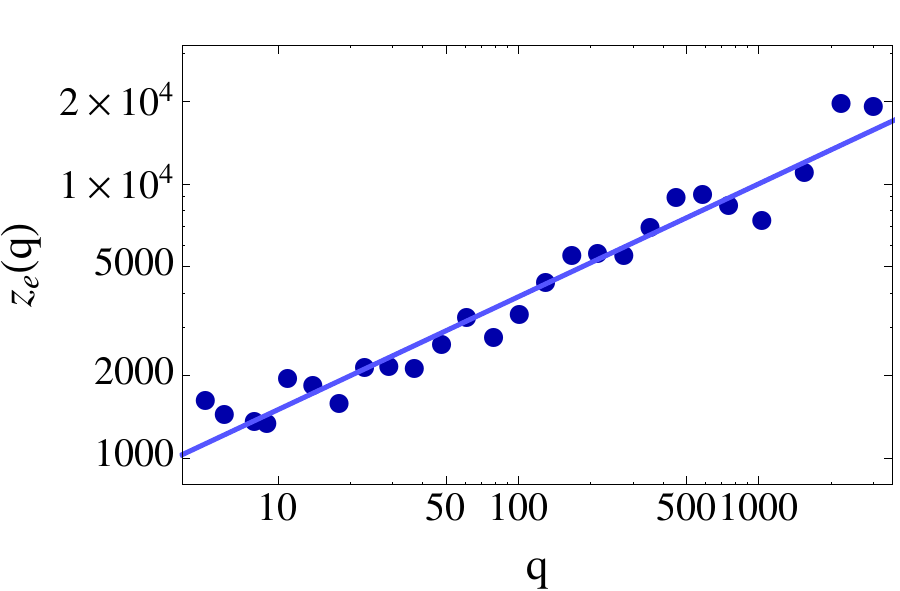}
    \end{center}
    \caption{ Top panel: Filled points show the $z_i(q)$ times obtained from simulations with an effective gauge boson $S_\mu$, whereas empty points show the analogous results from simulations with a single $W_\mu$ boson shown in Fig.~\ref{fig:gauge-zgq}. Blue squares and purple diamonds correspond to $q$ values inside and outside a resonance band of the Lam\'e equation. Bottom: points represent the $z_e(q)$ values obtained for the effective $S_\mu$ boson, whereas the blue line corresponds to the phenomenological fit of Eq.~(\ref{fit-gaugedecayII}). } \label{fig:zi-3gauge}
\end{figure}

It is crucial, then, that we figure out the importance of these differences. If they are irrelevant, we can then simply use the results presented so far for a single gauge field, for describing the real case of three gauge bosons. In order to find this out, we have compared the results for $z_i$ and $z_e$ from simulations with only one gauge boson, identical $q$ and $\beta$, but different $\lambda$ ( = $\lambda_{1b}$ and $\lambda_{3b}$, according to the discussion above). 
In the new simulations with one effective gauge boson $S_\mu$, we have observed the same dynamics as in the case of only one $W$ boson. We first observe a stage in which the volume-averaged Higgs amplitude $|h|$ shows a $plateau$. In this regime, as before, the gauge energies grow very fast, but their contribution is still not important enough to affect the Higgs condensate. The times $z_i$ at which the plateau ends are reduced slightly with respect to the $W$-boson case when $q$ is outside a resonance band, but they are almost identical when it is within a band; see Fig.~\ref{fig:zi-3gauge}. There are, however, virtually no differences in the time scale $z_e$, which signals again both the end of the Higgs decay and the onset of equipartition. The new fit of $z_e$ from the simulations with an effective gauge boson $S_\mu$ is
\begin{eqnarray}\label{fit-gaugedecayII}
z_e (q) = 581 {q}^{0.42} = 581 (q_Z + 2 q_W)^{0.42}\,,
\end{eqnarray}
very similar to the old fit Eq.~(\ref{fit-gaugedecay}). Note that, as mentioned before, this fit is done for a RD universe with $\beta=0.01$. Anticipating again the results that we will explain in Section \ref{sec:vi}, the generalization of this fit to other $\beta$ values and expansion rates (characterized by $\omega$) is
\be z_e (q) \approx 58.1 \beta^{\frac{-(1 + 3 \omega)}{3 (1 + \omega)}} (2q_W+q_Z)^{0.42} \label{fit-gaugedecayII-extrapol} \ ,\ee
This equation probably represents the most relevant result of our paper. We see that the real decay time $z_e$ of the Higgs into the three gauge bosons $W^\pm, Z$ is, using again the approximate high-energy relation $q_Z \approx 2 q_W$, a factor $((q_Z + 2 q_W)/q_W)^{0.42} \approx 4^{0.42} \approx 1.79$ times longer than if we only considered the decay of the Higgs into a single $W$ boson [equivalently, a factor $ ((q_Z + 2 q_W)/q_Z)^{0.42} \approx 2^{0.42} \approx 1.34$ longer if we considered the decay of the Higgs into a $Z$ boson]. 

It is perhaps worth noticing that it seems surprising at first glance, that the decay takes longer when the resonance parameter is effectively larger, $q = 2q_W + q_Z > q_W$; naively one would expect a faster decay if there are more bosons into which to decay. This is, however, a reflection again of the nonlinear behavior of the system at $z \gtrsim z_i$, responsible for the previously discussed counterintuitive growth of $z_e(q)$ with $q$.

We also expect the energy equipartition not to change with respect to the single $W$ boson case, simply because the way in which the gauge fields are initially excited, should not affect the late-time dynamics when the nonlinearities are important. As confirmed by the lattice simulations, this is indeed the case. We have checked that the final equipartition state is identical to the previously studied case of one single boson, reaching at late times, 
\be\frac{E_{\rm K}}{E _t} \approx 0.3 \ , \hspace{0.1cm} \frac{E_{\rm GD}}{E _t} \approx 0.3 \ , \hspace{0.1cm} \frac{E_{\rm E}}{E _t} \approx 0.2 \ , \hspace{0.1cm} \frac{E_{\rm M} }{E _t} \approx 0.2 \ , \ee
and $E_{\rm V} / E _t \ll 1$. Note that, in the case of three gauge bosons, we have three different electric and magnetic fields. From the relation $Z_i (z) = 2 W_i (z)$ (valid at high energies), and given the definition of the electric and magnetic energies (\ref{eq:electric-energy})-(\ref{eq:magnetic-energy}), we see that 50\% of the total electric energy corresponds to the $Z$ boson, while the other 50\% is divided equally between the other two $W$ bosons. The same distribution takes place for the magnetic energy.

As a final remark, let us note that before nonlinear effects become important, the behavior of the three gauge fields is described by the same Lam\'e equation with resonance parameter $q = 2q_W+q_Z$. Due to this, we have observed that for $z \lesssim z_i(q)$, the spectra of the three fields excite the same range of momenta, corresponding to the resonance band of the Lam\'e equation for such resonance parameter. This is important, as one could naively think that the spectra of $W^{\pm}$ and $Z$ are independent, with different rangesof excited momenta accordingly to their different resonance parameters $q_W$ and $q_Z$. On the contrary, the introduction of three gauge fields in the system makes them evolve, as seen, as a single effective gauge boson, with the same effective resonance parameter given by Eq.~(\ref{eq:qeff}).

\begin{figure*}[t]
    \begin{center}
        \includegraphics[width=8.4cm]{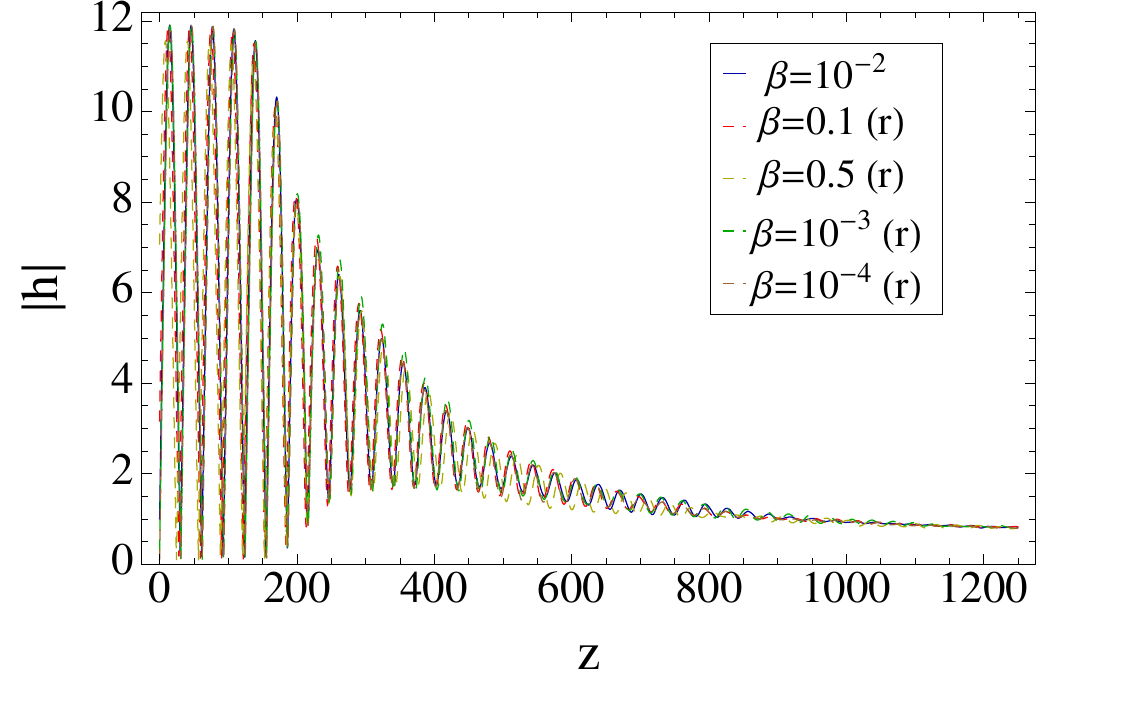}
        \includegraphics[width=8.4cm]{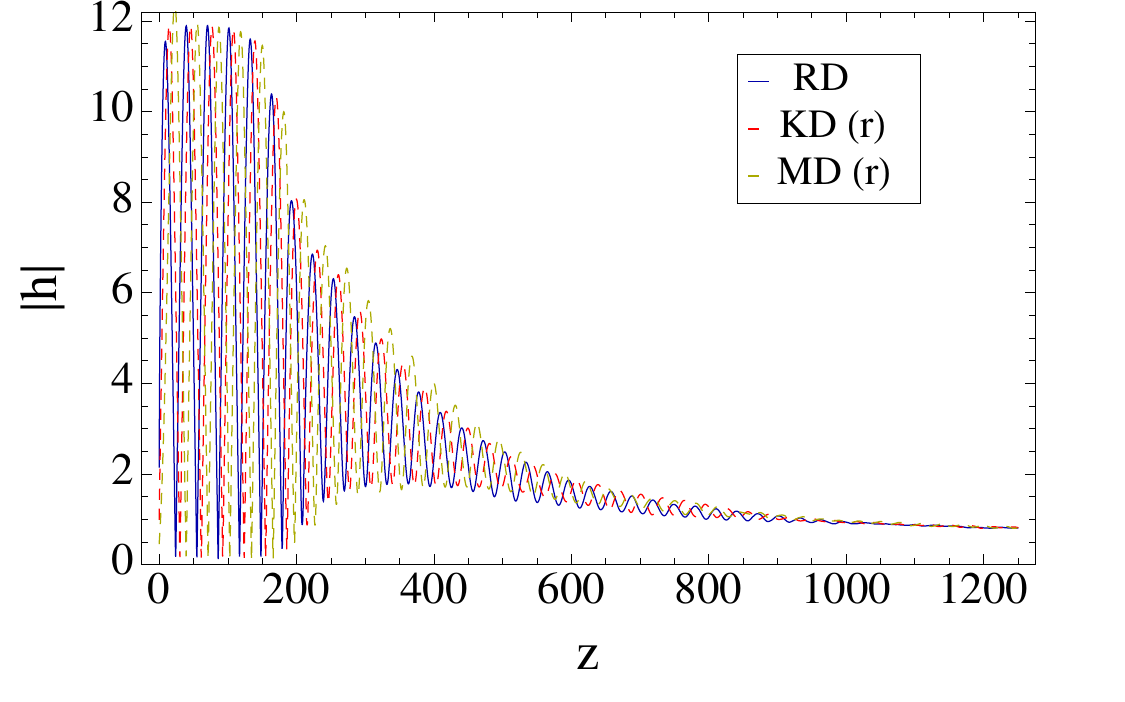}
        \includegraphics[width=8.4cm]{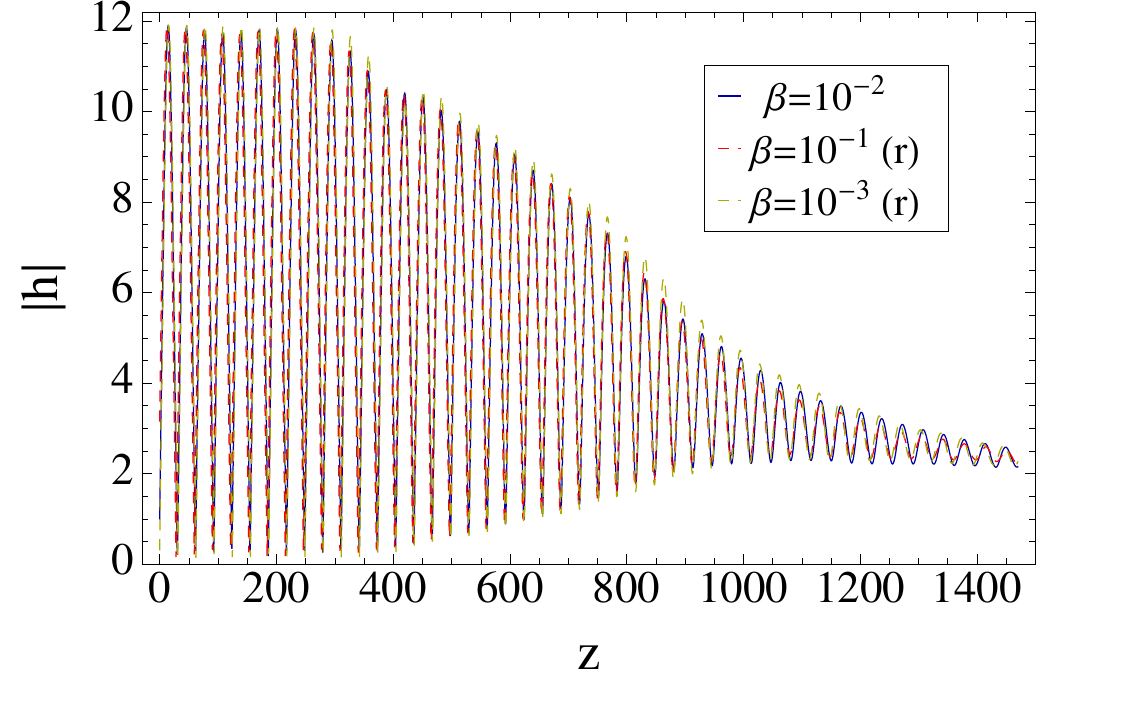}
        \includegraphics[width=8.4cm]{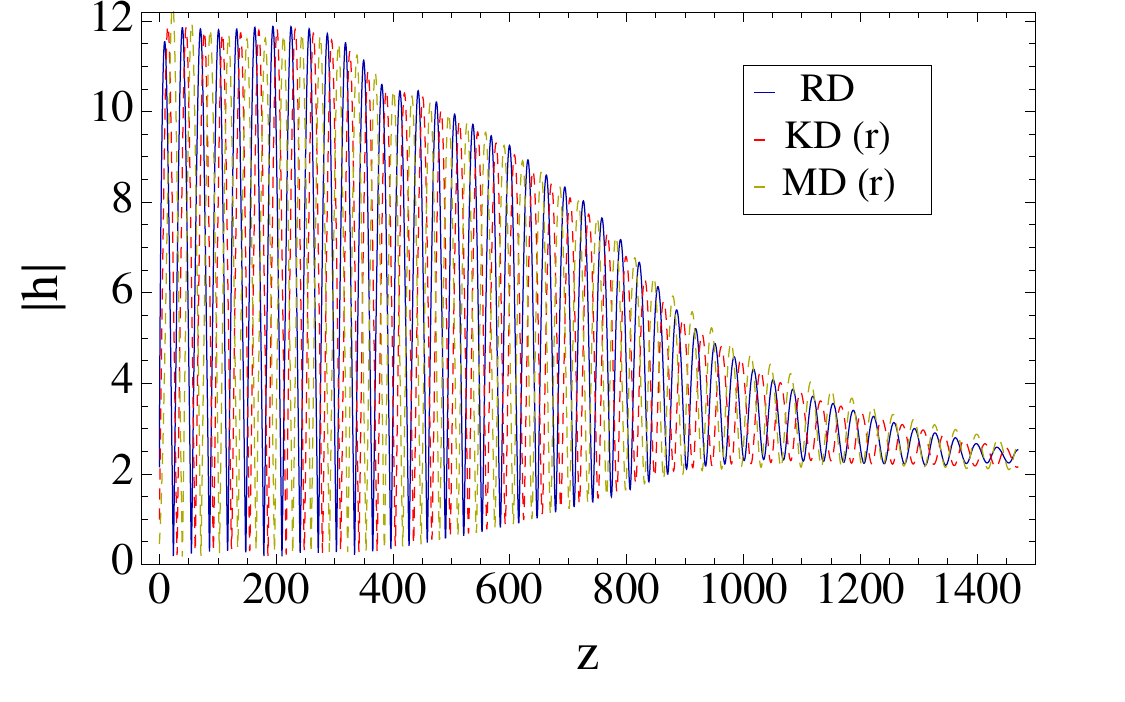}
    \end{center}
    \caption{We plot the volume-averaged value of the Higgs conformal field $|h|$ as a function of time, obtained directly from our simulations, for either different $\beta$ parameters or expansion rates. Lines with the symbol `(r)' have been extrapolated, using an inversion of Eqs.~(\ref{eq:extrapol-1}) and (\ref{eq:extrapol-2}), to obtain a theoretical prediction of the results of a RD universe ($\omega = 1/3$) with $\beta = 0.01$. The top figures correspond to global simulations with $q =8$, and the bottom figures correspond to Abelian-Higgs simulations with $q=6$. In the left figures, we vary $\beta$, while in the right figures, we vary $\omega$. We see that the lattice results for $(\omega, \beta) = (1/3,0.01)$ coincide quite well with the different theoretical extrapolations obtained from the lattice results for other $(\omega, \beta)$ parameters.} 
    \label{fig:bsf}
\end{figure*}

\section{Varying the Higgs initial amplitude and the expansion rate}
\label{sec:vi}

All results from sections \ref{sec:iv} and \ref{sec:v} have been presented for a scale factor evolving in a RD universe ($\omega = 1/3$), and for $\beta =0.01$. Naturally, in order to fully understand the dynamical properties of the Higgs decay after inflation, we have explored other $\beta$ parameters, and we have also considered other expansion rates such as MD ($\omega = 0$) or KD ($\omega = 1$). Fortunately, one can easily extrapolate the results from one particular set of parameters, say ($\beta_1$, $\omega_1$), to another set ($\beta_2$, $\omega_2$), using the analytical properties of the Higgs equation described in Section \ref{sec:II}. In other words, from the results obtained for ($\beta_1$, $\omega_1$), we can obtain a very good approximation to the ones for ($\beta_2$, $\omega_2$).

More specifically, we saw in Eq.~(\ref{eq:VariablesAtMax}) that in the case of no coupling to the gauge bosons, the conformal period $Z_T$ and the value of the transformed Higgs field at the first maximum $h(z_{\rm M})$, can be approximated as $Z_T = c_1 \beta^{\frac{-(1 + 3 \omega)}{3 (1 + \omega )}}$ and $h (z_{\rm M} ) = c_2 \beta^{-\frac{2}{3 (1 + \omega )}}$, where $c_1$ and $c_2$ are constants independent of $\omega$ and $\beta$. From these properties we can see that, if for a given set of values $(\omega_1,\beta_1)$, the volume-averaged Higgs field takes the value $h (\beta_1, \omega_1)$ at the time $z (\beta_1, \omega_1) $, then for $(\omega_2,\beta_2)$ the Higgs field at the time
\be z (\beta_2, \omega_2 ) \simeq \beta_2^{\frac{- (1 + 3 \omega_2)}{3 (1 + \omega_2 )}} \beta_1^{\frac{(1 + 3 \omega_1 )}{3 ( 1 + \omega_1)}} z (\beta_1, \omega_1 ) \ , \label{eq:extrapol-1} \ee
should take the value
\be h (\beta_2, \omega_2 ) \approx \beta_2^{\frac{-2}{3 (1 + \omega_2 )}} \beta_1^{\frac{2}{3 (1 + \omega_1)}} h (\beta_1, \omega_1 ) \label{eq:extrapol-2} \ . \ee

Notably, this property is maintained quite well even in the presence of a Higgs coupling to its decay products (either scalars in the global simulations or gauge bosons in the Abelian-Higgs simulations). This extrapolation is therefore very powerful. In Fig.~\ref{fig:bsf}, we have plotted the volume-averaged value of $|h|$ as a function of time, for both global (top figures) and Abelian-Higgs simulations (bottom figures). Let us focus for instance on the top-left figure. We have obtained for $q=8$ the behavior of $|h|$ as a function of time for $\beta= 10^{-4},10^{-3},10^{-2},10^{-1},$ and $0.5$, directly from the simulations. Using the outcome from these simulations with different $\beta$ parameters, we have then inverted Eqs.~(\ref{eq:extrapol-1}) and (\ref{eq:extrapol-2}), and obtained the (extrapolated) behavior corresponding to $\beta = 0.01$. These are different predictions for the Higgs decay when $\beta = 0.01$, but obtained from the real data from simulations with different $\beta$ values. We see that the four different extrapolated theoretical predictions obtained for $\beta= 10^{-4},10^{-3},10^{-1}$ and $0.5$ coincide very well with the real simulation for $\beta =0.01$.

The same is done in the top-right figure, but changing the scale factor instead of $\beta$ (which we fix in this figure as $\beta=0.01$). There, we compare the result of the Higgs decay for a RD universe, on one hand obtained directly from simulations with $\omega = 1/3$, and on the other hand from the corresponding extrapolated predictions from the lattice simulations with $\omega = 0$ (MD) and $\omega = 1$ (KD). The three lines also coincide very well. The same analysis is repeated for Abelian-Higgs simulations in the two bottom figures, with identical conclusions.  

This property allows us to extrapolate easily the results for the Higgs decay time for a RD universe with $\beta=0.01$, presented in the last two sections, to another set of $(\omega,\beta)$ parameters. In particular, from Eq.~(\ref{eq:zs(q)}) we obtain Eq.~(\ref{eq:zs(q)-extrap}), from Eq.~(\ref{fit-gaugedecay}) we obtain Eq.~(\ref{fit-gaugedecay-extrap}), and from Eq.~(\ref{fit-gaugedecayII}) we obtain Eq.~(\ref{fit-gaugedecayII-extrapol}).

\section{Summary and discussion} 
\label{sec:vii}
The recent measurements of the Higgs boson mass~\cite{ATLAS2012,CMS2012} imply a relatively slow rise of its effective potential at high energies. In the regime where the EW vacuum is stable with the Higgs self-coupling kept positive, the Higgs develops a large VEV during inflation, representing a classical condensate, homogeneous over scales exponentially larger than the inflationary radius $1/H_*$. In this paper we have studied the relaxation of the Higgs, i.e.~its decay, during the stages following immediately after inflation. In reality, the origin of the VEV during inflation, which sets up the initial condition for the decaying process, is not particularly relevant for our study. If another mechanism (different than quantum fluctuations) is responsible for the development of the Higgs VEV during inflation, our calculations and results would be equally applicable. The case considered in the paper, with the initial amplitude of the Higgs condensate dictated by the equilibrium distribution Eq.~(\ref{eq:ProbEQ}), due to the stretching of its quantum vacuum fluctuations, simply serves as a starting and practical point, to assess the typical Higgs amplitudes at the end of inflation.

The decay of the Higgs condensate during the early postinflationary stages constitutes an important event in the evolution of the Universe, which might have interesting cosmological consequences. In this article we have focused on the details of the Higgs decay process itself. We have used different methods of progressive complexity, accuracy and proximity to the real case of the SM. We have modeled the SM interactions in a two-step manner. First, considering a global scenario, ignoring the gauge structure of the SM, representing the gauge fields as a collection of scalar fields appropriately coupled to the Higgs. Secondly, we have considered an Abelian gauge scenario, with the gauge fields and the Higgs embedded within an Abelian-Higgs framework, ignoring the nonlinearities due to the truly non-Abelian nature of the SM. For the global model we have presented both analytical (Section~\ref{subsec:III.a}) and lattice calculations (Section~\ref{sec:iv}), whereas in the most precise and involved gauge modeling, we have just presented the outcome from lattice simulations (Section~\ref{sec:v}). 

The analytical results of the global modeling estimate correctly the right order of magnitude of the Higgs decay time. When studying such a scenario in the lattice, including all nonlinearities within such a scheme, we find that the actual Higgs decay takes longer, typically a factor $z_e/z_{\rm eff} \sim 3.17 q^{0.44}$ larger: see Eq.~(\ref{eq:zs(q)-extrap}) for $z_e$ and Eq.~(\ref{eq:EffEnergyTransferTimeScaleApprox}) for $z_{\rm eff}$. This is because the analytical calculations are only capable of estimating the order of magnitude of the time scale when sufficient energy has been transferred into the extra scalar fields (mimicking the EW gauge bosons). However, that time only signals the moment $z = z_i(q)$ when the Higgs condensate really starts noticing that it is coupled to extra species. From then on, at times $z \gtrsim z_i(q)$, the Higgs energy density begins to decrease in a noticeable manner, being transferred to the most strongly coupled species, the EW gauge bosons. It is this decrease of the energy of the Higgs that should be interpreted as the decay of the Higgs. Eventually, the Higgs energy density saturates to an approximately constant value, at some moment $z_e(q) > z_i(q)$. Around the same time, the energy of the species coupled to the Higgs has also stopped growing, and saturates into slowly evolving magnitudes.

Very interestingly, the same pattern and time scales are observed in the gauge scenario, though the final fractions of energies are different. The time scale $z_e (q)$ that characterizes the end of the Higgs decay in the gauge case is given by Eq.~(\ref{fit-gaugedecay-extrap}), which represents a factor $z_e/z_{\rm eff} \sim 3.68 q^{0.42}$ larger than the analytical prediction $z_{\rm eff}$ of Eq.~(\ref{eq:EffEnergyTransferTimeScaleApprox}). We see therefore that, at the end, the differences between the global and gauge modelings are not so relevant, at least in terms of the estimation of the Higgs decay time $z_e(q)$. It is worth stressing that $z_e(q)$ grows with $q$ (both in the global and gauge scenarios), which could be thought as being a counter-intuitive fact. This is due to the nonlinearities characteristic of the system, which become relevant from $z \gtrsim z_i$ onwards.

One of our more interesting results is the extrapolation laws Eqs.~(\ref{eq:extrapol-1}),(\ref{eq:extrapol-2}). We have seen that the dynamics of the system depend basically on three parameters: $q$, $\beta$, and the expanding background equation of state $\omega$. Eqs.~(\ref{eq:extrapol-1}),(\ref{eq:extrapol-2}) allow us to extrapolate the lattice results for parameters $(\omega_1,\beta_1)$ into a very good approximation to the results of another set of parameters $(\omega_2,\beta_2)$. This technique works very well indeed for both global and Abelian-Higgs simulations (see Fig.~\ref{fig:bsf}). This happens because the properties of Eqs.~(\ref{eq:VariablesAtOsc}) and (\ref{eq:VariablesAtMax}) derived in Section III, also hold quite well in the presence of a coupling of the Higgs to its decay products. This has led us to obtain the generic formula for the Higgs decay time $z_e$, Eq.~(\ref{fit-gaugedecayII-extrapol}), as a function of $\beta$, $q$ and $\omega$.

Remarkably, we have also shown that the case of the SM, where the Higgs is coupled simultaneously to the three EW gauge bosons $W^+$, $W^-$ and $Z$, behaves identically to the case in which the Higgs is only coupled to one {\it effective} gauge boson, with resonance parameter $q = q_Z + 2 q_W$. We have found that when the three gauge bosons are considered, $z_e (q) = 581 (q_Z + 2 q_W)^{0.42}$ [Eq.~(\ref{fit-gaugedecayII})]. The decay of the Higgs takes then a factor $(2+q_Z/q_W)^{0.42}$ larger than if the Higgs were coupled to only one $W$ boson, or equivalently $(1+2q_W/q_Z)^{0.42}$ times larger than if it were coupled to only $Z$ gauge bosons. Again, this counterintuitive result is due to the nonlinearities that dominate the system at $z \gtrsim z_i$. 

Interestingly, at the time $z \approx z_e (q)$, in both in the global and gauge scenarios, we see that the distributions of fields reach equipartition. In the global model we find that the kinetic energy of the Higgs becomes equal to the sum of the gradient energy of the Higgs plus the interaction with the $\chi_i$ fields, $E_{\rm K}^{\varphi} \simeq E_{G}^{\varphi} + E_{\rm int}$. This equality holds to better than 1 \% from $z \gtrsim z_e$ onwards. In the gauge scenario, we find that the kinetic energy of the Higgs becomes equal to the sum of the covariant gradient energy (which includes the Higgs-gauge bosons interactions) plus the Higgs potential, $E_{\rm K} \simeq E_{\rm GD}+ E_{\rm V}$. This equality also holds to better than 1\% from $z \gtrsim z_e (q)$ onwards. At some later time $z \gtrsim z_e$, the electric and magnetic energy densities also reach equipartition to better than 1\%, $E_{\rm E} \simeq E_{\rm M}$. The distribution of energy in the gauge scenario is actually universal, since the system always reaches equipartition, with $E_{\rm K} \simeq E_{\rm GD}$ representing 30\% of the total energy, and $E_{\rm E} \approx E_{\rm M}$ representing 20\% each. In both global and gauge scenarios, once in the stationary equipartitioned regime, the potential energy becomes gradually more and more irrelevant.

Before we conclude, let us note that the postinflationary decay of the Higgs analyzed here is very similar to the analogous decay during reheating after Higgs-inflation~\cite{Bezrukov:2007ep,Bezrukov:2008ut,GarciaBellido:2008ab,Figueroa:2009jw,GarciaBellido:2011de}. The contexts are, however, very different. In Higgs-inflation the Higgs plays the role of the inflaton and dominates the energy budget of the Universe, so the decay of the Higgs after inflation truly represents the actual reheating of the Universe. In the case we have studied in this paper, the Higgs is simply a spectator field during inflation, and its energy density is only a marginal fraction of the inflationary one; see Eq.~(\ref{eq:higgs-energy-infl}). In Higgs-inflation, a nonminimal coupling $\xi\varphi^2R$ to gravity is required, with $\xi \sim \mathcal{O}(10^4)\sqrt{\lambda}$. The resonance in both Higgs-inflation and Higgs-spectator scenarios is dominated by the decay into the gauge bosons $W^{\pm}, Z$. The resonance parameter, however, scales as $q \sim {g^2\over \lambda}\xi$ in Higgs-inflation, versus $q \sim {g^2\over \lambda}$ in our Higgs spectator scenario. Therefore, the resonance is $\sim 10^{3}\sqrt{\lambda_{001}}$ times broader in Higgs-inflation than in the Higgs-spectator case. However, in Higgs-inflation, the nonperturbatively produced gauge bosons (at each Higgs zero crossing), decay very fast into the SM fermions via perturbative decays. So for around $\sim 100$ oscillations of the Higgs, the resonance is blocked in Higgs-inflation, simply because the occupation numbers of the gauge bosons do not pile up~\cite{GarciaBellido:2008ab}. This phenomenon is called {\em combined preheating}, and it is absent (or in general it is expected to be only a marginal effect) in the Higgs spectator scenario studied here, as shown in~\cite{Enqvist:2014bua}. 

To conclude, let us note that our paper is intended to be the first one of a series, where we plan to analyze further the details of the Higgs decay (i) and its cosmological consequences (ii). In particular,\\

(i) The results obtained here have gone far beyond the analytical ones available in the literature~\cite{Enqvist:2013kaa,Enqvist:2014tta}. We have presented different approaches to the nonperturbative and nonlinear dynamics of the decay process. Our most precise results are the outcome from our simulations in Section~\ref{sec:v}, corresponding to an Abelian-gauge model mimicking the structure of the SM interactions. Even though there is a good motivation to neglect  the truly non-Abelian nature of the interactions, only lattice simulations which fully incorporate the non-Abelian $SU(2)\times U(1)$ structure of the SM will really tell us about the (un)importance of the corrections due to the nonlinearities in the gauge sector. Besides, the details of the stationary stage might very well (and indeed most likely will) change when the full non-Abelian structure of the SM is restored. Therefore, even if the time scales of the start of the Higgs decay and onset of stationary regime may (expectedly) not change much, the fine details can only be quantified in light of such non-Abelian simulations, which are beyond our present work. Moreover, in order to assess with even a higher degree of realism the final outcome of the energy distribution among fields, thermal corrections~\cite{Enqvist:2012tc,Enqvist:2013qba,Lerner:2015uca} and fermions~\cite{Borsanyi:2008eu,Saffin:2011kc,Saffin:2011kn} should be effectively incorporated into such simulations. 

(ii) The postinflationary decay of the SM Higgs may have several observable consequences. The possibility has been recently proposed~\cite{Kusenko:2014lra,Pearce:2015nga} of realizing baryogenesis via leptogenesis, thanks to the Higgs oscillatory behavior. The time dependence of the Higgs condensate oscillations can create an effective chemical potential for the lepton number, which could lead to the generation of a lepton asymmetry in the presence of right-handed Majorana fermions with sufficiently large masses. The electroweak sphalerons would then redistribute such asymmetry among leptons and baryons. Second, the fields excited from the decay of the Higgs may act as a source of gravitational waves~\cite{GWs2006EastherLim,GWsPRL2007FigueroaGarciaBellido,GWsPRD2007FigueroaGarciaBellidoSastre,GWs2007DufauxEtAl,
GWsAbelianHiggsHybridPreheatingDufauxEtAl2010,Enqvist:2012im,Figueroa:2013vif}. The case of the charged fermions of the SM was considered~\cite{Figueroa:2014aya}, but it is expected that the background of gravitational waves from the EW gauge bosons contributes to a much larger signal~\cite{Figueroa:2014aya}. Besides, the fact that the Higgs is a condensate varying at superhorizon scales may give rise to interesting anisotropic effects~\cite{Bethke:2013aba,Bethke:2013vca} in the amplitude of such a background of gravitational waves. Thirdly, it is indeed possible that the gauge field production that we have described in this paper could provide the necessary conditions for primordial magnetogenesis. Although it might be challenging to obtain a sufficiently large correlation length, it is conceivable that an inverse cascade process provides the appropriate mechanism for the growth of an initially small correlation length~\cite{DiazGil:2007dy,DiazGil:2008tf}. Finally, if dark matter is a gauge singlet field coupled to the Higgs, it is also possible that the Higgs oscillations could produce the right amount of dark matter, such that its distribution could account for the correct relic abundance \cite{Enqvist:2014zqa}.

{\it Note added. -} After completion of this work, the preprint~\cite{Enqvist:2015nab} by Enqvist et al was uploaded to the {\tt ArXiv}, presenting lattice simulations of the same process analyzed in this paper, but considering the non-Abelian structure of the standard model. Only low-resonance parameters with $q\leq 20$ were considered, and for a fixed initial amplitude of the Higgs and postinflationary expansion rate. The expected broadening of the gauge field spectra, due to the nonlinearities introduced by the non-Abelian terms, is indeed clearly observed after some time, as compared to the Abelian simulations. However, for the lowest case of $q\approx6$, where the effects of such nonlinearities are expected to be maximum, only a factor $\sim 2$ of difference in the estimation of $z_i$ is observed, as compared to the analogous Abelian simulation. For higher-resonance parameters, the Abelian approximation becomes better and better, as the correction due to non-Abelian terms become more and more irrelevant, see Eq.~(\ref{eq:nonab}). Besides, in the context of a large inflationary energy scale (close to its upper bound), it is rather expected that $q \gg 10$, as $q \sim \mathcal{O}(10)$ requires an excessively large Higgs self-coupling. Therefore, we are positive that the work we have developed here is a very good approximation to the real, non-Abelian dynamics. We plan to study this issue in a future publication.

\acknowledgments
We thank Mustafa Amin for illuminating discussions on the initial conditions. We would like to thank Fedor Bezrukov for making publicly available the package to compute the running of the Higgs self-coupling in the website \url{http://www.inr.ac.ru/~fedor/SM/}. F.T. acknowledges the CERN Theory Division for kind hospitality.  This work is supported by the Research Project of the Spanish MINECO FPA2012-39684-C03-02 and the Centro de Excelencia Severo Ochoa Program SEV-2012-0249. F.T. is supported by the FPI-Severo Ochoa Ph.D. fellowship SVP-2013-067697.  We acknowledge the use of the IFT Hydra cluster for the development of this work.

\appendix

\section{Lattice formulation}
\label{App-Lattice}

In this appendix, we provide a more detailed discussion of the lattice formulations for both the global and the Abelian-Higgs simulations. Let us start by writing the action for both scenarios in the continuum. In the global case, the continuous action Eq.~(\ref{eq:action-scalar}) can be written in our natural variables as (from now on $a_* = 1$)
\bea S &=& \int \frac{d^4 z}{a^4} \left[ \frac{\beta^2}{2 \lambda} \Bigg( - (h' - \mathcal{H} h )^2  + \partial_i h \partial_i h  \right. \Big. \nonumber \\
&+& \left. \left.  \sum_j \left\lbrace - \left( X_j^{'} - \mathcal{H} X_j\right)^2 + \partial_i X_j \partial_i X_j \right\rbrace \right) \right.  \nonumber \\
&+& \left. \frac{\beta^4}{4 \lambda} h^4 + \frac{e^2 \beta^2}{2} h^2 \sum_i {X_j}^2 \right] \ . \label{eq:action1} \eea
Varying this action, we find the continuum EOMs of the system 
\bea h'' -\nabla^2 h + \beta^2 h^3 + e^2 h \sum_{j} X_j^2 = \frac{a''}{a} h\,, \\ 
X''_j - \nabla^2 X_j + q \beta^2 h^2 X_j = \frac{a''}{a} X_j \ , \eea
which are Eqs.~(\ref{eq:eoms1a}) and (\ref{eq:eoms2}) of the main text.

 We now want to write the equivalent of these equations in the lattice. We will work in a lattice cube of length $L$ with $N^3$ points. We take the time step to be $d_0$ and the lattice spacing to be $d_i = d \equiv L/N$ ($i=1,2,3$). We write
 \bea 
 \Delta_0^{-}\Delta_0^{+} h - {\Delta}_i^-{\Delta}_i^+
 h + \beta^2 h^3 + e^2 h \sum_{j} X_j^2 = \frac{a''}{a} h\,, \label{eq:eomsL1} \\ 
 \Delta_0^{-}\Delta_0^{+}
 X_j - {\Delta}_i^-{\Delta}_i^+
 X_j + q \beta^2 h^2 X_j = \frac{a''}{a} X_j \ , \label{eq:eomsL2}  \eea
 where we have defined the discrete derivatives $\Delta_{\mu}^{+} \phi \equiv \frac{1}{d_{\mu}}(\phi  (\hat{n} + \hat{\mu} ) - \phi)$ and $\Delta_{\mu}^{-} \phi \equiv \frac{1}{d_{\mu}}(\phi - \phi (\hat{n} - \hat{\mu} ))$. Normally, one obtains the operators $\Delta_{\mu}^{-} \Delta_{\mu}^{+}$ from discretizing the continuum action (\ref{eq:action1}) and then minimizing it with respect the lattice field variables (which live in the lattice sites $\hat{n}$). However, since we are not treating at the same level in the lattice the scale factor $a(t)$ (which we do not discretize) and the field variables $h$, $X_j$ (which are discretized), we prefer to proceed by simply substituting the continuum operators $ \partial_{\mu} \partial_{\mu}$ for the lattice equivalent $\Delta_{\mu}^{-} \Delta_{\mu}^{+}$. Actually, if we indeed proceeded by discretizing the action and then finding the lattice EOM, we would of course obtain the lattice operators $\Delta_{\mu}^{-} \Delta_{\mu}^{+}$ on the left-hand sides of Eqs.~(\ref{eq:eomsL1}) and (\ref{eq:eomsL2}). However, on the right-hand side, the term $\frac{a''}{a} h$ would be more involved in discrete derivatives of time. Since we know that such a term decays very fast, we have simply introduced the term $\frac{a''}{a} h$ on the right-hand side of the EOM by using a continuous function $a(t)$ evaluated at the appropriate discrete times.
 
With respect to the Abelian-Higgs simulations, let us first write its continuous action Eq.~(\ref{action}) in terms of natural variables (where $h \equiv h_1 + i h_2$):
\bea S =&& \frac{1}{\lambda} \int  \frac{d^4 z}{a^4} \left( \frac{\beta^2}{2} \left[ - \left| h' - \mathcal{H} h \right|^2 \right. \Big. + | D_i h |^2  \right]  \nonumber \\ 
 && + \left.\frac{1}{4 q} \sum_{i \neq j} G_{i j}^2 - \frac{1}{2 q} \sum_i G_{0 i}^2 + \frac{\beta^4}{4} (h^* h)^2 \right) \ , \eea
where $G_{\mu \nu} \equiv \partial_{\mu} V_{\nu} - \partial_{\nu} V_{\mu}$. Varying this action, we obtain the EOM in the continuum 
\bea 
h''  - D_{i} D_{i} h  + \beta^2 |h|^2 h  &=& h \frac{a''}{a} \ ,  \\
V_j'' + \partial_{j} \partial_{i} V_i - \partial_{i} \partial_{i} V_j &=&  j_i (z) \ , \\
\partial_i V_{i}' &=&  j_0 (z) \ , 
\eea
where the current $j_{\mu} (x)$ is defined as $j_{\mu} (x) \equiv q \beta^2  \mathfrak{Im} [ (h_1 - i h_2) D_{\mu} (h_1 + i h_2) ]$. These are precisely equations (\ref{eomb1})-(\ref{eomb4}) of Section \ref{sec:v}. Again, as in the global case, the standard procedure would be to discretize the continuum action such that the covariant derivatives are substituted for the standard lattice ones defined in terms of links $U_i = e^{-i V_i d_i}$. However, this would introduce an unnecessary complication for describing the term $\frac{a''}{a} h$ in the EOM. Therefore, we proceed again by simply discretizing directly the EOM with the correct lattice operators on the left-hand side of the equations coming from the discretization of the lattice gauge invariant action, whereas we maintain again the term $\frac{a''}{a} h$ with the scale factor given by a continuous function evaluated at the discrete times. The lattice EOMs then look as follows: 
\bea
\Delta_0^{-} \Delta_0^{+} h - \sum_i D_i^{-} D_i^{+} h + \beta^2 |h|^2 h &=& \frac{a''}{a} h \ ,  \nonumber \\
\Delta_0^{-} \Delta_0^{+} V_i  - \sum_j ( \Delta_j^{-} \Delta_j^{+} V_i - \Delta_i^{+} \Delta_j^{-} V_j   ) 
&=& \frac{q \beta^2}{d_ i} \mathfrak{Im} [ h^* U_i h_{+ i}  ] \ , \nonumber \\
\sum_i \Delta_i^{-} \Delta_0^{+} V_i &=&  J_{\hat{n}} \ , \\ \nonumber \label{eq:discrete-3} \eea
where we have defined $J_{\hat{n}}$ at the lattice point $\hat{n}$ as
\be J_{\hat{n}}  \equiv \frac{q \beta^2}{d_0}  \mathfrak{Im} [h^* U_0 h_{+ 0} ] \ ,\ee
and the lattice covariant derivatives as $D_{\mu}^{+} \phi = \frac{1}{d_{\mu}} (U_{\mu} \phi  (\hat{n} + \hat{\mu} )- \phi)$ and $D_{\mu}^- \phi = \frac{1}{d_{\mu}} ( \phi - U_{ \mu}^* (\hat{n} - \hat{\mu} ) \phi (\hat{n} - \hat{\mu} ))$.

One needs to check that for all times, the discrete Gauss law (\ref{eq:discrete-3}) is conserved. In particular, we require that for all times
\be \Delta_{\rm G} \equiv \frac{1}{N^3} \sum_{\tilde{n}} \frac{|\sum_i \Delta_i^{-} \Delta_0^{+} V_i - J_{\hat{n}}|}{|\sum_i \Delta_i^{-} \Delta_0^{+} V_i + J_{\hat{n}}|} \ll 1 \ . \label{eq:Gauss-discrete} \ee
We have checked that this is indeed the case. In particular, we find that depending on the simulation, at the end of the running time the Gauss law is in fact only marginally broken, with $ \Delta_{\rm G} \lesssim { \rm 10^{-12} - 10^{-15} } $.

All results presented in this work have been obtained for $N=128$ points for both global and Abelian-Higgs simulations. Apart from $N$, we also need to fit the range of momenta that we want to cover in our simulations. This is a crucial step, as this range must be chosen carefully in order to capture all the relevant phenomenology of the Higgs decay. Let us call $p_{\rm min}$ the minimum momentum covered by the lattice. We can then fix the length of the cube $L$ and the maximum momentum covered by the lattice, $p_{\rm max}$, in terms of $N$ and $p_{\rm min}$ as
\be p_{\rm max} = \frac{\sqrt{3} N}{2} p_{\rm min} \ , \hspace{0.5cm} L = \frac{2 \pi}{p_{\rm min}} \ . \ee

Note that in this appendix, $k$ refers to physical momentum and $p$ to lattice momentum. As discussed in section \ref{sec:III}, the Higgs EOMs possess a well-known structure of resonance bands, which can be either of the form $0< k<k_{\rm *}$ or of the form $k_{\rm min}<k<k_{\rm *}$. We expect these momenta to be physically excited, at least at the first stages of the Higgs decay. Therefore, we must have a good coverage of this range of momenta. Let us define the coefficient $\alpha_c$
\be \alpha_c \equiv \frac{k_{\rm *}}{p_{\rm min}} \ . \ee
The larger $\alpha_c$ is, the better the infrared coverage of the resonance band, but the worse the ultraviolet scales are captured. In order to probe well the posterior displacement of the spectra to higher momenta when the system becomes nonlinear, we need to choose $\alpha_c$ judiciously. With this idea in mind, we have determined for each $q$, the $\alpha_c$ parameter that ensures a good infrared coverage without spoiling the ultraviolet part of the spectra. For simulations with $N=128$ points, we have fixed $\alpha_c$ typically within the range $4 \lesssim \alpha_c \lesssim 11$.

\begin{figure}[t]
    \begin{center}
        \includegraphics[width=8.5cm]{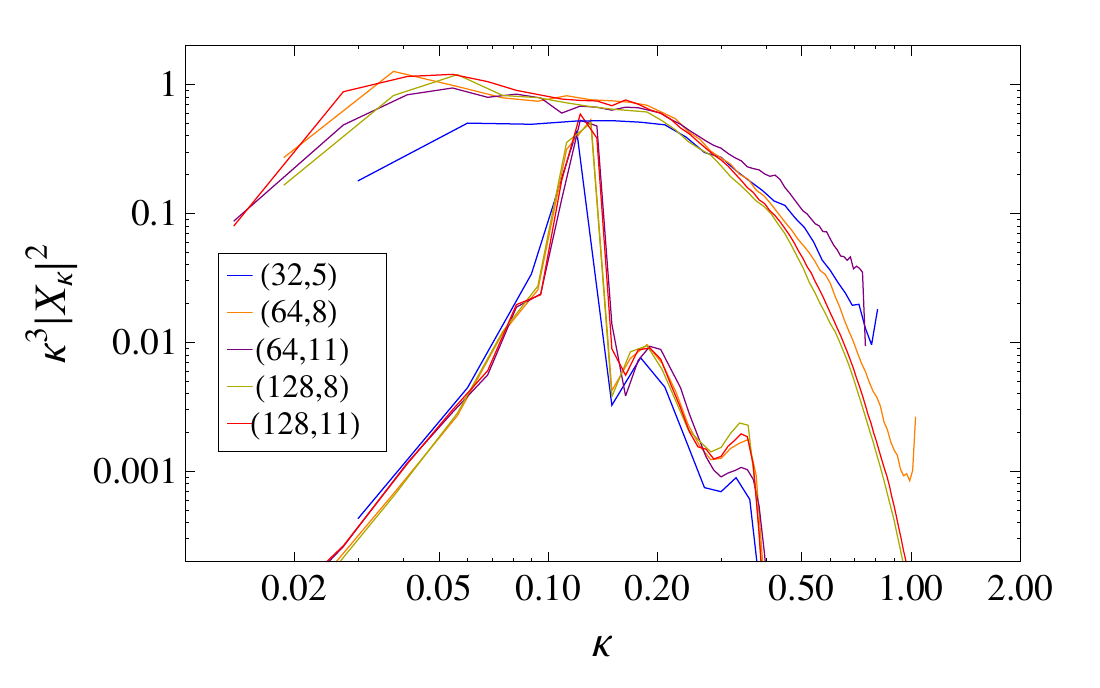}
                \includegraphics[width=8.5cm]{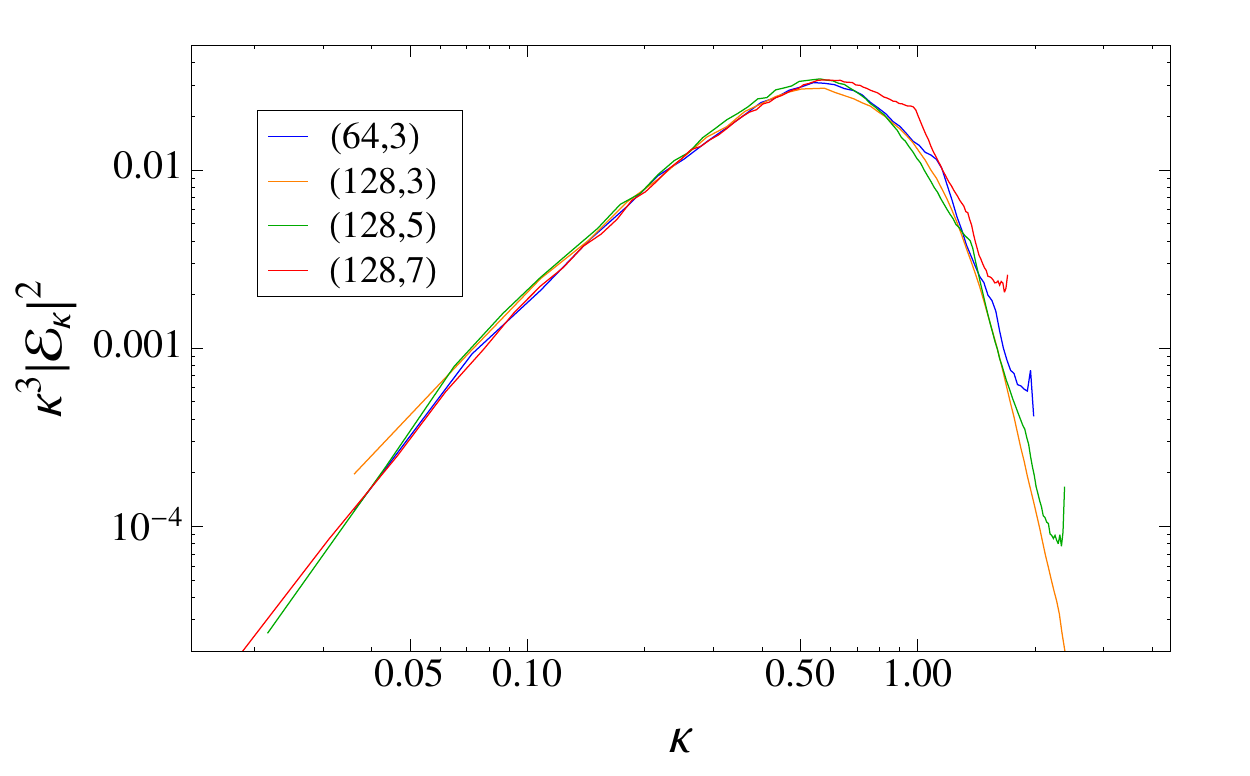}
    \end{center}
    \caption{Top: The spectra of one scalar field $\kappa^3 |X_{\kappa} |^2$ ($\kappa \equiv k / H_*$) in the global modeling for two different times, $z = 173$ and $z=996$, and for different sets of $(N,\alpha_c)$. We have taken $\beta=0.01$, $q=14$, and a scale factor evolving as a RD universe. Bottom: The electric spectra $\kappa^3 |\mathcal{E}_{\kappa} |^2$ in the Abelian-Higgs modeling for the time $z=615$, for different $(N,\alpha_c)$, and for $\beta=0.01$, $q=101$ and a RD universe. } 
    \label{fig:lattice-dependence}
\end{figure}

We show in Fig.~\ref{fig:lattice-dependence} two particular spectra obtained from simulations of the global scenarios at two different times, for different $(N,\alpha_c)$ parameters. Apart from a better or worse coverage of the ultraviolet or infrared regimes, the main physical results are well captured in all simulations, and are also consistent between them. The same consistency is observed in the Abelian-Higgs simulations, making our results robust versus lattice artifacts.

\section{Initial conditions}\label{App-Init}

In this section, we discuss in more detail the initial conditions of our lattice simulations for both the global and the Abelian-Higgs models. As has already been mentioned in the main text, we start our simulations (in both the global and Abelian-Higgs models) just after inflation ends, which we take as the time $z = 0$. From $z=0$ to $z=z_{\rm osc}$, we keep the gauge bosons deactivated, solving only the Higgs equation for the homogeneous mode. Therefore, at $z=z_{\rm osc}$, we have $h(z_{\rm osc})$ given by Eq.~ (\ref{eq:VariablesAtOsc}), and the rest of scalar/gauge fields set to 0. It is at this time that our lattice simulations truly begin, because at this moment we put quantum fluctuations over the homogeneous modes of both the Higgs and the decay product fields. We now explain how these fluctuations are set in both models.

\subsection{Global model}

For sake of clarity, let us come back temporarily to physical variables. Let us use $f({\bf x})$ to denote the quantum fluctuations of a field in position space, and $f_k$ its Fourier transform, defined as
\be f({\bf x}) = \frac{1}{(2\pi)^3} \int d^3 {\bf k} f_k e^{i {\bf k \cdot x}} \ . \ee

At $z = z_{\rm osc}$ we set, over the homogeneous mode of the different fields, a spectrum of quantum fluctuations corresponding to the probability distribution of the ground state of a scalar field in a FLRW universe
\be {\rm P} (|f_k|) d |f_k| = \frac{2 |f_k|}{\langle | f_k|^2 \rangle} e^{ - \frac{|f_k|^2}{\langle |f_k|^2 \rangle} } d |f_k| \ , \label{quantum-fluct} \ee
where we have
\be \langle |f_k|^2 \rangle = \frac{1}{2 a_{\rm osc}^2 \omega_{k,{\rm osc}}} \ .\ee
Here, $\omega_{k,{\rm osc}} \equiv \sqrt{k^2 + a_{\rm osc}^2 m^2_{\rm osc}}$ is the frequency of the field at the time $z_{\rm osc}$, and $m_{\rm osc}$ is the mass at this same time, $m^2_{\rm osc}= (\partial^2 V / \partial f^2) (z_{\rm osc})$ with $V$ the potential. The mode $f_k$ also contains an arbitrary random constant phase $ \forall {\rm Arg}(f_k)$. To maintain isotropy properties, we add both left-moving and right-moving waves, so that we take
\bea f_k &=& f_{k,l} + f_{k,r} \equiv \frac{|f_k|}{\sqrt{2}} ( e^{i \theta_1} + e^{i \theta_2} ) \ , \label{fluct1} \\ 
f_k' &=& i \omega_k a (f_{k,l} - f_{k,r} )  - \mathcal{H} f_k \label{fluct2} \ , \eea
where $\theta_1$ and $\theta_2$ are constants with $\theta_i \in [0,2 \pi) $. In the discrete lattice, we set the fluctuations in momentum space so that, from lattice point to lattice point, $|f_k|$  varies according to Eq.~(\ref{quantum-fluct}), and the phases $\theta_1$ and $\theta_2$ vary randomly within the interval $\theta_i \in [0,2 \pi) $.

From the properties of the Lam\'e equation discussed in Section \ref{sec:III}, we know that depending on the value of the resonance parameter $q \equiv g^2 / (4 \lambda)$, the Higgs equation has a certain structure of resonance bands. As we see in Fig. \ref{fig:FloquetVarious}, the most infrared band is always the one with the greatest Floquet index, and we hence expect that the Higgs decay will be dominated by this band, at least at initial times. It has a maximum at a given momentum, which we call $k_{\rm max}$. Therefore, this allows us to set a cutoff to the probability spectrum (\ref{quantum-fluct}), such that for $k > k_{\rm max}$, $|f_k| = 0$. As it should be, we have confirmed that changing this cutoff within a wide range of values does not significantly modify our results.

\subsection{Abelian-Higgs model: Gauss conservation law}

We now discuss how we set the initial quantum fluctuations in the Abelian-Higgs model. Caution must be taken in this step, because as we will see, we must ensure that the Gauss condition [Eq.~(\ref{eq:Gauss-discrete}) in the discrete] holds at the beginning of the simulations.

Let us come back to natural variables. In this section, we define $ h_j (\vec{z},z_{\rm osc})$ with $j=1,2$ to be the fluctuations of the two components of the conformally rescaled Higgs field at the time $z_{\rm osc}$. Let us also define $ h_j (k) \equiv  h_j (\vec{k},z_{\rm osc})$ to be their corresponding Fourier transforms. Following very closely our discussion of the initial conditions in the global modeling, we impose, over the two components of the Higgs, the spectra
\bea h_{1} (k) &=& \frac{|h_1|}{\sqrt{2}} \left( e^{i \theta_1} + e^{i \theta_2} \right) \ , \nonumber \\
h_{2} (k) &=& \frac{|h_2|}{\sqrt{2}}  \left(  e^{i \theta_3} +  e^{i \theta_4} \right) \label{fluct1} \ .\eea

Here, $|h_1|$ and $|h_2|$ are quantities that change, from point to point of the lattice in momentum space, according to the probability distribution function
\be P(|h_j|) d |h_j|= \frac{2 |h_j|}{\langle |h_j|^2 \rangle} e^{-\frac{|h_j|^2}{\langle |h_j|^2 \rangle}} d |h_j| \ , \label{fluct2}  \ee
($j = 1,2$) where 
\be \langle |h_j|^2 \rangle \equiv \frac{\lambda}{2 H_*^3 \beta^2 \omega_j} \ . \label{rms-hj}\ee 
The frequency is $\omega_j \equiv \sqrt{\kappa^2 + m_j^2}$, with $\kappa = k / H_*$ the natural momentum, and the natural masses defined as
\bea m_1^2 & \equiv & (3 h_{1 {\rm osc}}^2 + h_{2{\rm osc}}^2 ) \beta^2 = 3 h_{1 {\rm osc}}^2  \beta^2 \ ,\nonumber \\
m_2^2 & \equiv & (h_{1 {\rm osc}}^2 + 3 h_{2{\rm osc}}^2 ) \beta^2  = h_{1 {\rm osc}}^2 \beta^2 \ , \eea
where $h_{1 \rm {osc}} \equiv h_1 (z_{\rm osc})$ and $h_{2 \rm {osc}} \equiv h_2 (z_{\rm osc})$. For the last equality, we have used that, for the initial conditions of the Higgs homogeneous mode given in Eq.~(\ref{h-init}), we have $h_{\rm 2 osc} = 0$. From (\ref{fluct1}), the fluctuations of the Higgs derivatives are
\bea h'_{1} (k) &=& \frac{|h_1|}{\sqrt{2}} i \omega_1 \left( e^{i \theta_1} - e^{i \theta_2} \right) \ , \nonumber \\ 
h'_{2} (k) &=& \frac{|h_2|}{\sqrt{2}} i \omega_2 \left(  e^{i \theta_3} - e^{i \theta_4} \right) \ . \label{fluct3}  \eea

Also, the four different phases vary, in momentum space, from lattice point to lattice point. These phases would vary in principle randomly within the interval $\theta_i \in [0,2 \pi ) $, but as we are working also with gauge bosons, we need to preserve the Gauss law initially. Due to this, we thus may need to impose one simple constraint to the phases.

Let us discuss this in more detail. As mentioned before, we must ensure at the initial time the Gauss law (\ref{eomb4})
\be \partial_i V_{i}' =  j_0 (z) \ , \label{eq:gauss-app}\ee
with $j_{0} (z) \equiv q \beta^2  \mathfrak{Im} [ (h_1 - i h_2)  (h'_1 + i h'_2) ]$. Therefore, the quantum fluctuations we impose on the gauge fields at $z_{\rm osc}$ must preserve this condition. Let us write the Gauss law (\ref{eq:gauss-app}) in momentum space as
\bea V'_{i} (\vec{k}, z_{\rm osc} ) &=& i \frac{k_i}{k^2} j_0 (k) \ , \nonumber \\
V'_i (\vec{0},z_{\rm osc}) &=& 0 \ , \label{eq:si-init} \eea
where $j_0 (k)$ is the Fourier transform of $j_0 (z)$ at the time $z = z_{\rm osc}$, and $\vec{p}_{\rm min}$ is the minimum momentum of the lattice. This allows us to set fluctuations to the gauge fields in the following way: First, for a given lattice point in momentum space, we produce the Higgs fluctuations according to Eqs. (\ref{fluct1}) and (\ref{fluct3}). With these Higgs fluctuations, we obtain the correspondent fluctuations of $j_0 (z)$ and its corresponding Fourier transform $j_0 (k)$. Finally, we fix $V'_{i} (\vec{k}, z_{\rm osc} )$ according to Eq.~(\ref{eq:si-init}). We have then obtained a spectrum of initial gauge fluctuations.

However, in order for this procedure to be valid, we must ensure that our current $j_0 (k)$ does not possess a zero mode, i.e. $j_0 (\vec{k}=0) = 0$. This requirement can be clearly seen in Eq.~(\ref{eq:si-init}). This is equivalent to saying that there must not be a total electric charge in our lattice box. However, from the spectrum of Higgs fluctuations described above, we obtain
\bea j_0 (\vec{k} = 0 ) &=& \int d^3 \vec{z} j_0 (z) =   \label{zerocond} \\
&=&  \int d^3 \vec{k} \mathfrak{Re} [ h_1 (k)  h'_2 (k) -  h'_1 (k) h_2 (k)] \nonumber \eea
with
\bea & \mathfrak{Re} [  h_1 (k)  h'_2 (k) -  h_2 (k) h'_1 (k) ] = |h_1| |h_2|  q \beta^2 \times \nonumber \\
&  \cos{\left( \frac{\theta_3 + \theta_4 - \theta_1 - \theta_2}{2} \right)} \times \left[ \omega_2 \sin \left( \frac{\theta_3 - \theta_4}{2} \right) \cos \left( \frac{\theta_2 - \theta_1}{2} \right) \right. \nonumber \\
& \left. - \omega_1 \sin \left( \frac{\theta_1 - \theta_2}{2} \right) \cos \left( \frac{\theta_4 - \theta_3}{2} \right) \right] \ . \label{charge2} \eea
This quantity is not zero in general. There does not seem to be a particular reason why we should have a total electric charge in our box, so we should find a way of making Eq.~(\ref{charge2}) null. We have found two different ways of modifying slightly the initial quantum fluctuations of the Higgs field to make the integrand of Eq.~(\ref{charge2}) zero, which do not modify significantly the amplitude of the fluctuations with respect to the approach used in the global model. The first one is to impose, at each lattice point, the following constraint to the four arbitrary phases of the Higgs fluctuations
\be \theta_4 = \theta_1 + \theta_2 - \theta_3 + \pi \label{eq:phases} \ ,\ee
so that the phases $\theta_1$, $\theta_2$ and $\theta_3$ are randomly generated within the interval $\theta_i \in [0,2 \pi ) $, and $\theta_4$ is fixed through Eq.~(\ref{eq:phases}).
The second one is to leave the four phases totally random, but to perform, at each lattice point, the following shift to the Higgs fluctuations:
\be h_1' \rightarrow h_1' + \frac{J_0 h_2}{h_1^2 + h_2^2} \ ,  \hspace{0.3cm} h_2' \rightarrow h_2' - \frac{J_0 h_1}{h_1^2 + h_2^2} \ , \ee
where $J_0 \equiv (1 /N^3) \sum_{\hat{n}} (h'_2 h_1 - h'_1 h_2)$ is a sum over all lattice points. This shift eliminates by hand the zero mode of the current. One can easily confirm that both methods make zero the integrand of Eq.~(\ref{charge2}). In practice, we have confirmed that both methods produce almost identical results. This is normal, as in order to trust our lattice simulations, the way in which we set the initial fluctuations must not play any relevant role, as long as their amplitude does not significantly change.

\bibliography{HiggsReheating.bib}
\bibliographystyle{h-physrev4}

\end{document}